\documentclass[journal]{IEEEtran}

\usepackage{fixltx2e}

\usepackage{tikz}
\newcommand{\citep}[1]{\cite{#1}}

\usepackage{mathtools}
\usepackage{esvect}

\usetikzlibrary{fit}

\usepackage{wrapfig}
\usepackage[innercaption]{sidecap}
\usepackage{tikz,epstopdf,pgfplots,pdfpages}
\usepackage{framed}
\usepackage{marginnote}
\usepackage{blkarray}
\usepackage{enumerate}%
\usepackage{amsmath,amssymb,amsfonts}
\usepackage{enumitem}

\usepackage{mathtools}
\usetikzlibrary{decorations.shapes,shapes.geometric,shadows,arrows,automata,positioning,calendar,mindmap,backgrounds,scopes,chains,er,patterns,pgfplots.groupplots}
 \usetikzlibrary{calc}
\usepackage[nonumberlist]{glossaries}
\newglossary[slg]{symbolslist}{syi}{syg}{List of symbols}

\newtheorem{theorem}{Theorem}
\newtheorem{definition}{Definition}
\newtheorem{problem}{Problem}
\newtheorem{proposition}{Proposition}
\newtheorem{lemma}{Lemma}
\newtheorem{remark}{Remark}
\newtheorem{example}{Example}
\newtheorem{corollary}{Corollary}

\usepackage{xcolor}
\definecolor{lightblue}{rgb}{0.67, 0.9, 0.93}
\definecolor{lightgreen}{rgb}{0.67, 0.88, 0.69}
\definecolor{lightpink}{rgb}{1.0, 0.72, 0.77}
\definecolor{lightpurple}{rgb}{0.96, 0.73, 1.0}
\definecolor{lightyellow}{rgb}{0.98, 0.93, 0.37}

\usepackage{algorithm}
\usepackage{algpseudocode}

\newcommand{\word}{\boldsymbol{\omega}}
\newcommand{\wordt}[1]{{\omega}_{#1}}

\newcommand{\eps}{\epsilon}

\newcommand{\rel}{\mathcal{R}}

\newcommand{\Wt}{\mathbb{W}_{\Tr}}

\newcommand{\Lim}{\mathbf L}

\newcommand{\InF}{\mathcal{U}_{v}}

\newcommand{\ind}[2]{\mathbf{1}_{#1} \left(#2\right)}

\newcommand{\mathscr}[1]{\mathcal{#1}}
 
{} 

\newcommand{\po}{\mathbb{P}}     

\newcommand{\M}{\mathbf{M}}
\newcommand{\X}{\mathbb{X}}
\newcommand{\xs}{\mathbf{x}}
 				\newcommand{\x}[1]{{x}_{#1}} 
 				\newcommand{\xin}{x_0}
 				
 \newcommand{\xh}[1]{{\hat x}_{#1}} 
 \newcommand{\xinh}{\hat{x}_0}

\newcommand{\A}{\mathbb{U}}
   			\newcommand{\acs}[1]{\mathbf{u}}         
   			\newcommand{\ac}[1]{u_{#1}}         
			\newcommand{\acsh}[1]{\hat{\mathbf{u}}}         
   			\newcommand{\ach}[1]{\hat{u}_{#1}} 
\newcommand{\Y}{\mathbb{Y}}
		\newcommand{\y}[1]{{y}_{#1}}          
    	\newcommand{\ys}{\mathbf{y}}
\newcommand{\yh}[1]{{\hat y}_{#1}}          

\newcommand{\Tr}{\mathbf{t}}
\newcommand{\h}{h}
\newcommand{\Ca}{{\mathbf{C}}}
	\newcommand{\pcm}[2]{\po_{\scalebox{0.5}[.5]{$\!\!#1\!\!\times\!\!#2$}}}

\newcommand{\pol}[1]{\mu_{#1}}
\newcommand{\pols}{{\boldsymbol{\mu}}}

\newcommand{\Lab}{\mathsf{L}}   

\newcommand{\trans}{\tau}

\newcommand{\alphabeth}{\Sigma}

\newcommand{\AP}{\mathsf{AP}}            

\newcommand{\always}{\Box}
\newcommand{\eventually}{\Diamond}
\newcommand{\until}{\mathbin{\sf U}}

\newcommand{\nex}{\mathord{\bigcirc}}

\makeatletter
\newcommand\tabfill[1]{%
\dimen@\linewidth%
\advance\dimen@\@totalleftmargin%
\advance\dimen@-\dimen\@curtab%
\parbox[t]\dimen@{#1\ifhmode\strut\fi}%
}

\newcommand{\TheTitle}{Robust Dynamic Programming for Temporal Logic Control of Stochastic Systems}

\begin{document}
%
\title{\TheTitle}
%
%
%

\author{Sofie~Haesaert,~\IEEEmembership{Member,~IEEE,},
		Sadegh~Soudjani,~\IEEEmembership{Member,~IEEE}
		\thanks{S. Haesaert is with the Control Systems Group, Department of Electrical Engineering, Eindhoven University of Technology, The Netherlands. e-mail: S.Haesaert@tue.nl}
		\thanks{S. Soudjani is with the School of Computing, Newcastle University, United Kingdom. e-mail: Sadegh.Soudjani@ncl.ac.uk}
		}

\maketitle

\begin{abstract}
Discrete-time stochastic systems are an essential modelling tool for many engineering systems. We consider stochastic control systems that are evolving over continuous spaces. For this class of models, methods for the formal verification and synthesis of control strategies are computationally hard and generally rely on the use of approximate abstractions. Building on approximate abstractions, we compute control strategies with lower- and upper-bounds for satisfying unbounded temporal logic specifications. Firstly, robust dynamic programming mappings over the abstract system are introduced to solve the control synthesis and verification problem. These mappings yield a control strategy and a unique lower bound on the satisfaction probability for temporal logic specifications that is robust to the incurred approximation errors. Secondly, upper-bounds on the satisfaction probability are quantified, and properties of the mappings are analysed and discussed. Finally, we show the implications of these results for linear stochastic dynamic systems with a continuous state space. This abstraction-based synthesis framework is shown to be able to handle infinite-horizon properties. Approximation errors expressed as deviations in the outputs of the models and as deviations in the probabilistic transitions are allowed and are quantified using approximate stochastic simulation relations.  
\end{abstract}


%
\IEEEpeerreviewmaketitle

\section{Introduction}
\IEEEPARstart{T}{here} is an ever more ubiquitous embedding of digital components into physical systems. The scale of this embedding is currently creating the need 
for
new computationally efficient methods that assist in their \emph{verifiable (control) design}. 
An example of this is the digitalisation of biological processes, power networks, and smart housing. 
These applications are often safety-critical and cannot tolerate design errors. 
Generally, the needed verifiable design can be achieved by using formal methods that build on formal specifications such as those formulated in \emph{temporal logic} \cite{clarke1996formal,fainekos2006translating}. 
For stochastic systems, there is a lack of methods that assist with their verifiable design and that
work with  \emph{uncountable state spaces}. Still, physical systems in relevant application domains, whose variables evolve over continuous spaces, are inherently stochastic.

In this work, we are interested in the verified design of control strategies  for (unbounded) probabilistic linear temporal logic properties.
Such properties, defined over finite-state Markov processes,
can be verified using tools such as PRISM~\citep{KNP11} or MRMC~\citep{KATOEN201190}. These tools are also able to perform policy synthesis for controlling finite-state Markov decision processes (MDPs) such that the satisfaction probability of temporal properties is maximised.
For discrete-time stochastic models over uncountable state spaces, the computation of the satisfaction probability
can in general not be performed analytically \citep{Abate1}; thus the use of approximation techniques is inevitable.
A well-established approximation approach is to \emph{abstract} these models and replace them by simpler processes, such as finite-state MDPs \citep{SAID} or continuous-space reduced order models \citep{safonov1989schur}, that are prone to be mathematically analysed or algorithmically verified \citep{FAUST13}.

The use of abstractions for formal verification was introduced first for discrete-time stochastic models over countable state spaces \cite{larsen1991bisimulation}.  
For discrete-time stochastic linear control systems, \cite{pola2018bisimulation} leverages  geometrical conditions for the quantification of exact equivalence of stochastic dynamical systems. 

Allowing for probabilistic deviations, \cite{Blute1997,desharnais2004metrics} extend \cite{larsen1991bisimulation} to labelled Markov decision processes with possible continuous states. Finite abstractions for discrete-time stochastic models with continuous spaces are employed in \citep{Abate1}. Scalability of the abstraction algorithms has been improved in \citep{SAID,SSoudjani} and extended to partially observed models \citep{LO17}.

The use of metric distances between execution trajectories of stochastic systems was first introduced in
the paper \cite{Julius2009a}, which proposed approximate simulation functions to quantify the probability of the exceeding a maximum distance between trajectories of two stochastic systems.
More recently, the work in \cite{LSMZ17} has given a related notion of simulation functions that is similar to a super-martingale property. 
The existence of such simulation functions can be checked via matrix inequalities for particular classes of systems, but there is no guarantee for finding such a function. Though the use of simulation functions is generally applicable \cite{LSMZ17}, for general Markov decision processes with additive noise, it can only bound deviations of finite horizon properties.

As an alternative to the abstraction-based techniques, there are results \cite{svorevnova2017temporal,Vinod17,Vinod18} that directly approximate the solution of the verification and synthesis problems.  
The analysis in \cite{svorevnova2017temporal} is devoted to stochastic linear systems and is limited to almost sure satisfaction of a property, i.e., it verifies whether specifications are satisfied with probability one. 
The papers \cite{Vinod17,Vinod18} provide approaches for under-approximating verification that is applicable only to linear systems using respectively Fourier transform and polytopic representations.


Quantifying the abstraction errors in satisfaction probability is a challenging problem for \emph{infinite-horizon} specifications over continuous-space models \citep{tmka2013,tkachev2011infinite,TKACHEV20171}. Such a task requires knowledge on structural properties of the stochastic system, i.e., lack of absorbing sets, which are difficult to establish in general. Alternatively, martingale properties of the system can be exploited via the notion of barrier certificates \citep{huang2017probabilistic,4287147}. This approach suffers from the lack of guarantees on the existence of such barrier certificates while being restricted to a subset of temporal properties.

The lack of abstraction-based synthesis frameworks in the literature  that can seemingly address infinite-horizon properties for a wider set of stochastic models motivates this work.
We have recently proposed in \citep{DBLP:conf/qest/HaesaertAH16,haesaert2017verification} a new notion of approximate stochastic similarity relations. This notion has two precision parameters ($\epsilon,\delta$) that bounds the deviations between models in both the output signals ($\epsilon)$ and the transition probabilities ($\delta$).
For approximately similar models, a control policy synthesised on an abstract model can be refined to an approximately similar model with quantified precision.
Up to now, this can only be practically applied to temporal logic properties over bounded time, as it generally holds that the deviation in transition probability ($\delta$) induces a decrease in the satisfaction probability that increases with the time horizon.

In this work, we develop an approach to synthesise and verify control strategies for a larger set of temporal properties known as \emph{syntactically co-safe linear temporal logic} (scLTL) specifications \citep{KupfermanVardi2001}, which can be unbounded in time. To deal with  unbounded time properties, we define \emph{dynamic programming mappings} that are robustified to the introduced deviations ($\epsilon,\delta$). 
These mappings, first introduced in \cite{DBLP:conf/adhs/HaesaertSA18},  are proven to converge to a robust lower bound on the satisfaction probability. 
Next to the robust dynamic programming mappings, we also give their dual, optimistic mappings that allow for computing an upper bound on the satisfaction probability.\\
Finally, for the specific case of linear stochastic dynamical systems, we develop a discretisation of the continuous state space that can be efficiently solved with linear matrix inequalities.
This paper extends the preliminary work in \cite{DBLP:conf/adhs/HaesaertSA18} with previously omitted proofs and with more detailed analysis of the dynamics programming mappings and uniqueness.
%
%
%

The paper is organised as follows. Sec.~\ref{sec:Model_Spec} gives the problem statement by defining general Markov decision processes, control strategies, and the class of scLTL properties.
In Sec.~\ref{sec:relation}, we define approximate simulation relations that assess similarities between two stochastic models using their joint probabilistic evolution in their coupled spaces. Sec.~\ref{sec:prod} gives a characterisation of the satisfaction probability of scLTL properties and how similarity relations can help us connect this quantity computed over approximate models. We present the core contribution of this paper in Sec.~\ref{sec:fullcase}.  This includes a robust synthesis approach for the satisfaction of properties using similar models.
Finally, we detail in Sec.~\ref{sec:LTI} the synthesis procedure for linear stochastic dynamical systems and describe its application to case studies.



%

%
%
%
\section{Stochastic models and temporal logic}
\label{sec:Model_Spec}

We study the class of general Markov decision processes (gMDPs) featuring a non-deterministic evolution with \emph{uncertainties} modelled by probability distributions and with \emph{actions} to be synthesised.
These processes extend upon Markov decision processes \citep{Bible} by having an output map that generates an output sequence, over which the \emph{desired properties} are defined.
The actions are generated by \emph{control strategies} which are themselves gMDPs that receive the state of the gMDP and compute the actions using their internal states. We formally define gMDPs and control strategies in Subsec.~\ref{subsec:models} and specify the class of properties in Subsec.~\ref{subsec:spec}.

\subsection{Models: general Markov decision processes}
\label{subsec:models}
Denote a Borel measurable space as $(\X,\mathcal{B}(\X))$. A probability measure $\po$ over this space defines the probability space  $(\X,\mathcal{B}(\X),\po)$ and has realisations  $x\sim \po$.  We assume all such spaces $\X$ are Polish \citep{bogachev2007measure} since Polish spaces are closed under taking \emph{countable products}. Additionally, this yields 
well-defined measurable events over unbounded executions. 

\smallskip%
\begin{definition}[general Markov decision process (gMDP)]
	\label{def:MDP}
	A discrete-time gMDP is a tuple $\M=(\X,\A,\Y,x_0,\Tr,\h )$ with
	\begin{itemize}
		\item $\X$,  a Polish state space with states $x\in\X$; 
		 \item $\A$,  the set of inputs, which is a Polish  space;
 \item $\Y$, the output space decorated with metric $\mathbf d_\Y$;
 \item  $x_0\in\X$, the initial state;
		 \item $\Tr:\X\times\A\times\mathcal B(\X)\rightarrow[0,1]$, a conditional stochastic kernel assigning to each state $x\in \X$ and control $u\in \A$  a probability measure $ \Tr(\cdot\mid x,u)$ over $(\X,\mathcal B(\X))$; and
	 \item $\h:\X\rightarrow\Y$, a measurable output map.
 \end{itemize}
\end{definition}

\smallskip%
We denote the class of all gMDPs with the same metric output space $\Y$ as $\mathcal{M}_\Y$.
We indicate the input sequence of the gMDP $\M$ by $\acs:= \ac{0},\ac{1},\ac{2},\ldots$
 and we define its \emph{executions} as sequences of states $\xs=\x{0},\x{1},\x{2},\ldots$ initialised with the initial state  $\xin$ of $\M$ at $t=0$.
In an execution, each consecutive state, $x_{t+1}\in\X$,
is obtained as a realisation $x_{t+1}\sim\Tr\left(\cdot\mid x_t, u_t \right)$ of the controlled Borel-measurable stochastic kernel. 
By applying the output map $h(\cdot)$ over states in the execution, the associated \emph{output trace}
$\ys:=\y{0},\y{1},\y{2},\ldots$
with $\y{t} = \h\big(\x{t}\big)$ is obtained.

\begin{figure}[htp]\centering
 \includegraphics[width=.8\columnwidth, trim={5cm 12cm 40cm 13cm}, clip]{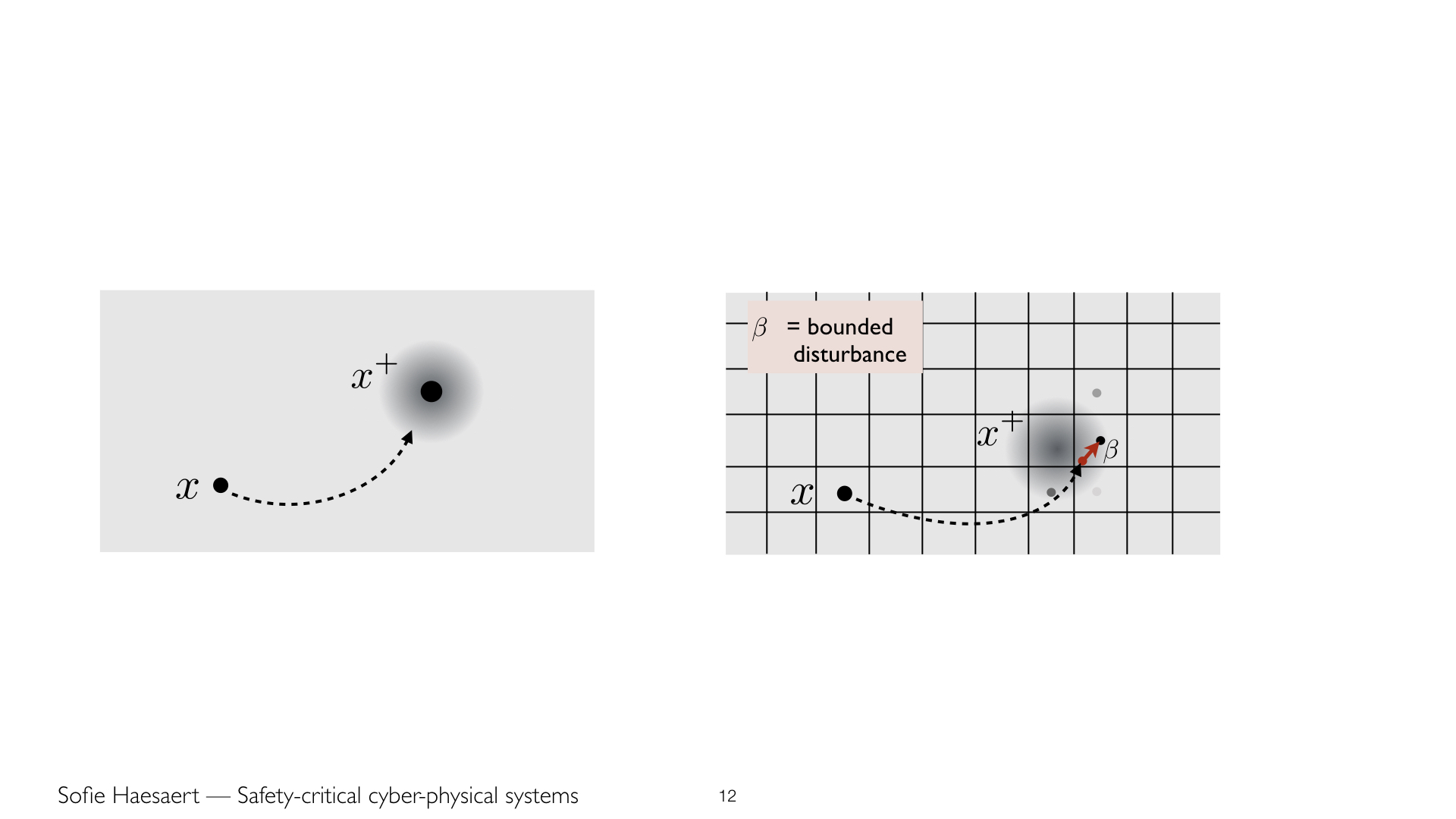}
		\caption{A probabilistic transition of the difference equation \eqref{eq:difference_equation} over two-dimensional Euclidean space: $x^+$ has Gaussian distribution with mean $f(x,u)$ and covariance matrix $\Sigma$.}
		\label{fig:stoch_tr}
	\end{figure}

\smallskip%
\begin{example}[Stochastic difference equations]
\label{ex:SDE}
	An example of a gMDP is a process with state space $\X = \mathbb R^n$ that is characterised by the stochastic difference equation
	\begin{equation}
	\label{eq:difference_equation}
	\begin{aligned}
		\x{t+1}&= f(\x{t},\ac{t})+w_{t}, \\
	     \y{t}&= h(\x{t}), \qquad \forall t\in\{0,1,2,\ldots\}=:\mathbb N.
	     \end{aligned}
	     \end{equation}
	     with functions $f:\X\times\A\rightarrow\X$ and $h:\X\rightarrow\Y$.
	    The stochastic disturbance $\{w_{t},\,t\in\mathbb N\}$ is a stationary process with $w_{t}$ having multivariate Gaussian distribution, i.e., $w_{t}\sim\mathcal{N}(0,\Sigma)$, and with $w_{t}$ and $w_{t'}$ independent for all $t\ne t'$.
	    This model can be written as a gMDP with
	    stochastic transition kernel
	    $ \Tr(d \x{t+1}|\x{t},\ac{t}) = \mathcal N(d \x{t+1}\mid  f(\x{t},\ac{t}), \Sigma)$.
	    A probabilistic transition of the state is illustrated in Fig.~\ref{fig:stoch_tr}.
\end{example}

\smallskip%

Denote the set of probability measures on the measurable space $(\A,\mathcal{B}(\A))$ as
$\mathcal P (\A,\mathcal{B}(\A)).$
When the control inputs are selected only based on the current states,
this is referred to as a Markov policy.

\smallskip%
\begin{definition}
A \emph{Markov policy} $\pols$ is a sequence $\pols=(\pol{0},\pol{1},\pol{2},\ldots)$ of universally measurable maps $\pol{t}:\X\rightarrow \mathcal P(\A,\mathcal B(\A))$, $t\in\mathbb N$, from the state space $\X$ to the set of controls.
	A \emph{Markov policy} $\pols$ is \emph{stationary} or time homogeneous  if  $\pols=(\pol{},\pol{},\pol{},\ldots)$ for some $\pol{}$.
	\end{definition}
\smallskip%

A more general set of control strategies are those that depend on the past history of states and controls.
We consider control strategies that depend on the past history via a memory state.
\smallskip%

\begin{definition}[Control strategy]
	\label{def:CS}
	A control strategy \[\Ca=(\X_\Ca,\A_\Ca,\Y_{\Ca},x_{{\Ca} 0},\Tr_\Ca,\h_\Ca)\]
	for a gMDP $\M=(\X,\A,\Y,\Tr,\h)$ is itself a gMDP  with 
\begin{itemize}
	\item $\X_\Ca$, the state space with states $x_{\Ca}\in \X_{\Ca}$ that act as the memory of the controller\,;
	\item $\A_\Ca:=\X$, the input space with the states of $\M$ as its elements;
	\item $\Y_\Ca:=\mathcal P(\A,\mathcal B(\A))$, the output space with probability distributions over the input space of $\M$ as its elements;
	\item $x_{\Ca 0}$, the initial state at $t=0$;
	\item $\Tr_\Ca$, the universally measurable kernel $\Tr_{\Ca}:\X_\Ca\times\A_\Ca\times\mathcal B(\X_\Ca)\rightarrow[0,1]$; and
	\item $\h_\Ca$, the universally measurable output map $\h_\Ca:\X_\Ca\rightarrow\Y_\Ca$.
\end{itemize}
\end{definition}
\smallskip%

The control strategy $\Ca$ is formulated as a gMDP that takes as its input the state of the to-be-controlled gMDP $\M$ and outputs probability measures on the input space $\A$. 
\begin{figure}[htp]
\centering
\resizebox{!}{2.5cm}{\begin{tikzpicture}[ node distance=1.6cm]
			\tikzset{cstate/.style={diamond,draw, fill=yellow!80!gray!40, inner sep=.1cm}}
			\tikzset{state/.style={circle, draw, fill=blue!40, inner sep=.1cm}}
			\node[cstate,label=left:{$x_{\Ca 0}$}] (cx0) {};
			\node[state,  below of = cx0, yshift=0.4cm] (x0){};

			\node[ left of =x0,xshift=.75cm] (x0pi) {$x_0$};
			\node[left of=x0pi, xshift=.2em, node distance=.75cm] (M_text) {$\M$};
			\node[above of=M_text, yshift=-0.4cm] (C_text) {$\Ca$};

			\node[cstate, right of= cx0] (cx1) {};
			\node[state,  right of = x0] (x1) {};

			\node[cstate, right of= cx1] (cx2){};
			\node[state,  right of = x1] (x2){};
			\node[cstate, right of= cx2] (cx3){};
			\node[  right of = x2, node distance=.75cm] (x3){$\ldots$};

			\path[->,draw] (x0pi)->(x0);
			\path[->,draw] (x0)edge node [below] {\small $x_0$} (x1);
			\path[->,draw] (x1)edge node [below] {\small$x_1$} (x2);

			\path[->,draw] (cx0) edge node [above] {\small$x_{\Ca,0}$} (cx1);
			\path[->,draw] (cx1)edge node [above] {\small$x_{\Ca,1}$} (cx2);
			\path[->,draw] (cx2)edge node [above] {\small$x_{\Ca,2}$} (cx3);
			\path[->,draw] (x0)edge node [state, fill=white ,draw=white, inner sep=0] {\small$x_0$} (cx1);
			\path[->,draw ] (x1)edge node [state, fill=white ,draw=white, inner sep=0] {\small $x_1$ }(cx2);
			\path[->,draw ] (x2)edge node [state, fill=white ,draw=white, inner sep=0] {\small $x_2$ }(cx3);
			\path[->,draw ] (cx1) edge node[state, fill=white ,draw=white, inner sep=0]{\small $u_0$} (x1);
			\path[->,draw ] (cx2) edge node[state, fill=white ,draw=white, inner sep=0]{\small $u_1$} (x2);
			\path[draw, ->] (x2)->(x3);
			\node[below of =x0,node distance=.8cm](y0){\small  $y_0$};
			\node[below of =x1, node distance=.8cm](y1){\small  $y_1$};
			\node[below of =x2,node distance=.8cm](y2){\small  $y_2$};
			\path[draw, ->] (x0)->(y0);
			\path[draw, ->] (x1)->(y1);
			\path[draw, ->] (x2)->(y2);
			\end{tikzpicture}}\\[-.3cm]
			\caption{ Execution semantics of the controlled gMDP  $\Ca\times \M$. }
			\label{fig:CM}
\end{figure}
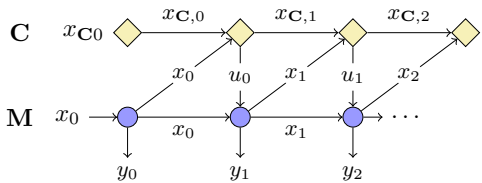
Consider an execution given in Fig.~\ref{fig:CM}. The sequence of states is obtained by drawing state transitions
 from the stochastic transition kernels $\Tr$ and $\Tr_\Ca$ and by drawing realisations from the output distribution of $\Ca$.
For a time horizon $N\in\mathbb N\cup\{\infty\}$, the execution \mbox{$\{(x_t,x_{\Ca,t})\}_{t\leq N}$} of a gMDP $\M$ controlled with strategy $\Ca$ (denoted by $\Ca\times \M$)
is defined on the canonical sample space $\Omega:= (\X\times\X_\Ca)^{N+1}$ endowed with its product topology $\mathcal B (\Omega)$ and with  a
unique probability measure $\pcm{\Ca}{\M}$ over the space of output traces.

\subsection{Specifications: probabilistic temporal logic for gMDPs}
\label{subsec:spec}

Consider a finite set of atomic propositions $\AP$ that defines the alphabet $\alphabeth := 2^{\AP}$. Thus, each letter $\alpha\in \alphabeth$ of this alphabet evaluates a subset of the atomic propositions as true. Composed as an infinite string, these letters form infinite words defined as
$\word=\wordt{0},\wordt{1},\wordt{2},\ldots\in\alphabeth^{\mathbb{N}}$.
These words are connected to output traces of gMDPs via
 a  measurable labelling function $\Lab:\mathbb Y\rightarrow \alphabeth$ that   assigns letters $\alpha =\Lab(y)$ to outputs $y\in\Y$.
That is, output traces $\ys\in \Y^{\mathbb N} $ are mapped to the set of infinite words $\alphabeth^{\mathbb N}$, as
$\word=\Lab(\ys) := \{\Lab(\y{t})\}_{t\geq0}$.\smallskip\\ 
\noindent{\bfseries Syntactically co-safe linear-time temporal logic.}
In this paper, we consider properties defined in a fragment of  linear-time temporal logic (LTL)  known as syntactically co-safe temporal logic (scLTL) \citep{KupfermanVardi2001}.
\smallskip%

\begin{definition}
	\label{def:scLTL}
	an scLTL formula over a set of atomic propositions $\AP$ has syntax
	\begin{equation*}
	\label{eq:scLTL}
	\psi ::=  p \,|\, \neg p \,|\,\psi_1 \wedge \psi_2\,|\,\psi_1 \vee \psi_2 \,|\, \nex \psi \,|\, \psi_1\until \psi_2
	\end{equation*}
	with $p\in \AP$.
	\end{definition}
\smallskip%

Let ${\word}_t=\wordt{t},\wordt{t+1},\wordt{t+2},\ldots  $ be a postfix of the word $ \word $, then
the satisfaction relation between $\word$ and a property $\psi$ is denoted by $ \word \vDash\psi$
(or equivalently $\word_0\vDash\psi$).
The semantics of the satisfaction relation are defined recursively over $\word_t$ as follows
\begin{itemize}
	\item An atomic proposition, $ \word_t\vDash   p$  for $ p\in \AP$ holds if $p \in\wordt{t}$.
	\item A negation,  $ \word_t\vDash \neg p$   holds if $ \word_t\nvDash p$.
\item A logical conjunction, $\word_t\vDash \psi_1\wedge\psi_2$ holds
if $ \word_t\vDash \psi_1$ and if $ \word_t\vDash \psi_2$.
\item The logical disjunction, $ \word_t\vDash \psi_1\vee\psi_2$  holds if $ \word_t\vDash \psi_1$ or $ \word_t\vDash \psi_2$.
\item A temporal next operator $\word_t\vDash\nex\psi $ holds if 
  $ \word_{t+1}\vDash \psi$.
\item A temporal until operator $\word_t\vDash \psi_1\until\psi_2$  holds if there exists an $i \in \mathbb{N}$ such that $\word_{t+i} \vDash \psi_2$, and for all
$j \in\mathbb{N}$, $0\leq j<i$, we have $\word_{t+j}\vDash \psi_1$.
\end{itemize}

In addition to aforementioned operators, we can also use
a temporal \emph{eventually} operator defined as $\lozenge\psi:=\left(\operatorname{true} \until \psi\right)$
and the time-bounded eventually or \emph{reachability} of $\psi$ as $\lozenge^N \psi$.
\smallskip%

\noindent{\bfseries Problem statement: Temporal logic control}.
\begin{problem}
\label{prob:TLC}
Given an scLTL property $\psi$ and probability $p\in[0,1]$, either 
 	find  a control strategy $\Ca$ for the gMDP $\M$ such that $\Ca\times \M$ satisfies $\psi$ with probability at least $p$:
\begin{equation}
\label{eq:sat_prob}
\pcm{\Ca}{\M} (\word\models\psi)\geq p\,;
\end{equation}
  or prove that a strategy $\Ca$ achieving \eqref{eq:sat_prob} does not exist.
\end{problem}
\smallskip%

%
%
A solution to Prob.~\ref{prob:TLC} can be characterised by quantifying the maximal value for the left-hand side probability of \eqref{eq:sat_prob}, $\pcm{\Ca}{\M}(\word\models\psi)$,
with respect to $\Ca$.
This temporal logic control problem is \emph{computationally} hard to solve, therefore it is generally split up into two sub-problems:
\begin{enumerate}[label=(\alph*)]
	\item
	\label{subproblem1}
	For a given concrete model $\M$, find an abstract model $\widehat \M$ with quantified deviations between the two models.
\item
\label{subproblem2}
Compute a control strategy $\hat\Ca$ over the abstract model $\widehat \M$ such that it can be refined to a strategy $\Ca$ for $\M$ while giving a guaranteed lower- and/or upper-bound  on the satisfaction probability $\pcm{\Ca}{\M} (\word\models\psi)$ in \eqref{eq:sat_prob}.
\end{enumerate}
%
%
%
In this paper, we provide a sound solution for Prob.~\ref{prob:TLC}. For this, we introduce an approximate simulation relation in the next section to quantify the deviation between the two models $\M$ and $\widehat \M$ in sub-problem~\ref{subproblem1}. This relation is founded on the notion of  $\delta$-lifting \citep{haesaert2017verification}. We characterise the satisfaction probability of scLTL properties in Sec.~\ref{sec:prod}, which paves the way to solving sub-problem~\ref{subproblem2} in Sec.~\ref{sec:fullcase}.

\section{Approximate simulation relations for gMDPs}
\label{sec:relation}

Consider a gMDP  $\M=(\X,\A,\Y,x_0,\Tr,\h)$  and its abstraction  $\widehat \M = (\hat\X,\hat\A,\Y,\hat x_0,\hat\Tr,\hat\h)$
 that shares with $\M$ the same output space $\Y$ with metric $\mathbf d_\Y$.
 Let
$\rel\subseteq \hat\X\times \X$ be a relation defined over the state spaces of the two models, and let this relation be measurable, i.e., $\rel\in \mathcal B(\hat\X\times \X).$
If a pair of states $(\hat x ,x)$ is in $\rel$, then we denote this as $\hat x\rel x$.
The relation $\rel$ can be extended to the pairs of transition kernels $\Tr(\cdot{\mid} x,u)$ 
and $\hat\Tr(\cdot{\mid} \hat x, \hat u)$
using the notion of $\delta$-lifting \citep{haesaert2017verification} and approximate stochastic simulation relations, as follows.

\smallskip%
\begin{definition}[$\delta$-lifting of $\rel$]
\label{def:del_lifting}
	Consider two measurable spaces $(\hat\X,\mathcal B(\hat\X))$ and $(\X,\mathcal B(\X))$, and a relation $\rel\in \mathcal B(\hat\X\times \X)$. The $\delta$-lifted relation $\bar\rel_\delta$ is defined as follows. Pairs of probability distributions $(\Delta, \Theta)$  belong to $\bar\rel_\delta$ 
	 if there exists a lifting $\mathbb W$ with probability space $(\hat\X\times \X,\mathcal B(\hat\X\times \X), \mathbb W)$  such that{ \setlength{\parskip}{1pt}\setlength{\parsep}{1pt}
		\begin{description}
			\item[\normalfont\textbf{L1.}] $\mathbb W(\hat A\times \X)=\Delta(\hat A)$ for all $\hat A\in \mathcal{B}(\hat\X)$;
			\item [\normalfont\textbf{L2.}] $\mathbb W(\hat \X\times A)=\Theta(A)$ for all $A\in \mathcal{B}(\X)$;
			\item[\normalfont\textbf{L3.}] 
			 $\mathbb{W}\left(\rel\right)\geq1-\delta$.
	\end{description}}%
We denote $\Delta \bar \rel_\delta \Theta$ if the pair of probability distribution belongs to the lifted relation: $(\Delta,  \Theta)\in\bar\rel_\delta$.
\end{definition}
Remark that $\delta$ quantifies the amount of the probability distribution that cannot be coupled into the relation $\rel$.
\smallskip%

Consider the notion of \emph{interface function} \citep{Girard2009} that refines control actions as follows
\begin{equation}
\label{eq:InF}
\InF: \hat \A\times \hat \X \times\X  \rightarrow \mathcal{P}(\A,\mathcal B(\A)).
\end{equation}
Intuitively, an interface function implements (or refines) any control action $\hat u$ synthesised over the abstract model $\widehat\M$ to an action for the concrete model $\M$.
This interface is used together with $\delta$-lifting to define an approximate stochastic simulation relation on gMDPs as defined next.
%
\smallskip%

\begin{definition}[$(\eps,\delta)$-stochastic simulation relation]
\label{def:apbsim}
	Let  $\M=(\X,\A,\Y,x_0,\Tr,\h)$ and  $\widehat \M =(\hat\X,\hat\A,\Y,\hat x_0,\hat \Tr,\hat\h)$ be two gMDPs in  $\M_\Y$ with metric $\mathbf{d}_\Y$.
	If there exist 
	\begin{itemize}
		\item an interface function $\InF$ as in \eqref{eq:InF},
		\item a measurable relation $\rel\subseteq \hat\X\times \X$, and
		\item a Borel measurable stochastic kernel $\Wt:  \hat \A\times\hat \X\times\X \rightarrow \mathcal P (\hat\X\times \X,\mathcal B(\hat\X\times \X))$ for the $\delta$-lifting,  
	\end{itemize}
	such that
	\begin{equation}(\hat x_0,x_0)\in\rel, \tag{\normalfont\textbf{A1.}}\end{equation}
	and for all $(\hat x,x)\in \rel$,
	{\setlength{\parskip}{0pt}\setlength{\parskip}{0pt}
		\begin{description}
			\item[\normalfont\textbf{A2.}] $ \mathbf{d}_\Y\left(\hat y,y\right)\leq \epsilon$  with $y = h(x)$ and $\hat y = \hat h(\hat x)$, and

			\item[\normalfont\textbf{A3.}]
			\(\Tr(\cdot| \hat x, \hat u)\ \bar \rel_\delta \  \Tr(\cdot| x, \InF(\hat u,\hat x,x))\) for all $\hat u\in\hat \A$
			 with the lifted kernel $\Wt(\,\cdot\,{\mid}\hat u,\hat x,x)$.
	\end{description}}
Then
	$\widehat\M$ is $(\epsilon,\delta)$-stochastically simulated by $\M$, and this simulation relation is denoted as $\widehat\M\preceq^{\delta}_\eps\M$.
\end{definition}\smallskip%

In the sequel, we will often use an \emph{auxiliary model} that couples the behaviour of the two models $\M,\widehat\M$, as defined next.\smallskip%

\begin{definition}[Coupling similar models]
\label{def:coupling}
	Suppose gMDPs $\M,\widehat\M\in\M_\Y$ are given for which $\widehat \M\preceq_\eps^\delta \M$ as defined in Def.~\ref{def:apbsim} with respect to the relation $\rel$, interface function $\InF$, and lifted kernel $\Wt$. We define the coupling gMDP
	\begin{equation*}
	\widehat\M ||_{\rel}\M:=(\X_{||}, \A_{||},\Y_{||},x_{0||},\Tr_{||}, \h_{||})
	\end{equation*}
	with
	\begin{itemize}
 \item $\X_{||}:=\hat \X\times \X$ the coupling state space;
 \item $\A_{||} := \hat\A$, the input of the abstract system $\widehat\M$;
 \item $\Y_{||} := \Y$, the common output space;
 \item $x_{0||}\!\!:=\!(\hat x_0, x_0)\!\!\in\! \X_{||}$ the pair of initial states of $\widehat \M$ and $\M$;
 \item $\Tr_{||}:=\Wt$, the stochastic transition kernel $\Wt$;
 \item $\h_{||}(\hat x,x):=\h(x)$, the output mapping of $\M$.
 \end{itemize}
	\end{definition}\smallskip%

The transitions between states of the coupling gMDP $\widehat\M ||_{\rel}\M$ are stochastically distributed according to the lifted kernel $\Wt$ specified in Def.~\ref{def:apbsim}.

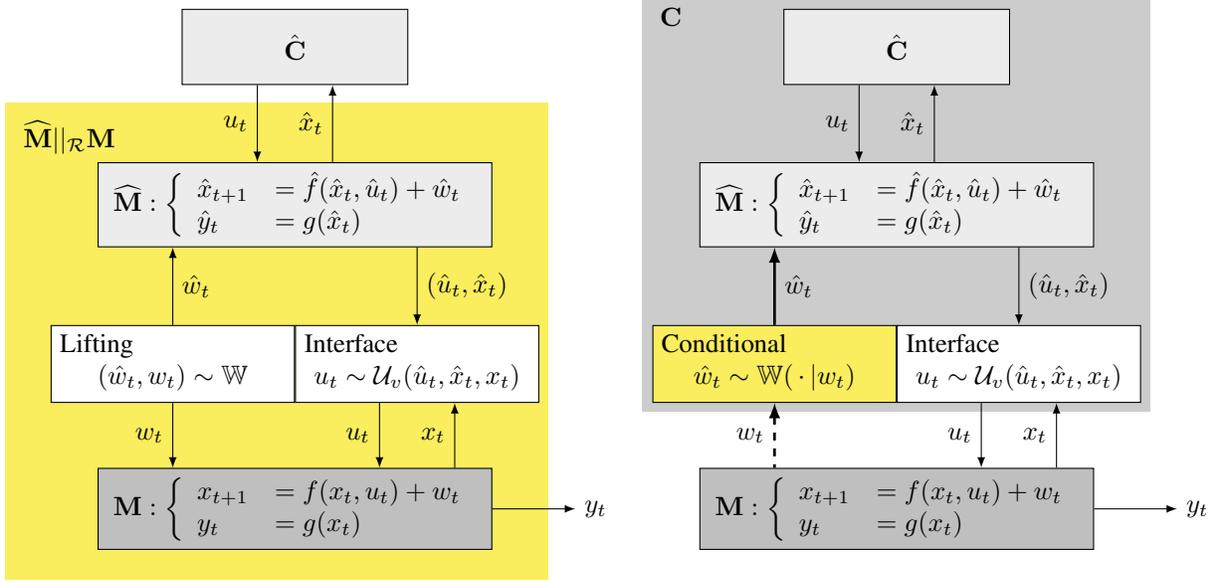
\begin{figure*}[htp]
 \centering
 \begin{tikzpicture}
 \node[rectangle, fill=gray!15,draw, anchor=south] (Mhat) {\begin{minipage}[c]{5cm}
	\begin{align*}\SwapAboveDisplaySkip
		\widehat \M:\left\{\begin{array}{ll}
 {\xh{t+1}}&= \hat f( {\xh{t}}, {\ach{t}})+\hat w_t\\
 {\yh{t}}&=g( {\xh{t}})
\end{array}
\right.
	\end{align*}
\end{minipage}
};
 \node[below of =Mhat ,draw,rectangle, anchor=north,fill=gray!50, node distance=3.5cm] (M) {\begin{minipage}[c]{5cm}
	\begin{align*}\SwapAboveDisplaySkip
	 \M:\left\{\begin{array}{ll}
\x{t+1}&= f(\x{t},\ac{t})+w_t\\
\y{t}&=g(\x{t})
\end{array}
\right.
	\end{align*}
\end{minipage}
};

 \node[below of =Mhat,draw ,rectangle, anchor=north west,fill=white, node distance=1.6cm] (if) {\begin{minipage}[c]{3cm}
 Interface
	\begin{align*}\SwapAboveDisplaySkip
	 \ac{t}\sim\InF(\ach{t}, \xh{t},\x{t})
	\end{align*}
\end{minipage}
};
 \node[below of =Mhat, draw,rectangle, anchor=north east,fill=white, node distance=1.6cm] (lif) {\begin{minipage}[c]{3cm}
 Lifting
	\begin{align*}\SwapAboveDisplaySkip
	 (\hat w_{t},w_t)\sim\mathbb W	\end{align*}
\end{minipage}
};
\node[right of =M, node distance= 4cm](out){$\y{t}$};
\path[draw] (if.north) edge[latex-,right] node{$(\ach{t},\xh{t})$}(if.north |- Mhat.south);
\path[draw] (lif.north) edge[-latex,right] node{$\hat w_t$}(lif.north |- Mhat.south);

\path[draw] (lif.south) edge[-latex,left] node{$ w_t$} (lif.south |- M.north);
\path[draw] (if.south) ++(0.5cm,0)  edge[latex-,left] node{$ \x{t}$} ([xshift=0.5cm]if.south |- M.north);
\path[draw] (if.south) ++(-0.5cm,0)  edge[-latex,left] node{$\ac{t}$} ([xshift=-0.5cm]if.south |- M.north);
\path[draw] (M) edge[-latex] (out);
\node[left of =Mhat, node distance=3cm,yshift=.9cm](MM) {$\widehat \M||_{\rel} \M$};
\node[right of =M, node distance=3cm,yshift=-.7cm](MM2) {};
\node[above of =Mhat,draw, fill=gray!15,rectangle, minimum width=3cm,node distance=2.1cm, minimum height=1cm] (Ca) {$\hat\Ca$};
\path[draw] (Ca.south) ++(-0.5cm,0)  edge[-latex,left] node{$ \ac{t}$} ([xshift=-0.5cm]Ca.south |- Mhat.north);
\path[draw] (Ca.south) ++(0.5cm,0)  edge[latex-,left] node{$ \xh{t}$} ([xshift=0.5cm]Ca.south |- Mhat.north);
\begin{scope}[on background layer]

\node[ fill=lightyellow, fit=(M) (MM) (MM2)(Mhat)(if)(lif)](FIt1) {};
\end{scope}

 \node[rectangle, right of= Mhat, node distance=8cm,fill=gray!15,draw] (Mhat2) {\begin{minipage}[c]{5cm}
	\begin{align*}\SwapAboveDisplaySkip
		\widehat \M:\left\{\begin{array}{ll}
 {\xh{t+1}}&= \hat f( {\xh{t}}, {\ach{t}})+\hat w_t\\
 {\yh{t}}&=g( {\xh{t}})
\end{array}
\right.
	\end{align*}
\end{minipage}
};
 \node[below of =Mhat2 ,draw,rectangle, anchor=north,fill=gray!50, node distance=3.5cm] (M2) {\begin{minipage}[c]{5cm}
	\begin{align*}\SwapAboveDisplaySkip
	 \M:\left\{\begin{array}{ll}
\x{t+1}&= f(\x{t},\ac{t})+w_t\\
\y{t}&=g(\x{t})
\end{array}
\right.
	\end{align*}
\end{minipage}
};

 \node[below of =Mhat2,draw ,rectangle, anchor=north west,fill=white, node distance=1.6cm] (if2) {\begin{minipage}[c]{3cm}
 Interface
	\begin{align*}\SwapAboveDisplaySkip
	 \ac{t}\sim\InF(\ach{t}, \xh{t},\x{t})
	\end{align*}
\end{minipage}
};
 \node[below of =Mhat2, draw,rectangle, anchor=north east,fill=lightyellow, node distance=1.6cm] (lif2) {\begin{minipage}[c]{3cm}
 Conditional
	\begin{align*}\SwapAboveDisplaySkip
	 \hat w_{t} \sim\mathbb W(\,\cdot\,|w_{t})	\end{align*}
\end{minipage}
};
\path[draw] (if2.north) edge[latex-,right] node{$(\ach{t},\xh{t})$}(if2.north |- Mhat2.south);
\path[draw, line width=1pt]  (lif2.north) edge[-latex,right] node{$\hat w_t$}(lif2.north |- Mhat2.south);
\node[right of =M2, node distance= 4cm](out2){$\y{t}$};

\path[draw, line width=1pt, dashed] (lif2.south) edge[latex-,left] node{$ w_t$} (lif2.south |- M2.north);
\path[draw] (if2.south) ++(0.5cm,0)  edge[latex-,left] node{$ \x{t}$} ([xshift=0.5cm]if2.south |- M2.north);
\path[draw] (if2.south) ++(-0.5cm,0)  edge[-latex,left] node{$\ac{t}$} ([xshift=-0.5cm]if2.south |- M2.north);
\node[left of =Mhat2, node distance=3cm,yshift=2.5cm](MM2) {$\Ca$};
\node[above of =Mhat2,draw,fill=gray!15, rectangle, minimum width=3cm,node distance=2.1cm, minimum height=1cm] (Ca2) {$\hat\Ca$};
\path[draw] (Ca2.south) ++(-0.5cm,0)  edge[-latex,left] node{$ \ac{t}$} ([xshift=-0.5cm]Ca2.south |- Mhat2.north);
\path[draw] (Ca2.south) ++(0.5cm,0)  edge[latex-,left] node{$ \xh{t}$} ([xshift=0.5cm]Ca2.south |- Mhat2.north);
\begin{scope}[on background layer]

\node[ fill=gray!40, fit= (MM2)(Mhat2)(if2)(lif2)(Ca2)](FIt1) {};
\path[draw] (M2) edge[-latex] (out2);

\end{scope}
 \end{tikzpicture}
 \caption{For the stochastic difference equations of Example~\ref{ex:SDE}, the behaviour of the abstract gMDP $\widehat \M$ and the concrete gMDP $\M$ can be modelled via the coupled gMDP $\widehat\M ||_{\rel}\M$ (cf. Def.~\ref{def:coupling}) with the lifted probability distribution $\mathbb W$ for the disturbances $(\hat w_{t},w_t)$ and with the interface function $\InF$. \textbf{Left:} coupling gMDP $\widehat\M ||_{\rel}\M$. \textbf{Right:} refining controller $\hat\Ca$ to $\Ca$ on $\M$ that preserves satisfaction probabilities.}
 \label{fig:aux_model}
 \end{figure*}

	Fig.~\ref{fig:aux_model} (left) represents this coupling gMDP for two stochastic difference equations as given in Example \ref{ex:SDE} driven by noise signals $\hat w_{t} $ and $w_t$. In this case, 
	the stochastic simulation relation can be constructed by coupling the noise sources first, i.e., $(\hat w_{t},w_t)\sim\mathbb W$.
	By using this joint distribution, or in general the stochastic kernel $\Wt(d\xh{t+1}d\x{t+1}\mid\ach{t},\xh{t},\x{t})$, any controller for  $\widehat\M ||_{\rel}\M$ can be written as a controller for $\M$ as depicted in Fig.~\ref{fig:aux_model} (right). This is formalised in the next proposition.
	 
\smallskip%
\begin{proposition}
\label{prop:lifting_prob}
	Suppose $\M,\widehat\M\in\M_\Y$ with $\widehat \M\preceq_\eps^\delta \M$. For every control strategy $\hat \Ca$ (cf. Def.~\ref{def:CS}) for $\widehat\M ||_{\rel}\M$, there exists a control strategy $\Ca$ for $\M$, such that
  their respective probability distributions over the space of output traces are exactly the same. This implies that for any temporal specification $\psi$ (c.f. Def.~\ref{def:scLTL}), we have
	 \[\pcm{\hat\Ca}{(\widehat\M ||_{\rel}\M)}(\word\vDash \psi)=\pcm{\Ca}{\M}(\word\vDash \psi).\]
\end{proposition}
\smallskip%

\begin{remark}
The central requirement for Prop.~\ref{prop:lifting_prob} to hold is the existence of the lifted kernel $\Wt$ that lifts stochastic kernels of $\M$ and $\widehat\M$
to the coupling state space. Thus the results of this proposition hold regardless of the values of $\eps\ge 0$ and $\delta\ge 0$.
\end{remark}
 
\begin{remark}
Controller $\hat\Ca$ in Prop.~\ref{prop:lifting_prob} is defined on the coupling gMDP $\widehat\M ||_{\rel}\M$ and can take the pair of states $(\xh{t},\x{t})$ as its input. In practice, it is more efficient to design $\hat\Ca$ only for $\widehat\M$ and refine it to a controller $\Ca$ for $\M$.
This is depicted in Fig.~\ref{fig:aux_model} (left) by having $\hat\Ca$ only receiving $\xh{t}$ as its input.
\end{remark}

We present in the next section the characterisation of the satisfaction probability \eqref{eq:sat_prob} for a generic $\M$. Then we define the robust controller synthesis (Sec.~\ref{sec:fullcase}) that enables us to refine any synthesised controller $\hat\Ca$ for $\widehat\M$ to a controller for $\M$.



\section{Satisfaction probability of scLTL properties}
\label{sec:prod}
%
The verification of scLTL properties defined over the alphabet $\alphabeth$ is formulated using deterministic finite-state automata (DFAs), as defined next.
\smallskip%
\begin{definition}[DFA]
	A DFA is a tuple $\mathcal A = \left(Q,q_0,\Sigma,F,\trans\right)$, where
	$Q$ is a finite set of locations,
	$q_0\in Q$ is the initial location,
	$\Sigma$ is a finite alphabet,
	$F\subseteq Q$ is a set of accepting locations, and
	$\trans: Q\times\Sigma\rightarrow Q$ is a transition function.
\end{definition}
\smallskip%

A word $\word = \wordt{0},\wordt{1},\wordt{2},\ldots\in\Sigma^{\mathbb N}$
is accepted by a DFA $\mathcal A$ if there exists a finite run $q =(q_0,\ldots,q_n)\in Q^{n+1}$ such that $q_0$ is the initial location,
$q_{i+1} =\trans(q_,\wordt{i})$ for all $0\le i< n$ and $q_n\in F$.
The set of all words accepted by $\mathcal A$ defines the accepted language of $\mathcal A$, denoted as $\mathcal L(\mathcal A)$.\smallskip\\
\noindent{\bfseries DFA reachability versus scLTL properties.}
For every scLTL property $\psi$ as in Def.~\ref{def:scLTL} and  \cite{KupfermanVardi2001}, there exists a DFA $\mathcal A_{\psi}$ such that
\begin{equation*}
\word\vDash\mathcal \psi \,\,\Leftrightarrow\,\,   \word\in \mathcal L(\mathcal A_\psi).
\end{equation*}
As a result, the satisfaction of the property $\psi$ is equivalent to the \emph{reachability} of the accepting states $F$ in the DFA $\mathcal A_\psi$.
 
%
This reachability probability $\pcm{\Ca}{\M}
(\word \in\mathcal L(\mathcal A_\psi))$  over the traces $\word$ of $\M$, which is equal to
$\pcm{\Ca}{\M}(\word\models\psi)$, can be explicitly written out over  product of the gMDP $\M$ and the automaton $\mathcal A_\psi$, which is denoted as $\M\otimes\mathcal A_\psi$. This product was originally derived in \citep{tmka2013} for MDPs without output spaces. We give a similar product construction for gMDPs.
For a gMDP  $\M$, 
a DFA $\mathcal A_\psi$, 
and a labelling function $\mathsf L:\Y\rightarrow\Sigma$,
 the product between $\M$ and $\mathcal A_\psi$ is a gMDP
	\begin{equation*}
	\label{eq:MxA}
			\M\otimes\mathcal A_\psi =
			(\bar\X,\A,\Y,\bar x_0 ,\bar\Tr,\bar\h )
	\end{equation*}
	with 
	\begin{itemize}
	    \item $\bar{\X} := \X\times Q$, the state space;
	    \item $\Y$, the output space of $\M$;
	    \item  $ \bar x_0 := (\xin,\bar q_0)$, the initial state with $\bar q_0= \trans(q_0,\mathsf L\circ h(\xin))$;
	    \item $\bar{\Tr}(d x' \times\{q'\}|x,q,u)$, the stochastic kernel, that assigns  for any $u\in\A$ and $(x,q)\in\bar{\X}$, probability equal to
	    	\begin{equation*}
	\bar{\Tr}(d x' \times\{q'\}|x,q,u) :=  \mathbf{1}_{\{q'\}}(q^+) \Tr(d x'|x,u),
	\end{equation*}
		where $q^+ 
	=\trans(q,\mathsf L\circ h(x'))$ and where $\mathbf{1}_{A}(\cdot)$ is the indicator function for the set $A$, i.e., if $q\in A$, then $\mathbf{1}_{A}(q)=1$, else $\mathbf{1}_{A}(q)=0$. 
	    \item $\bar h(x,q) := h(x)$, the output map.
	\end{itemize}
%

Next proposition shows how policies on the product gMDP $\M\otimes \mathcal{A}_\psi$ are connected with control strategies on $\M$.
\smallskip%
\begin{proposition}
\label{prop:refine_controller}
For every 
Markov policy $\pols$ 
on the product gMDP $ \M\otimes\mathcal{A}_\psi$, there exists a control strategy $\Ca(\pols,\psi)$ for $\M$, 
such that
	\begin{equation}
	\label{eq:prod_spec}
\pcm{\Ca}{\M}(\word\models\psi)
= \mathbb P_{\pols\times (  \M\otimes \mathcal{A}_\psi)}
	(\exists t:q_t\in F).
	\end{equation}
\end{proposition}
\smallskip%
The proposition can be proved by taking steps similar to the ones in \citep{tmka2013}.
Eq.~\eqref{eq:prod_spec}
enables us to rewrite the probability of satisfying an scLTL property as the probability that the set of accepting states $F$ is reached in the product gMDP $\M\otimes\mathcal A_\psi$.
Additionally, for any 
Markov policy $\pols$ on $\M\otimes \mathcal{A}_\psi$ that ensures reachability with probability $p$, there exists a control strategy $\Ca(\pols,\psi)$ for $\M$, such that the controlled gMDP denoted as $\Ca(\pols,\psi)\times\M$ satisfies $\psi$ with the same probability $p$.


Probabilistic unbounded reachability in the right-hand side of \eqref{eq:prod_spec}  
can be computed as the limit of the probabilistic bounded reachability, that is,
\begin{align*}
	&\mathbb P_{\pols\times (  \M\otimes \mathcal{A}_\psi)}
	(\exists t:q_t\in F)\notag \\&\qquad :=\lim_{N\rightarrow\infty }  \mathbb P_{\pols\times (  \M\otimes \mathcal{A}_\psi)}
	(\exists t\leq N:q_t\in F).
\end{align*}
The limit in the above equation converges as it is non-decreasing and upper-bounded by one.
%


 For a given Markov policy $\pols$ and time horizon $N\in \mathbb N\cup \{\infty\}$, the probability of reaching $F$ while $t\leq N$, 
 can be written as
\begin{align*}
\label{eq:reach_prob}
&\mathbb P_{ \pols\times ( \M\otimes\mathcal{A}_\psi )}(\exists t\leq N\!\!:q_t\in F)
= \mathbb E_{\pols{}}\bigg[\max_{t\leq N}\ind{F}{q_t}{\mid} x_0,q_0\bigg],
\end{align*}
where the expectation is with respect to the state transitions of $ \M\otimes \mathcal{A}_\psi $ starting from $(x_0,q_0)$ and controlled with the Markov policy $\pols$. 
 For bounded horizons, $N\in \mathbb N$, this represents the probability that words $\word$ generated by $\Ca(\pols,\psi)\times\M$ have a satisfying prefix word \cite{Bible}.
\smallskip\\
\noindent{\bfseries Dynamic programming mappings.}
The probability can be computed recursively via value functions. For the bounded horizon $N$ and Markov policy $\pols$, define value functions $V_{k}^{\pols }:\X\times Q\rightarrow[0,1]$, $k\in[0,N]$, as the probability that the set of accepting states $F$ is reached within $k$ time steps starting from the state $(x,q)$ at time $N-k$. This yields the following explicit expression
\[V_{k}^{\pols}(x,q):=  \mathbb E\bigg[\max_{N-k< t\leq N}\ind{F}{q_t}\mid x_{N-k}=x,q_{N-k}=q\bigg].\]
These value functions $V_{k}^{\pols}$ can be recursively computed  with $V_0^{\pols} =0$ as
\begin{equation}
V_{k+1}^{\pols}(x,q) = \mathbf T^{\pol{l}} (V_{k}^{\pols})(x,q),
\,\,\, k\in[0,N-1],
\label{eq:V_rec}
\end{equation}
with $l:={N-(k+1)}$ and
where $\mathbf T^{\pol{}}$ is the Bellman operator  defined  as
\begin{align}\label{eq:T_op}
&\mathbf T^{\pol{}} (V)(x,q):= \int_{\X\times Q} \!\!\!  \max \left\{\mathbf 1_{F}(q'), V(x', q' )\right\}\notag\\ &\hspace{3cm}
\times  	\bar{\Tr}(d x' \times\{q'\}|x,q,\mu(x,q)).
\end{align}

%

%
\smallskip%
\begin{proposition}
\label{thm:Bellman1}
For a bounded horizon $N\in \mathbb N$ and a  Markov policy $\pols$ on the product gMDP $\M\otimes \mathcal{A}_\psi$ (with initial state $(x_0,\bar q_0)$),
the bounded reachability probability is
\begin{align*}
&\mathbb P_{\pols\times (  \M\otimes \mathcal{A}_\psi)}(\exists t\leq N:q_t\in F)\\&\hspace{2.5cm}= \max\left\{\mathbf 1_{F}( \bar q_0), V_N^{\pols}( x_0,  \bar q_0)\right\},
\label{eq:prop:reach2value}
\end{align*}
where $V_N^{\pols} (x,q)$ is computed with the Bellman operator as given in \eqref{eq:V_rec} and \eqref{eq:T_op}.  

\end{proposition}
 \smallskip%

We now consider the optimal reachability. 
Let $V_N^\ast(x,q)$ be computed using the optimal value functions $V_k^*:\bar\X\rightarrow[0,1]$, $k\in[0,N]$, defined inductively with $V_0^*=0$ and with the optimal Bellman recursion
\begin{equation}
V_{k+1}^*(x,q) =\mathbf T^{\ast}  (V^*_{k})(x,q),\quad k\in[0,N-1].
\label{eq:V_recoptT}
\end{equation}
The optimal Bellman operator is $\mathbf T^{\ast}(\cdot):=  \sup_{\mu}\mathbf T^{\mu}(\cdot)$ with $\mathbf T^{\mu}$ defined in \eqref{eq:T_op}.
Moreover, the optimising Markov policy $\pols^*=  (\pol{0}^*,\pol{1}^*,\pol{2}^*,\ldots)$ is computed as
\begin{equation}\textstyle
\pol{k}^\ast(x,q) \in \arg\sup_{\pol{k}}\mathbf T^{\pol{k}}  (V^*_{N-k})(x,q).
\label{eq:mu:opt}
\end{equation}

The next proposition characterises the bounded optimal reachability probability. 
\smallskip%
\begin{proposition}
\label{thm:Bellman2}
For a bounded horizon $N\in \mathbb N$,
the optimal bounded reachability probability of the product gMDP $\M\otimes \mathcal{A}_\psi$ (with initial state $(x_0,\bar q_0)$) is
\begin{align}\notag
&\sup_{\pols} \mathbb P_{\pols\times ( \M\otimes \mathcal{A}_\psi )}(\exists t\leq N:q_t\in F) \\&\hspace{2.5cm}=  \max\left(\mathbf 1_{F}(\bar q_0), V_N^\ast(x_0,\bar q_0)\right),
\end{align}
with $V_N^{\pols} (x,q)$  computed with the Bellman operator as given in \eqref{eq:V_recoptT}. The corresponding optimising Markov policy is given in \eqref{eq:mu:opt}. 
\end{proposition}
\smallskip%



\begin{proposition}\label{prop:sup_satisfaction}
Given the product gMDP
$\M\otimes \mathcal{A}_\psi$ with initial state $(x_0,\bar q_0)$, the infinite horizon reachability probability is    \begin{equation}
\label{eq:V_recopt_inf}
\sup_\Ca \pcm{\Ca}{\M}
	(\word \vDash \psi)=
	\max\{\ind{F}{\bar q_0}, V^*_\infty (\xin,\bar q_0)\},
\end{equation}
where $V^*_\infty (\x{},q)$ is the converged value function defined as
\begin{equation*}
V^*_\infty (\x{},q):= \lim_{N\rightarrow\infty} (\mathbf T^{\ast})^{N}  (V_0^{\ast})(\x{},q),\quad V_0^{\ast}=0.
\end{equation*}
\end{proposition}
\begin{IEEEproof}
	The optimal value functions $V_k^*$ in Prop.~\ref{thm:Bellman2} are monotonically increasing and bounded by one, thus converging for  $N\rightarrow \infty$ to a unique function  $V^*_\infty (\x{},q)$.
	This yields that
	\begin{equation}
	\sup_{\pols} \mathbb P_{\pols\times ( \M\otimes \mathcal{A}_\psi )}(\exists t:q_t\in F) =
	\max\{\ind{F}{\bar q_0}, V^*_\infty (\xin,\bar q_0)\}.\notag \end{equation}
	 In combination with Prop.~\ref{prop:refine_controller}, this concludes the proof. 
\end{IEEEproof}
\smallskip

\begin{figure}[htp]\centering
\begin{tikzpicture}
	\begin{scope}
		\draw[fill=green!60, fill opacity=0.5] plot [smooth,tension=1] coordinates { (1,0) (1.14,-0.6) (0.5,-0.5) (0.5,0.5)  (1,0) };
				\draw[dashed]  plot [smooth,tension=1]   coordinates { (1.2,0) (1.3,-0.7)  (0.5,-0.75)  (0.3,0.45) (0.75,0.6) (1.2,0)};
		\draw  plot coordinates { (0.1,-1.1)   (0.1,0.7) (1.5,0.9)  (1.5,-0.9)   (0.1,-1.1)};
\path[->, draw]  (0.1,-1.5) -- (5,-1.5);
\node at (4.45,-1.3) {$t$};

	\end{scope}
	\begin{scope}[xshift=0.9cm]
		\draw[fill=green!60, fill opacity=0.5] plot [smooth,tension=1] coordinates { (1,0) (1.14,-0.6) (0.5,-0.5) (0.5,0.5)  (1,0) };
				\draw[dashed]  plot [smooth,tension=1]   coordinates { (1.2,0) (1.3,-0.7)  (0.5,-0.75)  (0.3,0.45) (0.75,0.6) (1.2,0)};
		\draw  plot coordinates { (0.1,-1.1)   (0.1,0.7) (1.5,0.9)  (1.5,-0.9)   (0.1,-1.1)};
	\end{scope}
		\begin{scope}[xshift=1.8cm]
		\draw[fill=green!60, fill opacity=0.5] plot [smooth,tension=1] coordinates { (1,0) (1.14,-0.6) (0.5,-0.5) (0.5,0.5)  (1,0) };
				\draw[dashed]  plot [smooth,tension=1]   coordinates { (1.2,0) (1.3,-0.7)  (0.5,-0.75)  (0.3,0.45) (0.75,0.6) (1.2,0)};
		\draw  plot coordinates { (0.1,-1.1)   (0.1,0.7) (1.5,0.9)  (1.5,-0.9)   (0.1,-1.1)};
	\end{scope}
			\begin{scope}[xshift=2.7cm]
		\draw[fill=green!60, fill opacity=0.5] plot [smooth,tension=1] coordinates { (1,0) (1.14,-0.6) (0.5,-0.5) (0.5,0.5)  (1,0) };
				\draw[dashed]  plot [smooth,tension=1]   coordinates { (1.2,0) (1.3,-0.7)  (0.5,-0.75)  (0.3,0.45) (0.75,0.6) (1.2,0)};
		\draw  plot coordinates { (0.1,-1.1)   (0.1,0.7) (1.5,0.9)  (1.5,-0.9)   (0.1,-1.1)};
	\end{scope}
				\begin{scope}[xshift=3.6cm]
		\draw[fill=green!60, fill opacity=0.5] plot [smooth,tension=1] coordinates { (1,0) (1.14,-0.6) (0.5,-0.5) (0.5,0.5)  (1,0) };
				\draw[dashed]  plot [smooth,tension=1]   coordinates { (1.2,0) (1.3,-0.7)  (0.5,-0.75)  (0.3,0.45) (0.75,0.6) (1.2,0)};
		\draw  plot coordinates { (0.1,-1.1)   (0.1,0.7) (1.5,0.9)  (1.5,-0.9)   (0.1,-1.1)};\node at  (1.3,0.7)  {$\Y$};
	\end{scope}
\end{tikzpicture}
\caption{A tube-shaped set $A$ and its expansion $A_{\eps}$ (dashed) defined via the Minkowski sum in \eqref{eq:plus}.}
\label{fig:Minkowski}
\end{figure}
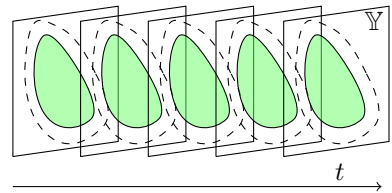

 
 \noindent{\bfseries Exact computation with correction.}
The computations of the Bellman recursions \eqref{eq:V_rec} and \eqref{eq:V_recoptT} are generally only tractable for finite state-space models \citep{Abate1}. This motivates us to construct an abstract model $\widehat{\M}$ for the concrete model $\M$ that satisfies the simulation relation of Def.~\ref{def:apbsim}, perform computations on $\widehat{\M}$, and refine the results to the $\M$.
Therefore,
%
for simple properties such as safety and reachability, the following proposition of \citep{haesaert2017verification} relates the computed probabilities over the two models with the $\epsilon$ and $\delta$ deviations.
  Suppose that $\Y$ is a vector space and 
denote a closed set that includes the $\eps$ neighbourhood of $A$ in $\Y^{N+1 }$ as
\begin{equation}
\label{eq:plus}
A_\epsilon\supseteq A\oplus\{y\in\Y^{N+1 }|\max_t\|y_t\|\leq \eps\},
\end{equation}
where $\oplus$ is the Minkowski addition.
The set $A_\epsilon$ is pictorially presented in Fig.~\ref{fig:Minkowski}.
Similarly, based on the Minkowski difference $\ominus$, we define a closed set
\begin{equation}
A_{-\epsilon}\subseteq A\ominus\{y\in\Y^{N+1 }|\max_t\|y_t\|\leq \eps\}.
\label{eq:min}
\end{equation}
%
%
\begin{proposition}
  	\label{thm:cr}
Suppose $\widehat\M\preceq_\eps^\delta \M$.
For any control strategy $ \hat{\Ca}$ on $\widehat\M$ there exists a control strategy $\mathbf C$ on $\M$ such that,
for all measurable events $A\subset \Y^{N+1 }$
\begin{align} \pcm{\hat{\Ca}}{\widehat{\M}}\!\left( \hat y  \in  A_{-\eps}\right)-\gamma&
 \leq \pcm{\Ca }{\M} \!\left( y   \in  A\right)\nonumber\\&
\leq\pcm{\hat{\Ca}}{\widehat{\M}}\!\left(\hat y  \in A_{\eps}\right)+\gamma,
\label{eq:linear_bound}
\end{align}
%
with sets $A_{\eps}$ and $A_{-\eps}$ as defined in \eqref{eq:plus}-\eqref{eq:min} and with constant $1-\gamma:=(1-\delta)^{N}$.
\end{proposition}
For small values of $\delta$, the probability deviation
can be approximated linearly as $\gamma \approx N\delta$. Clearly, $\gamma$ is composed of the probabilistic deviation incurred in $N$-transitions. 

\begin{figure}[htp]
\centering
	\includegraphics[width=.4\textwidth]{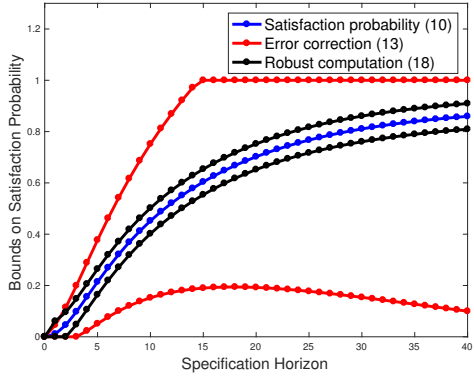}
	\caption{Bounds on the satisfaction probability of a specification as a function of the specification horizon computed by different approaches. The red curves are computed via \eqref{eq:linear_bound} and the black curves are computed via robust Bellman operators (cf. Sec.~\ref{sec:fullcase}).}
	\label{fig:bound1}
\end{figure}
Fig.~\ref{fig:bound1} compares bounds on satisfaction probability of a specification obtained from different approaches as a function of the specification horizon. The  blue  curve is the probability computed over the abstract model. The red curves are the bounds \eqref{eq:linear_bound}; the lower bound is first increasing then converging to zero, thus not useful for infinite-horizon specifications. The black curves are computed via robust mappings and they will be discussed in the next section.

\section{Robust control for scLTL properties}
\label{sec:fullcase}

\begin{definition}[($\eps,\delta$)-Robust satisfaction]
\label{def:epdelrob}
Given the gMDP $\widehat\M\in\mathcal M_\Y$, we say that the control strategy $\hat\Ca $ for $\widehat\M$ $(\eps,\delta)$-robustly satisfies $\psi$ with probability $p$ if for every $\M\in\mathcal M_\Y$ with $\widehat\M\preceq^\delta_\epsilon \M$ there exists a control strategy $\Ca$ such that
 \(\pcm{\Ca}{\M}(\word\vDash\psi)\geq p.\)
\end{definition}
We first consider in the next subsection the case where the output deviation is zero, i.e., $\eps=0$. This prepares us to tackle the full ($\eps,\delta$)-robust satisfaction in Subsec.~\ref{sebsec:robust}.
%
%


\subsection{$\delta$-Robust satisfaction of scLTL properties}

In this subsection, we provide a method to compute the robust satisfaction for scLTL specifications with respect to $(0,\delta)$-errors.\smallskip\\
\noindent{\bfseries $\delta$-Robust mapping.}
For any universally measurable map $\pol{}:\hat\X \times Q\rightarrow \mathcal P(\hat\A,\mathcal B(\hat\A))$,
we define the robust operator $\mathbf T_\delta^{\pol{}}$ as a modification of $\mathbf T^{\pol{}}$ in \eqref{eq:T_op}, i.e.,
\begin{equation}
 \mathbf T_\delta^{\pol{}} (V)(\hat x,q) :=\Lim\Big(\mathbf T^{\pol{}} (V)(\hat x ,q)-\delta\Big),
\label{eq:kronmap}
\end{equation}
with $\Lim:\mathbb R\rightarrow [0,1]$ being the truncation function  $\Lim(\cdot):=\min(1,\max(0,\cdot))$.
Similarly, we define the optimal $\delta$-robust operator $\mathbf T_\delta^{*}(V)$ as
\begin{equation}
\mathbf T_\delta^{*}(V)(\hat x,q):= \sup_{\pol{}}\mathbf T_\delta^{\pol{}}(V)(\hat x,q).
\label{eq:kronmap2}
\end{equation}
Notice that for $\delta=0$, the operators are the same: $\mathbf T_\delta^{\pol{}} = \mathbf T^{\pol{}}$ and $\mathbf T_\delta^\ast = \mathbf T^\ast$.
Next lemma establishes properties of the robust operator $\mathbf T_\delta^{\pol{}}$.

\smallskip
\begin{lemma}[Properties of the $\delta$-robust operator]
\label{lem:rob_op}
The robust operator $\mathbf T_\delta^{\pol{}}$ defined in \eqref{eq:kronmap} is monotonically increasing for any $\pol{}:\hat\X\times Q\rightarrow \mathcal P(\hat\A,\mathcal B(\hat\A))$. Namely, for any two functions $V$ and $W$ satisfying  $V(\hat x,q)\geq W(\hat x,q)$ for all $(\hat x,q)\in  \hat \X\times Q$,
	 it holds that
	 \begin{equation}\mathbf T_\delta^{\pol{}}(V)(\hat x,q)\geq \mathbf T_\delta^{\pol{}}(W)(\hat x,q).
	 \label{eq:mono}
	 \end{equation}
Moreover, the series
	\(\{(\mathbf T^{\pol{}}_\delta)^{l}(V_0)\}_{l\ge 0}\)
	with $V_0 =0$
	 is \emph{monotonically increasing} and
	 \emph{point-wise converging}. Additionally, the fixed-point equation
	 \begin{equation}
	 \label{eq:fixed_point}
	V_{\infty}^{\pol{}} = \mathbf T^{\pol{}}_{\delta} (V_{\infty}^{\pol{}}),
	\end{equation}
	has a unique solution for $\delta>0$, which is
	  \begin{equation}
	  \label{eq:sat_rob_sol}
	  V_\infty^{\pol{}} :=\lim_{l\rightarrow\infty} (\mathbf T^{\pol{}}_\delta)^{l}(V_0) \quad \mbox{with }V_0 =0.
	  \end{equation}
\end{lemma}\smallskip
 The proof of Lem.~\ref{lem:rob_op} 
 has been relegated to the appendix.

The optimal robust operator $T_\delta^{*}$ has the same properties. 
More precisely, $T_\delta^{*}$ defined in \eqref{eq:kronmap} is monotonically increasing. Namely, for any two functions $V$ and $W$ satisfying  $V(\hat x,q)\geq W(\hat x,q)$ for all $(\hat x,q)\in \hat \X\times Q$, it holds that
\begin{equation}
\label{eq:mono_rob}
\mathbf T_\delta^{*}(V)(x,q)\geq \mathbf T_\delta^{*}(W)(x,q).
\end{equation}
Moreover, the series $\{(\mathbf T^{\ast}_\delta)^{l}(V_0)\}_{l\ge 0}$ with $V_0 =0$
is \emph{monotonically increasing} and \emph{point-wise converging}.
Additionally, from the fixed-point equation
\begin{equation*}
	V_\infty^\ast =  \mathbf T^*_\delta (V_\infty^\ast)
	\label{eq:fixed_point_opt}
\end{equation*}
with solution 
\begin{equation}
\label{eq:sat_rob_sol_opt}
	V_\infty^\ast = \lim_{l\rightarrow\infty} (\mathbf T^*_\delta)^l(V_0),\quad V_0 =0,
\end{equation}
we can get the maximising policy as a stationary Markov policy. This is formalised next.
\smallskip
\begin{theorem}\label{thm:opt:mu}
The stationary Markov policy $\pols{}^\ast = (\mu^\ast,\mu^\ast,\ldots)$ given as
\begin{equation}
\label{eq:robust_opt_pol}
	\mu^\ast \in \arg\sup_{\pol{}}  \mathbf T^{\pol{}} _\delta (V_\infty^\ast)
\end{equation}
is the \emph{optimal $(0,\delta)$-robust policy}. That is, it holds that  
\begin{equation*}
\label{eq:thm_eq_mu_ast}
	V_\infty^{\ast} = 
\lim_{l\rightarrow\infty}(\mathbf T^{\mu^\ast}_\delta)^l (V_0)\mbox{ with }V_0=0.
\end{equation*}
\end{theorem}

\smallskip


\noindent{\bfseries $\delta$-Robust satisfaction probability.}
For a given stationary Markov policy $\pols{} = (\mu,\mu,\ldots)$, we define \emph{($0,\delta$)-robust satisfaction probability} as
\begin{equation}
\label{eq:sat_rob}
\mathcal S_{\delta}^{\pols{}}
 := 
\max\left\{\mathbf 1_{F}(\bar q_0),V_\infty^{\mu}(\hat x_0,\bar q_0)\right\}
\end{equation} with $ \bar q_0 =\trans(q_0,\mathsf L( \hat h(\hat x_0)))$
	and $V_\infty^{\mu}$ in \eqref{eq:sat_rob_sol}.	
Similarly, we define the \emph{optimal ($0,\delta$)-robust satisfaction probability} as
\begin{equation}
\label{eq:opt_sat_rob}
\mathcal S_{\delta}^{\ast}
:= \max\left\{\mathbf 1_{F}(\bar q_0),V_\infty^{\ast}(\hat x_0,\bar q_0)\right\}
\end{equation}
with $V_\infty^{\ast}$ in \eqref{eq:sat_rob_sol_opt}.
The \emph{optimal $(0,\delta)$-robust policy}
computed as
\eqref{eq:robust_opt_pol} achieves \eqref{eq:opt_sat_rob}.

Quantities \eqref{eq:sat_rob}-\eqref{eq:opt_sat_rob} satisfy the condition of Def.~\ref{def:epdelrob} and enable us to refine strategies from the abstract model to the original model. This is formally stated in the next theorems.

\begin{theorem}
\label{thm:delreach_infty}
Let a gMDP $ \widehat\M $, an scLTL specification $\psi$, 
and a stationary  Markov policy $\pols$ on $ \widehat\M\otimes\mathcal{A}_\psi$ be given.   
Then the robust satisfaction probability $\mathcal S_{\delta}^{\pols{}}$ as given in \eqref{eq:sat_rob} is the \emph{($0,\delta$)- robust satisfaction probability} defined in Def.~\ref{def:epdelrob}.
Moreover, for any gMDP $\M$ with $\widehat\M\preceq^{\delta}_0\M$, we can \emph{construct} $\Ca$ such that $\psi$ is satisfied by $\Ca\times\M$ with probability at least $\mathcal S_{\delta}^{\pols{}}$.
\end{theorem}
In a similar fashion to the above theorem, we can now state the following corollary.
\begin{corollary}
\label{thm:delreach_infty_v2}
Let $\widehat\M\preceq^{\delta}_0\M$ and scLTL specification $\psi$ be given. We can construct $\Ca$ such that $\psi$ is satisfied by $\Ca\times\M$ with probability at least $\mathcal S_{\delta}^{\ast}$ in \eqref{eq:opt_sat_rob}.
\end{corollary}
The proofs of these two statements are relegated to the appendix.
\subsection{Analysis: Connection with the mean trace length}
Recall the results of Prop.~\ref{thm:cr} that shows the probabilistic deviation  converges to 1 and grows close to linearly with the horizon for small values of $\delta$. 
In this subsection, we show that
for the $\delta$-robust computations \eqref{eq:kronmap} 
the probabilistic deviation is now relative to \emph{the mean trace length}.
 \\
For executions 
of the gMDP $\widehat\M\otimes \mathcal A$, 
the first \emph{hitting time} of a set $A\subset\hat\X\times Q$ is a random variable  conditioned on the initial state $(\hat x_0,q_0)$ and given as
\begin{equation*}
	H_A(\hat x_0,q_0) \!:=\left\{\inf\{t\!\in\!\mathbb N\!\cup\!\{\infty\}\!:(\hat x_t,q_t)\in A\}|(\hat x_0,q_0)\right\}\!.
\end{equation*}
%
The support of the first hitting time is $\mathbb N\cup\{\infty\}$. It can be infinity if the execution does not hit the set $A$. It is zero with probability one if $(\hat x_0,q_0)\in A$, otherwise positive.



\begin{theorem}[First hitting time]\label{thm:1sthit}
The loss in probability by the sequential application of the robust operator 
\eqref{eq:kronmap}
is related to the first hitting time of $(\widehat \M\otimes\mathcal{A}_\psi )$ as follows 
\begin{align}
	\big(\mathbf{T}^\mu_{\delta}\big)^l (V_0)(\hat x,q) &\geq \big({\mathbf{T}}^\mu\big)^l (V_0)(\hat x,q) \notag \\&\quad 
	- \delta \sum_{n=1}^{l} \mathbb P\left(H_{\hat\X\times F}(\hat x,q)\ge n\right),
	\label{eq:hitting_time_bound}
\end{align}
for any $l\geq 1$ with $V_0 = 0$. 
\end{theorem}
The proof of this theorem, which can be found in the appendix, relies on a rewrite of the $\delta$ correction to be limited to the domain that excludes $F$.
We define the \emph{mean hitting time} of a set $A$ as
\begin{align}
	h(A) :=& \mathbb E\,[H_A(\hat x_0,q_0)]
	= \hspace{-0.4cm}\sum_{n\in\mathbb N\cup\{\infty\}}\hspace{-0.4cm} n\,\, \mathbb P[H_A(\hat x_0,q_0) = n]\nonumber\\
	=& \sum_{n=1}^\infty \mathbb P[H_A(\hat x_0,q_0)\ge n].
	\label{eq:mean_hit}
\end{align}
 
 \begin{corollary}
 For a given stationary Markov policy $\pols{}$ and the corresponding control strategy $\hat\Ca$ as defined in Prop.~\ref{prop:refine_controller}, 
 the robust satisfaction is lower bounded as
\begin{equation}
\pcm{\hat\Ca}{\widehat\M}
	(\word \vDash \psi)\geq\,\mathcal S_{\delta}^{\pols{}} \,\geq
	\pcm{\hat\Ca}{\widehat\M}
	(\word \vDash \psi)
	- \delta h(\hat\X\times F),
	\label{eq:hitting_time}
\end{equation}
where $h(\hat\X\times F)$ \eqref{eq:mean_hit} is the mean length of traces satisfying the property $\psi$.
 \end{corollary}

The mean hitting time $h(A)$ in \eqref{eq:mean_hit} is bounded only when executions of $\widehat \M\otimes\mathcal{A}_\psi$ starting from $(\hat x_0,q_0)$ reach $A$ with probability one. Thus the lower bound in \eqref{eq:hitting_time} is only informative for initial states $(\hat x_0,q_0)$ reaching $F$ for sure. This inequality can be tightened  using the notion of \emph{absorbing sets}.

\smallskip

\noindent{\bfseries Absorbing set.} For a stationary Markov policy $\pols:=(\pol{},\pol{},\pol{},\ldots)$ on the product gMDP $ \widehat\M\otimes\mathcal{A}_\psi$, define the \emph{absorbing set} $I\subset \hat\X\times (Q\setminus F)$ with the property
\begin{equation}
    \mathbb P_{\pols\times (\widehat\M\otimes \mathcal{A}_\psi)}\left[(\xh{t+1},q_{t+1})\in I |\xh{t},q_t\right]=1,\, \forall (\xh{t},q_t)\in I.
    \label{eq:absorbing}
\end{equation}
This property means that if the gMDP starts from a state in $I$, it will remain inside $I$ for all future time instances.

In case that $\widehat \M\otimes\mathcal{A}_\psi$ has absorbing sets, $\mathcal S_{\delta}^{\pols{}}$ will be zero over the absorbing sets  whereas $h(\hat\X\times F)$ will be infinity. Therefore, the right inequality in \eqref{eq:hitting_time} is trivial. The proof of Theorem \ref{thm:1sthit} can be extended for states outside the absorbing sets, this gives the next corollary. 
 
\begin{corollary}[Extension to absorbing sets]
For a given Markov policy $\pols{}$ and the corresponding control strategy $\hat\Ca$ as defined in \eqref{prop:lifting_prob}, the robust satisfaction is lower bounded as
\begin{equation}
\mathcal S_{\delta}^{\pols{}} \,\geq \pcm{\hat\Ca}{\widehat\M}
	(\word \vDash \psi)
	- \delta h(I\cup (\hat\X\times F)),
	\label{eq:hitting_time1}
\end{equation}
where $I$ is any absorbing set for the given policy $\pols{}$.
 \end{corollary}
 \begin{remark}
 By taking $I$ in \eqref{eq:hitting_time1} to be the largest absorbing set of $\widehat \M\otimes\mathcal{A}_\psi$, the provided lower bound is non-trivial 
 states outside the absorbing set.
 \end{remark}
 
 \color{black}
 
\begin{figure}[htp]
\centering
	\includegraphics[width=.4\textwidth]{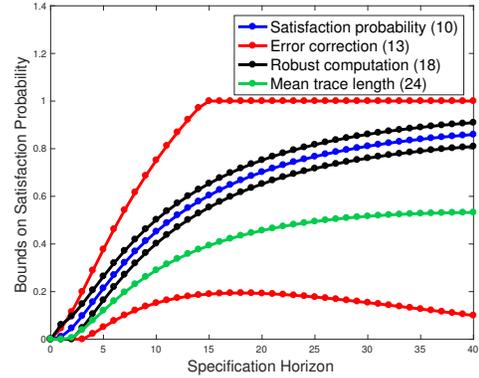}
	\caption{Bounds on the satisfaction probability of a specification computed by different approaches as a function of the specification horizon. The green curve is computed via the mean trace length in Thm.~\ref{thm:1sthit}.}
	\label{fig:bound2}
\end{figure}
Fig.~\ref{fig:bound2} complements Fig.~\ref{fig:bound1} with the green curve, which is the bound provided by Thm.~\ref{thm:1sthit}.

\subsection{$(\eps,\delta)$-Robust satisfaction of scLTL properties}
\label{sebsec:robust}
In  this  subsection, we handle the full case of having deviations in both probability and in output maps. We provide a  method  to  compute  the robust satisfaction for scLTL specifications with respect to $(\eps,\delta)$-errors.

Denote the $\epsilon$-neighbourhood of an element $y\in\Y$ as
\[\mathcal N_\epsilon(\hat y) :=\{y \in \Y|\,  \mathbf{d}_\Y\left(y,\hat y\right)\leq \epsilon\} .\] 
Then for $\widehat\M\preceq_\eps^\delta\M$, it holds that
$\Lab(y)\in \Lab (\mathcal N_\epsilon(\hat y) )$ for $y = h(x)$, $\hat y = \hat h(\hat x)$ and  $ (\hat x,x)\in\rel$.
 \\
 
\noindent{\bfseries $(\eps,\delta)$-Robust mapping.}
 For any universally measurable map $\pol{}:\hat\X\times Q\rightarrow \mathcal P(\hat\A,\mathcal B(\hat\A))$,
the $\delta$-robust operator in Eq.~\eqref{eq:kronmap} is modified to the \emph{$(\eps,\delta)$-robust operator}
\( \mathbf T_{\eps,\delta}^{\pol{}}\)  defined as
\begin{align}
\notag \mathbf T_{\eps,\delta}^{\pol{}} (V)(\hat x,q)\!:=\!\Lim&\Big(\!\!\int_{\hat\X} \min_{q'\in \bar\trans(q,\hat x')}\!\!\max\left\{\mathbf 1_{F}(q'), V(\hat x',q')\right\}\label{eq:rob_operator2v}\\
&\hspace{1cm}\times \Tr(d\hat x'|\hat x,\mu)-\delta\Big),
\end{align}
with $\bar\trans(q,\hat x'):=\{\trans(q,\alpha)\mbox{ with }\alpha \in\Lab(\mathcal N_\epsilon(\hat h (\hat x')))\}$.

Similarly, we define the \emph{optimal $(\eps,\delta)$-robust operator} $\mathbf T_{\eps,\delta}^{\ast}$ as
\begin{equation}
    \textstyle\mathbf T_{\eps,\delta}^{\ast}(V):= \sup_{\mu}\mathbf T_{\eps,\delta}^{\mu}(V).
    \label{eq:opt_rob_operator2v}
\end{equation}
Notice that for $\epsilon=0$, we retrieve the same operators as in \eqref{eq:kronmap}-\eqref{eq:kronmap2}, namely \(\mathbf T_{0,\delta}^{\pol{}} = \mathbf T_{\delta}^{\pol{}}\) and \(\mathbf T_{0,\delta}^\ast = \mathbf T_{\delta}^\ast\).

The addition of a $\min$ operator does not change the   properties given in Lem.~\ref{lem:rob_op}.
\\

\noindent{\bfseries $(\eps,\delta)$-Robust satisfaction probability.}
For a given stationary Markov policy $\pols{}$, 
 we define \emph{($\epsilon,\delta$)-robust satisfaction probability} as
\begin{equation}
\label{eq:sat_rob_2v}
\mathcal S_{\epsilon,\delta}^{\pols{}} 
 := 
\min_{\bar q_0\in \bar\trans(q_0,\hat x_0)}\!\!\max\left(\ind{F}{\bar q_0}, V_\infty^{\mu}(\hat x_0,\bar q_0)\right), 
\end{equation}
	where $V_\infty^{\mu}$ is the unique solution of the fixed-point equation $V_{\infty}^{\pol{}} = \mathbf T^{\pol{}}_{\eps,\delta} (V_{\infty}^{\pol{}})$ obtained from $V_\infty^{\pol{}} :=\lim_{l\rightarrow\infty} (\mathbf T^{\pol{}}_{\epsilon,\delta})^{l}(V_0)$ with $V_0 =0$.
	Similarly, we define the \emph{optimal ($\epsilon,\delta$)-robust satisfaction probability} as
\begin{equation}
\label{eq:opt_sat_rob_2v}
\mathcal S_{\epsilon,\delta}^\ast
:=\min_{\bar q_0\in \bar\trans(q_0,\hat x_0)}\!\!\max\left(\ind{F}{\bar q_0}, V_\infty^{\ast}(\hat x_0,\bar q_0)\right),
\end{equation}
where $V_\infty^{\ast}$ is the unique solution of the fixed-point equation $V_{\infty}^\ast = \mathbf T^\ast_{\eps,\delta} (V_{\infty}^\ast)$
obtained from $V_\infty^\ast :=\lim_{l\rightarrow\infty} (\mathbf T^\ast_{\epsilon,\delta})^{l}(V_0)$ with $V_0 =0$.
 The optimal $(\epsilon,\delta)$-robust policy is the stationary Markov policy $\pols{}^\ast = (\mu^\ast,\mu^\ast,\ldots)$ defined as
\begin{equation}
\label{eq:robust_opt_pol_2v}
	\mu^\ast \in \arg\sup_{\pol{}}  \mathbf T^{\pol{}} _{\epsilon,\delta} (V_\infty^\ast).
\end{equation}


\begin{theorem}
\label{thm:delepsreach_infty}
Let $ \widehat\M\otimes\mathcal{A}_\psi$ and a stationary  Markov policy $\pols$ be given.  
Then for any $\M$ with $\widehat\M\preceq^{\delta}_\epsilon\M$, we can construct $\Ca$ such that $\psi$ is satisfied by $\Ca\times\M$ with probability at least $\mathcal S_{\epsilon,\delta}^{\pols{}}$.
\end{theorem}

\begin{corollary}
\label{thm:delepsreach_infty_ast}
Let $ \widehat\M\otimes\mathcal{A}_\psi$ be given for which $\mathcal S_{\epsilon,\delta}^{\ast}$ in \eqref{eq:opt_sat_rob_2v} has been computed.  
Then for any gMDP $\M$ with $\widehat\M\preceq^{\delta}_\epsilon\M$, we can construct $\Ca$ using $\mu^\ast$ in \eqref{eq:robust_opt_pol_2v} such that $\psi$ is satisfied by $\Ca\times\M$ with probability at least $\mathcal S_{\epsilon,\delta}^{\ast}$.
\end{corollary}


In conclusion, we have shown that we can leverage approximate stochastic simulation relations to synthesise control strategies for the approximate model that robustly satisfy an scLTL specification, and refine these strategies to ones over concrete models with guaranteed lower bounds on the satisfaction probability of the specification.

\subsection{$(\eps,\delta)$-Optimistic satisfaction of scLTL properties}

In contrast to the robust satisfaction of a property in Def.~\ref{def:epdelrob} that gives a lower bound, we now define \emph{optimistic satisfaction} that quantifies an upper bound on the satisfaction probability of an scLTL property using the approximate model $\widehat\M$.

\begin{definition}[($\eps,\delta$)-Optimistic satisfaction]\label{def:opt_sat}
Given the gMDP $\widehat\M\in\mathcal M_\Y$, we say that a control strategy $\hat\Ca$ for $\widehat\M$ $(\eps,\delta)$-optimistically satisfies $\psi$ with probability $p$ if for every $\M\in\mathcal M_\Y$ with $\M\preceq^\delta_\epsilon \widehat\M$  and for all control strategies $\Ca$ for  $\M$ it holds that \[\pcm{\Ca}{\M}(\word\vDash\psi)\leq p.\]
\end{definition}

\noindent{\bfseries $(\eps,\delta)$-Optimistic mapping.}
We define the $(\eps,\delta)$-optimistic operator
\( \mathbf T_{-\eps,-\delta}^{\ast}(V)\) as
\begin{align*}
&\notag \mathbf T_{-\eps,-\delta}^\ast (V)(\hat x,q):= \sup_\mu\Lim \Big(\int_{\hat \X} \max_{q'\in \bar\trans(q,\hat x')}\!\!\max\left\{\mathbf 1_{F}(q'), \right.\\
&\hspace{2.5cm}\left.V(\hat x',q')\right\}  \Tr(d\hat x'|\hat x,\mu)+\delta\Big).
\end{align*}

\begin{theorem}
	\label{thm:maxprob}
Given a gMDP $\widehat\M$ and an scLTL specification $\psi$, a control strategy $\hat\Ca$ computed based on the $(\eps,\delta)$-optimistic operator $\mathbf T_{-\eps,-\delta}^{\ast}$  satisfies $\psi$ $(\eps,\delta)$-optimistically as defined in Def.~\ref{def:opt_sat}.
\end{theorem}

\section{Linear Time-Invariant Systems}\label{sec:LTI}

\subsection{Abstraction of linear gMDPs}
\label{sec:abstr}

In this section, we show how to compute an abstract gMDP for a set of linear stochastic difference equations.
To obtain an abstract model, one can use model-order reduction techniques \citep{LSMZ17} or employ lumping based abstractions that result in discrete-state MDPs \citep{SAID}. We present an approach in the sequel that combines both model order reduction and abstraction to a discrete-state model.
Our error quantification approach utilises the disturbance induced in the state trajectory, which is initially proposed in \cite{DBLP:conf/adhs/HaesaertSA18,LSZ18_ADHS}.

\smallskip

\noindent\textbf{Concrete model.} Consider the following linear gMDP $\M$:
\begin{equation}
\begin{aligned}
\x{t+1} &= A \x{t} + B \ac{t} + B_{w} w_t,\\
\y{t} &= C \x{t},\quad \xin \in\X,
\end{aligned}\label{eq:M2}
\end{equation}
where $\x{t}\in\X\subset\mathbb R^n$, $\ac{t}\in \A\subset \mathbb R^m$, and $\y{t}\in\Y\subset\mathbb R^p$. Matrices $A$, $B$, $B_{w}$, and $C$ have appropriate dimensions and $\{w_t,t\in\mathbb N\}$ is an i.i.d. sequence with standard Gaussian distributions $w_t\sim\mathcal N(0,\mathbb I)$.

\smallskip

\noindent\textbf{Constructing the abstract model.}
We will construct the abstract model in two steps. First, we construct a reduced order model $\M_s$, which is then used to construct a discrete state abstract model $\widehat\M$.
The reduced order model $\M_s$ has the state space $\X_s\subset \mathbb R^{n_s}$ with $n_s<n$. Let the dynamics of $\M_s$ be given as
\begin{equation}
\begin{aligned}
x_{s,t+1}& = A_s x_{s,t}+B_su_{t}+B_{sw} w_{t},\\
\y{s,t}& = C_s	x_{s,t}.
\end{aligned}
\label{eq:Ms}
\end{equation}
The stochastic kernels $\Tr,\Tr_s$ for $\M$ and $\M_s$ are obtained similar to the one in Example~\ref{ex:SDE} using difference equations \eqref{eq:M2}-\eqref{eq:Ms}.
The construction of $\widehat\M$ relies on partitioning this new space $\X_s$ with regions $\mathbb A_i\subset \X_s$, indexed by $\varGamma$, such that $\bigcup_{i\in\varGamma} \mathbb A_i = \X_s$ and $\mathbb A_i \cap\mathbb A_j =\emptyset $ for $i\not = j$.
For each region, we select one representative point $\hat x^i\in \mathbb A_i$. The collection of representative points defines the set of abstract states of $\widehat \M$, i.e.,
 $\hat\X:=\{\hat x^i\in \mathbb A_i,\,\,i\in\varGamma\}$.
 States $x_s\in\X_s$ are mapped to abstract states $\hat x\in\hat\X$ using the
the operator $\Pi:\X_s\rightarrow\hat\X$ that assigns to any $x_s\in  \X_s$,  the representative point $\hat x^i$ iff  $x_s\in \mathbb A_i$ for
$i\in\varGamma$.
As the abstract action space, we select $\hat\A \subset\A$, a (finite) subset of $\A$.
We define the stochastic transitions of $\widehat \M$ as
\begin{align*}
\hat\Tr(\hat x^i |\xh{t},\ach{t})&:=\Tr_s(\mathbb A_i|\xh{t},\ach{t})\notag\\&=\mathcal N(\mathbb A_i|A_s \xh{t}+B_s\ach{t}, B_{sw}B_{sw}^T).
\end{align*}
The abstract gMDP $\widehat \M:=(\hat\X,\hat\A,\Y,\xinh, \hat\Tr,\hat\h)$ is fully characterised by choosing an initial state $\xinh$,
the output map $\hat\h(\xh{}):=C_s\xh{t}$, and the output space $\Y$.

\begin{figure}[htp]\centering
 \includegraphics[width=.8\linewidth, trim={33cm 12cm 10cm 13cm}, clip]{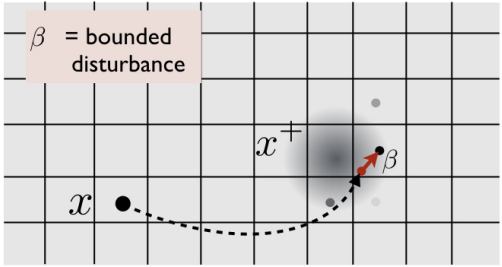}
		\caption{From linear dynamics to finite state dynamics by choosing the appropriate $\beta\in\Delta$.} \label{fig:m1_dyn}
	\end{figure}

\noindent\textbf{Computation of the ($\eps,\delta$)-simulation relation.}
To quantify the difference between $\widehat\M$ and $\M$, we start by analysing the abstract gMDP $\widehat \M$.
Executions of $\widehat\M$ can be obtained via the stochastic difference equations
\begin{equation}\begin{aligned}
\xh{t+1} &= \Pi\left( A_s \xh{t} + B_s\ach{t} + B_{s w} w_t\right),\\
\yh{t} &= C_s \xh{t},\ \,   \xinh \in \X,
\end{aligned}\label{eq:M1}\end{equation}
that are disturbed with the noise term $w_t\sim \mathcal N(0,\mathbb I)$. This noise $w_t$ is exactly the one affecting $\M$, thereby allowing us to define an intuitive lifting $\mathbb W_{\Tr}$ based on $\mathcal N(0,\mathbb I)$. Consider the set
\begin{equation*}
\Delta:=\{\Pi(x_s)-x_s\mid x_s\in \X_s\}.
\end{equation*}
Then the state transition \eqref{eq:M1} can be rewritten as a transition into a bounded set \begin{equation}
	\xh{t+1} \in A_s \xh{t} + B_s\ach{t} + B_{s w} w_t +\Delta.
	\label{eq:beta_m1}
\end{equation}
The state transitions of \eqref{eq:beta_m1} is illustrated in Fig.~\ref{fig:m1_dyn}. We assume that the partition sets $\{\mathbb A_i,i\in\varGamma\}$ are selected such that $\Delta$ is bounded, namely there exists a vector $\boldsymbol{\delta}$ such that $|\beta|\le \boldsymbol{\delta}$ element-wise for all $\beta\in\Delta$.

 Consider the linear interface function
\begin{equation}\label{eq:interface}
\ac{} = R \ach{} + Q \xh{} + K(\x{}-P\xh{}),
\end{equation}
for some matrices $P,Q,R,K$ such that the Sylvester equation
\begin{equation*}
  PA_s = AP + BQ
  \end{equation*}
holds.
Define the relation $(\xh,\x{})\in\rel$ as
\begin{equation*}
	\rel:=\{(\xh{},\x{})\mid\|\x{}-P\xh{}\|_M\le \epsilon\},
\end{equation*}
with $\|\x{}\|_M:=\sqrt{\x{}^T M \x{}}$.
We check conditions {\bfseries A1-3} of Def.~\ref{def:apbsim} under which $\widehat\M\preceq_\eps^\delta \M$.

\noindent{\itshape \bfseries Condition A1.}
For the pair of initial states $(\xinh,\xin)$, this condition requires that
$
\|\xin{}-P\xinh{}\|_M\le \epsilon.
$
One choice is to find an $x_{s,0}$ that minimises $\|\x{0}-P x_{s,0}\|_M$ and then project it into the representative points, i.e., $\xinh:= \Pi(x_{s,0})$ with $x_{s,0}:=(P^T M P)^{-1}\! P^T\!M\xin$.

\noindent{\itshape \bfseries Condition A2.}
This condition $\mathbf d_\Y(\yh{},\y{})=\|\yh{}-\y{}\|\le \epsilon$ for any $(\xh{},\x{})\in\rel$ is achieved if $ C_s = CP$, and $C_s^TC_s \le M$.
\noindent{\itshape \bfseries Condition A3.}
This condition holds if the following inequality
	\begin{equation}
	\label{eq:case_study}
	\|\bar A\bar x+\bar B \ach{} + \bar B_w w + P\beta\|_M\le \epsilon
	\end{equation} with matrices 
	defined as $\bar A := A+BK$, $\bar B := BR-PB_s$, $\bar B_{w} := B_{w}-PB_{ws}$ holds with probability at least $(1-\delta)$ for all bounded values of $\bar x,\ach{},$ and $\beta$.
	That is, \eqref{eq:case_study} has to be satisfied  \begin{itemize}
	\item for all $\beta\in \Delta$,
		\item for all $\bar x$ with $\|\bar x\|_M\le \epsilon$,
		\item for all $w\in \mathcal C_w$, where $\mathcal C_w$ is a set with $\mathbb P(w\in \mathcal C_w)\ge 1-\delta$,
		\item and for all $\ach{}\in \hat\A$.

	\end{itemize}
%
%
Condition {\bfseries A3.} in~\eqref{eq:case_study} can be checked using LMIs and S-procedure \citep{Boyd2004}.
Notice that the output deviation $\epsilon$ depends on the attenuation of the disturbance inputs $\bar B \ach{} + \bar B_w w + P\beta$.
Additionally, when there is no order reduction, the disturbance input reduces to $P\beta$, i.e., the impact of $\ach{}$ and $w$ in \eqref{eq:case_study} equate to zero and the resulting approximate simulation relation does not have a deviation in probability: $\delta=0$.

\noindent{\bfseries \itshape Interface condition \eqref{eq:InF}.}
This condition holds if the interface defined in \eqref{eq:interface}, is an interface of the form \eqref{eq:InF}. 
More specifically, we require that over the domain $\hat\A\times \rel$ the interface takes values in $\mathcal P(\mathbb U,\mathcal B(\mathbb{U}))$. Since \eqref{eq:interface} yields a deterministic value, this condition can be written as
\begin{align} 
 R \ach{} + Q \xh{} + K\bar x  \in \A,
\end{align}
for all $\ach{}\in \hat\A$, $\xh{}\in \hat\X$, and for all $\bar x$ with $\|\bar x\|_M\le \epsilon$. This constraint can be verified with linear matrix inequalities when $\A$ is bounded.


\subsection{Case studies}
\label{sec:case_study}
\begin{figure}[htp]
\centering
	\includegraphics[width=.2\textwidth, trim={0 0 0 6.5cm},clip]{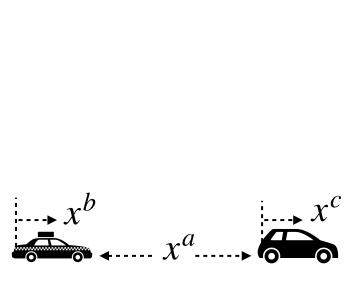}
	\includegraphics[width=.25\textwidth,trim={0 0 0 6.5cm},clip]{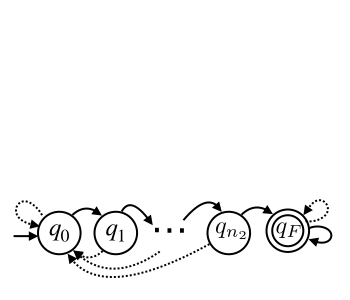}
	\caption{The specification $\psi = \eventually\always^{\le n_2}\{x^a\in K_s\}$ with the DFA for $\psi$
	(right).}\label{fig:gametag}
\end{figure}

\noindent\textbf{Toy example.} We consider the specification $\psi = \eventually\always^{\le n_2} K_s$ which encodes reach and stay over bounded time intervals. The associated DFA is given in Fig.~\ref{fig:gametag}, together with an illustration of a car following another as a potential application of this toy example.

Consider the original model $\M$, which is a 3-dimensional linear system with output $y_t=x_t^a$ and
\begin{align*} \displaybreak[3]
x^a_{t+1}&=x^a_t- a_1(x_t^b-x_t^c)-a_2u_t+a_3w_t\notag
\\ \displaybreak[3]
x_{t+1}^b&=b x_t^b+u_t\notag\\
x_{t+1}^c&= c_1x_t^c+c_2w_t
\end{align*}
with $a_1=0.3$, $a_2=0.03$, $a_3=0.006$, $b=c_1=0.8$ and $c_2=0.1$.
For the specification we select $n_2=3$.
%
%
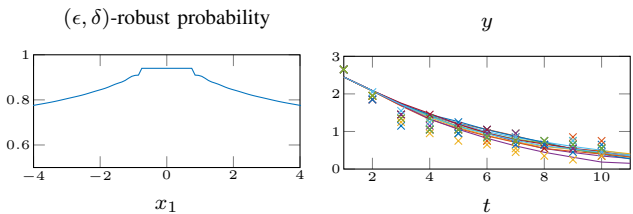
\begin{figure}
\centering
\definecolor{mycolor2}{rgb}{0.85000,0.32500,0.09800}%
\definecolor{mycolor3}{rgb}{0.92900,0.69400,0.12500}%
\definecolor{mycolor4}{rgb}{0.49400,0.18400,0.55600}%
\definecolor{mycolor5}{rgb}{0.46600,0.67400,0.18800}%
\definecolor{mycolor6}{rgb}{0.30100,0.74500,0.93300}%
\definecolor{mycolor7}{rgb}{0.63500,0.07800,0.18400}%

\definecolor{mycolor1}{rgb}{0.00000,0.44700,0.74100}%
\begin{tikzpicture}
\begin{axis}[%
x label style={at={(axis description cs:0.5,0.05)},anchor=north},
 y label style={at={(axis description cs:0.2,0.5)}}, 
every x tick label/.append style={font=\tiny, yshift=0.5ex},
every y tick label/.append style={font=\tiny, xshift=0.5ex},
width=.4\columnwidth,
height=1.5cm,  
    xlabel near ticks,
scale only axis,
xmin=-4,
xmax=4,
ymin=0.5,
ymax=1,
axis background/.style={fill=white},title={\footnotesize $(\eps,\delta)$-robust probability},xlabel={\footnotesize $x_1$},ylabel style={at={(0.07,0.5)}, anchor=south}
]
\addplot [color=mycolor1, forget plot]
  table[row sep=crcr]{%
-9.95	0.677144666152418\\
-9.85	0.678305421981977\\
-9.75	0.679478388359084\\
-9.65	0.680661257440357\\
-9.55	0.681851665446273\\
-9.45	0.683047342009643\\
-9.35	0.684246268721207\\
-9.25	0.685446837811119\\
-9.15	0.686648000684952\\
-9.05	0.687849395269387\\
-8.95	0.689051440846107\\
-8.85	0.69025538959911\\
-8.75	0.691463325743889\\
-8.65	0.692678105755258\\
-8.55	0.693903236682071\\
-8.45	0.695142693816228\\
-8.35	0.696400684004567\\
-8.25	0.697681366311799\\
-8.15	0.698988547058625\\
-8.05	0.700325370980316\\
-7.95	0.701694033921699\\
-7.85	0.703095544668799\\
-7.75	0.704529563717926\\
-7.65	0.705994344421724\\
-7.55	0.7074867966637\\
-7.45	0.709002685070524\\
-7.35	0.710536962956783\\
-7.25	0.712084229809002\\
-7.15	0.713639284799467\\
-7.05	0.715197733186703\\
-6.95	0.716756588408367\\
-6.85	0.718314801923146\\
-6.75	0.719873647593133\\
-6.65	0.721436890361526\\
-6.55	0.723010682168515\\
-6.45	0.724603151887945\\
-6.35	0.72622368982775\\
-6.25	0.727881968954847\\
-6.15	0.729586790277606\\
-6.05	0.731344882751639\\
-5.95	0.73315982184389\\
-5.85	0.735031248144028\\
-5.75	0.736954561099739\\
-5.65	0.738921227876835\\
-5.55	0.740919780625773\\
-5.45	0.742937477254825\\
-5.35	0.744962477601931\\
-5.25	0.746986253840666\\
-5.15	0.749005834201104\\
-5.05	0.751025402977378\\
-4.95	0.753056782942857\\
-4.85	0.755118439597248\\
-4.75	0.757232881390596\\
-4.65	0.759422668655986\\
-4.55	0.761705636131197\\
-4.45	0.764090296204619\\
-4.35	0.7665726186821\\
-4.25	0.769135376462761\\
-4.15	0.771750922236965\\
-4.05	0.774387574913598\\
-3.95	0.777018779294415\\
-3.85	0.779633018413099\\
-3.75	0.782241435843712\\
-3.65	0.784879766872202\\
-3.55	0.787602016799537\\
-3.45	0.790465618351263\\
-3.35	0.793511206697827\\
-3.25	0.796743630861334\\
-3.15	0.800122856048606\\
-3.05	0.803572496094751\\
-2.95	0.807008821923898\\
-2.85	0.810384048310837\\
-2.75	0.813726308813308\\
-2.65	0.817150644684607\\
-2.55	0.820819922048811\\
-2.45	0.82485852142605\\
-2.35	0.829258169462602\\
-2.25	0.833841884672989\\
-2.15	0.838344117021138\\
-2.05	0.842609843784117\\
-1.95	0.846808088803896\\
-1.85	0.851438258192076\\
-1.75	0.856955591684379\\
-1.65	0.863221580619784\\
-1.55	0.869373639877674\\
-1.45	0.874548455559123\\
-1.35	0.879304163631906\\
-1.25	0.886134685959379\\
-1.15	0.895871226231319\\
-1.05	0.904509041660057\\
-0.95	0.908727712011937\\
-0.850000000000001	0.909833408427297\\
-0.75	0.93997618858648\\
-0.649999999999999	0.939999054673196\\
-0.55	0.93999997723423\\
-0.45	0.939999999506008\\
-0.350000000000001	0.939999999984391\\
-0.25	0.939999999999453\\
-0.149999999999999	0.939999999999982\\
-0.0499999999999998	0.939999999999988\\
0.0499999999999998	0.939999999999988\\
0.149999999999999	0.939999999999982\\
0.25	0.939999999999453\\
0.350000000000001	0.939999999984391\\
0.45	0.939999999506008\\
0.55	0.93999997723423\\
0.649999999999999	0.939999054673196\\
0.75	0.93997618858648\\
0.850000000000001	0.909833408427297\\
0.95	0.908727712011937\\
1.05	0.904509041660058\\
1.15	0.895871226231319\\
1.25	0.886134685959379\\
1.35	0.879304163631906\\
1.45	0.874548455559123\\
1.55	0.869373639877674\\
1.65	0.863221580619784\\
1.75	0.856955591684379\\
1.85	0.851438258192076\\
1.95	0.846808088803896\\
2.05	0.842609843784117\\
2.15	0.838344117021138\\
2.25	0.833841884672989\\
2.35	0.829258169462602\\
2.45	0.82485852142605\\
2.55	0.820819922048811\\
2.65	0.817150644684607\\
2.75	0.813726308813308\\
2.85	0.810384048310838\\
2.95	0.807008821923898\\
3.05	0.803572496094752\\
3.15	0.800122856048606\\
3.25	0.796743630861334\\
3.35	0.793511206697826\\
3.45	0.790465618351263\\
3.55	0.787602016799537\\
3.65	0.784879766872202\\
3.75	0.782241435843712\\
3.85	0.779633018413099\\
3.95	0.777018779294415\\
4.05	0.774387574913598\\
4.15	0.771750922236965\\
4.25	0.769135376462761\\
4.35	0.7665726186821\\
4.45	0.764090296204619\\
4.55	0.761705636131197\\
4.65	0.759422668655986\\
4.75	0.757232881390596\\
4.85	0.755118439597248\\
4.95	0.753056782942857\\
5.05	0.751025402977378\\
5.15	0.749005834201104\\
5.25	0.746986253840667\\
5.35	0.744962477601931\\
5.45	0.742937477254825\\
5.55	0.740919780625773\\
5.65	0.738921227876835\\
5.75	0.736954561099739\\
5.85	0.735031248144029\\
5.95	0.73315982184389\\
6.05	0.731344882751639\\
6.15	0.729586790277606\\
6.25	0.727881968954847\\
6.35	0.72622368982775\\
6.45	0.724603151887944\\
6.55	0.723010682168515\\
6.65	0.721436890361526\\
6.75	0.719873647593133\\
6.85	0.718314801923146\\
6.95	0.716756588408368\\
7.05	0.715197733186704\\
7.15	0.713639284799467\\
7.25	0.712084229809002\\
7.35	0.710536962956784\\
7.45	0.709002685070524\\
7.55	0.7074867966637\\
7.65	0.705994344421724\\
7.75	0.704529563717926\\
7.85	0.703095544668799\\
7.95	0.701694033921699\\
8.05	0.700325370980316\\
8.15	0.698988547058625\\
8.25	0.697681366311799\\
8.35	0.696400684004567\\
8.45	0.695142693816228\\
8.55	0.693903236682071\\
8.65	0.692678105755258\\
8.75	0.691463325743889\\
8.85	0.69025538959911\\
8.95	0.689051440846107\\
9.05	0.687849395269387\\
9.15	0.686648000684952\\
9.25	0.685446837811119\\
9.35	0.684246268721207\\
9.45	0.683047342009644\\
9.55	0.681851665446273\\
9.65	0.680661257440358\\
9.75	0.679478388359084\\
9.85	0.678305421981977\\
9.95	0.677144666152418\\
};
\end{axis}
\end{tikzpicture} \begin{tikzpicture}
\begin{axis}[%
x label style={at={(axis description cs:0.5,0.05)},anchor=north},
 y label style={at={(axis description cs:0.2,0.5)}}, 
every x tick label/.append style={font=\tiny, yshift=0.5ex},
every y tick label/.append style={font=\tiny, xshift=0.5ex},
width=.43\columnwidth, 
    xlabel near ticks,
height=1.5cm, 
scale only axis,
xmin=1,title={\footnotesize $y$},xlabel={\footnotesize $t$},y label style={at={(0.05,0.5)}},
xmax=11,
ymin=0,
ymax=3,
axis background/.style={fill=white}
]
\addplot [color=mycolor1, draw=none, mark=x, mark options={solid, mycolor1}, forget plot]
  table[row sep=crcr]{%
1	2.65\\
2	1.95\\
3	1.45\\
4	1.35\\
5	1.25\\
6	1.05\\
7	0.75\\
8	0.649999999999999\\
9	0.45\\
10	0.649999999999999\\
};
\addplot [color=mycolor2, forget plot]
  table[row sep=crcr]{%
1	2.45\\
2	2.08281825883856\\
3	1.72297166354556\\
4	1.43203695709684\\
5	1.21690498767597\\
6	1.05454638058171\\
7	0.880512997649619\\
8	0.689425568595503\\
9	0.533871365035334\\
10	0.414391332403024\\
11	0.391748675132592\\
};
\addplot [color=mycolor3, draw=none, mark=x, mark options={solid, mycolor3}, forget plot]
  table[row sep=crcr]{%
1	2.65\\
2	1.95\\
3	1.45\\
4	1.25\\
5	1.05\\
6	0.75\\
7	0.55\\
8	0.55\\
9	0.55\\
10	0.45\\
};
\addplot [color=mycolor4, forget plot]
  table[row sep=crcr]{%
1	2.45\\
2	2.0871113747558\\
3	1.7428081605892\\
4	1.43212362636695\\
5	1.19378301181628\\
6	0.955399219287496\\
7	0.7311315249561\\
8	0.546350199258822\\
9	0.439466169410851\\
10	0.347023167036659\\
11	0.282554458630428\\
};
\addplot [color=mycolor5, draw=none, mark=x, mark options={solid, mycolor5}, forget plot]
  table[row sep=crcr]{%
1	2.65\\
2	1.95\\
3	1.45\\
4	1.35\\
5	1.05\\
6	0.95\\
7	0.95\\
8	0.75\\
9	0.75\\
10	0.55\\
};
\addplot [color=mycolor6, forget plot]
  table[row sep=crcr]{%
1	2.45\\
2	2.08437242029748\\
3	1.72964250424555\\
4	1.43256266099907\\
5	1.21546678473743\\
6	0.992270428898916\\
7	0.839742570952465\\
8	0.72036722767742\\
9	0.600287536685836\\
10	0.50011691129157\\
11	0.379700889442169\\
};
\addplot [color=mycolor7, draw=none, mark=x, mark options={solid, mycolor7}, forget plot]
  table[row sep=crcr]{%
1	2.65\\
2	1.95\\
3	1.45\\
4	1.45\\
5	1.15\\
6	1.05\\
7	0.850000000000001\\
8	0.55\\
9	0.55\\
10	0.350000000000001\\
};
\addplot [color=mycolor1, forget plot]
  table[row sep=crcr]{%
1	2.45\\
2	2.08325216234967\\
3	1.72644664053003\\
4	1.44940682976872\\
5	1.26307576843685\\
6	1.04449841608698\\
7	0.871148580223421\\
8	0.710470267671952\\
9	0.530648338704551\\
10	0.398506689964389\\
11	0.27030802594444\\
};
\addplot [color=mycolor2, draw=none, mark=x, mark options={solid, mycolor2}, forget plot]
  table[row sep=crcr]{%
1	2.65\\
2	1.85\\
3	1.35\\
4	1.05\\
5	0.95\\
6	0.75\\
7	0.75\\
8	0.75\\
9	0.850000000000001\\
10	0.75\\
};
\addplot [color=mycolor3, forget plot]
  table[row sep=crcr]{%
1	2.45\\
2	2.0772947215596\\
3	1.68891743467978\\
4	1.3717482384539\\
5	1.10213105092041\\
6	0.89691729369732\\
7	0.716580065595192\\
8	0.597768286705844\\
9	0.523060979159613\\
10	0.487065777871116\\
11	0.405949067969639\\
};
\addplot [color=mycolor4, draw=none, mark=x, mark options={solid, mycolor4}, forget plot]
  table[row sep=crcr]{%
1	2.65\\
2	1.85\\
3	1.45\\
4	1.15\\
5	1.05\\
6	0.95\\
7	0.95\\
8	0.649999999999999\\
9	0.55\\
10	0.45\\
};
\addplot [color=mycolor5, forget plot]
  table[row sep=crcr]{%
1	2.45\\
2	2.07758538156955\\
3	1.69809006231344\\
4	1.41307036584877\\
5	1.13738862532075\\
6	0.952014365546549\\
7	0.797246278063067\\
8	0.688231149487747\\
9	0.524584574827064\\
10	0.410446498436523\\
11	0.315811946468197\\
};
\addplot [color=mycolor6, draw=none, mark=x, mark options={solid, mycolor6}, forget plot]
  table[row sep=crcr]{%
1	2.65\\
2	2.05\\
3	1.55\\
4	1.25\\
5	0.95\\
6	0.850000000000001\\
7	0.850000000000001\\
8	0.75\\
9	0.75\\
10	0.45\\
};
\addplot [color=mycolor7, forget plot]
  table[row sep=crcr]{%
1	2.45\\
2	2.08757422741894\\
3	1.7549242426729\\
4	1.47743988099213\\
5	1.19200132120817\\
6	0.959745423464698\\
7	0.780160877253685\\
8	0.665133984079329\\
9	0.550736642387298\\
10	0.463973836593935\\
11	0.330710595483775\\
};
\addplot [color=mycolor1, draw=none, mark=x, mark options={solid, mycolor1}, forget plot]
  table[row sep=crcr]{%
1	2.65\\
2	1.85\\
3	1.15\\
4	1.05\\
5	0.95\\
6	0.850000000000001\\
7	0.649999999999999\\
8	0.649999999999999\\
9	0.649999999999999\\
10	0.55\\
};
\addplot [color=mycolor2, forget plot]
  table[row sep=crcr]{%
1	2.45\\
2	2.07768139490263\\
3	1.68104374922527\\
4	1.32335770020432\\
5	1.07622543290764\\
6	0.879841144887135\\
7	0.70973102546008\\
8	0.547717930397934\\
9	0.470549027258106\\
10	0.398957098087294\\
11	0.326394933925277\\
};
\addplot [color=mycolor3, draw=none, mark=x, mark options={solid, mycolor3}, forget plot]
  table[row sep=crcr]{%
1	2.65\\
2	1.95\\
3	1.25\\
4	0.95\\
5	0.75\\
6	0.649999999999999\\
7	0.45\\
8	0.350000000000001\\
9	0.25\\
10	0.350000000000001\\
};
\addplot [color=mycolor4, forget plot]
  table[row sep=crcr]{%
1	2.45\\
2	2.08105094144378\\
3	1.69914927389061\\
4	1.33438418443958\\
5	1.05752257857596\\
6	0.818949754117422\\
7	0.615968525378736\\
8	0.446628460712029\\
9	0.314315876454544\\
10	0.185747938625319\\
11	0.153017814510057\\
};
\addplot [color=mycolor5, draw=none, mark=x, mark options={solid, mycolor5}, forget plot]
  table[row sep=crcr]{%
1	2.65\\
2	1.95\\
3	1.35\\
4	1.05\\
5	1.05\\
6	0.850000000000001\\
7	0.75\\
8	0.75\\
9	0.649999999999999\\
10	0.55\\
};
\addplot [color=mycolor6, forget plot]
  table[row sep=crcr]{%
1	2.45\\
2	2.08250274694372\\
3	1.71766155327182\\
4	1.39866695369378\\
5	1.12459444091449\\
6	0.939007741931164\\
7	0.765132051656333\\
8	0.615380404070769\\
9	0.5375845282806\\
10	0.442918007569914\\
11	0.351034337112777\\
};
\end{axis}
\end{tikzpicture}\\[-1em]
\caption{ On the left: $(\eps,\delta)$-robust satisfaction probability of $\eventually\always^{\le n_2}\{y\in [-2,2]\}$ with $\eps= 1.2266$ and $\delta=0.03$.
	On the right: simulation runs for the original model  and the abstract model with the composed robust controller.}
\label{fig:simresult}
\end{figure}
We follow Sec.~\ref{sec:abstr} and compute a one-dimensional model $\M_s$ via the balanced truncation of the original controlled model with a suitable feedback gain $K=[-0.7738, 0.9369, -0.6829]$.
In Fig.~\ref{fig:simresult}, we give an example of such a robust temporal logic computation.  On the right side of the figure, 10 simulation runs are given that are initialised at $[x_a,\,x_b,\,x_c] = [2.45,\,2.5,\,1.3]$.
Crosses are the outputs of $\widehat\M$ whereas lines are the outputs of $\M$.

\begin{figure*}[t]
\definecolor{color1}{rgb}
{1,0.498039215686275,0.0549019607843137}
\definecolor{color0}{rgb}{0.12156862745098,0.466666666666667,0.705882352941177}
\definecolor{color3}{rgb}{0.83921568627451,0.152941176470588,0.156862745098039}
\definecolor{color2}{rgb}{0.1,0.4,0.08}
\definecolor{color5}{rgb}{0.549019607843137,0.337254901960784,0.294117647058824}
\definecolor{color4}{rgb}{0.580392156862745,0.403921568627451,0.741176470588235}
 \begin{minipage}[b]{\columnwidth}
\begin{tikzpicture}

\begin{axis}[
xlabel={\footnotesize $x_1$},
ylabel={\footnotesize $x_2$},
x label style={at={(axis description cs:0.5,0.05)},anchor=north},
 y label style={at={(axis description cs:0.05,0.5)}}, 
every x tick label/.append style={font=\tiny, yshift=0.5ex},
every y tick label/.append style={font=\tiny, xshift=0.5ex},
xmin=-11, xmax=11,
ymin=-11, ymax=11,
width=0.85\columnwidth,
height=0.85\columnwidth,
tick align=outside,
tick pos=left,
x grid style={lightgray!92.026143790849673!black},
y grid style={lightgray!92.026143790849673!black}
]
\addplot [only marks, mark options = {scale = 0.5}, mark options = {scale = 0.5},draw=black, fill=black]
table{%
x                      y
-5.000000000000000e+00 -7.500000000000000e+00
-5.725274551177081e+00 -7.274363091974448e+00
-6.916706085191789e+00 -5.895169153793956e+00
-7.054413488243962e+00 -5.548738601977000e+00
-6.617003864535690e+00 -4.635289598855503e+00
-6.270167131211649e+00 -3.639654643511439e+00
-6.700411212689070e+00 -2.828316280222954e+00
-6.786248116396539e+00 -1.385723631062333e+00
-6.945086945987489e+00 -3.057572100478813e-01
-7.468885125450747e+00 +3.825136300287954e-01
-7.536158552450306e+00 +1.597830905264234e+00
-7.345474530565864e+00 +2.458267040554325e+00
-6.897443574146584e+00 +3.606599551383728e+00
-5.858262423241204e+00 +4.442255377532495e+00
-5.138951136758745e+00 +4.856842517860159e+00
-3.617426074151047e+00 +4.487811509921817e+00
-2.036386777142462e+00 +4.610092425723711e+00
-7.393986366501411e-01 +3.470849974270804e+00
-1.003920473973694e-01 +4.203147856553773e+00
+9.775751735352661e-01 +4.527640487772146e+00
+2.007775137389935e+00 +4.192225791085999e+00
+1.362659000161790e+00 +3.664784614222673e+00
+2.327775591870608e+00 +4.115104153652663e+00
+2.580245176329460e+00 +2.636028913876157e+00
+2.735217748646167e+00 +1.283587306864419e+00
+1.635532114751951e+00 +2.386488241695575e-01
+1.982887835503985e+00 -8.468768027814521e-01
+2.552528502654961e+00 -2.907018306395518e-01
+2.301757539046146e+00 -2.215676175740186e-02
+2.130606124721249e+00 +3.625613711581309e-01
+2.176580847868236e+00 +6.463348231671654e-01
+2.638493586196688e+00 +1.316497015478817e+00
+3.066095077506222e+00 +2.580406875686362e+00
+2.459337404485105e+00 +3.521627388322125e+00
+1.815135953360716e+00 +3.992515124144457e+00
+1.211810288113829e+00 +4.797696176440447e+00
-5.300247851011997e-02 +3.821393764282960e+00
-1.146462595046015e+00 +4.401341508041059e+00
-2.263282715411896e+00 +4.929165421984651e+00
-3.263083249202902e+00 +4.238728352821157e+00
-4.152649238800384e+00 +4.975714582905404e+00
-5.192206048655768e+00 +3.917759320526970e+00
-6.463459101199735e+00 +4.243867540509451e+00
-7.320150371374558e+00 +3.262677540371834e+00
-7.222469348869547e+00 +2.135842555391880e+00
-7.978416766564415e+00 +1.128982441146786e+00
-7.298361400538812e+00 +6.144760049602621e-01
-7.563347147699741e+00 -1.255495534724380e-01
-7.498147386041391e+00 -9.622845785942504e-01
-7.814398849165864e+00 -1.785804126208810e+00
-7.610452912735227e+00 -2.771989678937776e+00
-7.626563902641481e+00 -3.464648468887892e+00
-7.249088181819166e+00 -4.264731538969516e+00
-7.656243723181650e+00 -4.810916110171738e+00
-6.661416399704164e+00 -5.802963165454329e+00
-7.266928825183555e+00 -6.337285293479979e+00
-6.103498716736530e+00 -7.240156002395449e+00
-5.127935682558647e+00 -7.391588623464089e+00
-3.606798951522518e+00 -7.414542536418556e+00
-2.270707631128309e+00 -7.193284893101087e+00
-2.123228564976691e+00 -7.535146333009081e+00
-9.453693238157175e-01 -6.810474960623736e+00
-4.388219408040323e-01 -6.798124666730367e+00
+4.734969559320978e-01 -6.815680126121477e+00
+1.270202417625269e+00 -6.752177150833894e+00
+2.267958030078067e+00 -6.764117668195766e+00
+2.864110892401889e+00 -6.838660057629107e+00
+4.015517632873808e+00 -6.562433266167627e+00
+5.059923457534653e+00 -7.270642232134364e+00
+4.291836591619865e+00 -7.484785621671303e+00
};
\addplot [only marks, mark options = {scale = 0.5}, draw=black, fill=black]
table{%
x                      y
-5.000000000000000e+00 -7.500000000000000e+00
-5.840129333706137e+00 -6.925436671843272e+00
-6.713047506787407e+00 -6.418821425825048e+00
-6.721883522891786e+00 -5.913142775311301e+00
-6.704441866292035e+00 -5.308884791659988e+00
-6.811231966716916e+00 -4.147874701926064e+00
-6.760850900442732e+00 -2.954191133630103e+00
-6.421558269916237e+00 -1.101623387808593e+00
-6.661812143116975e+00 -3.305385918992048e-01
-6.830446250381947e+00 +4.304040818775126e-01
-6.830054199661705e+00 +1.567375230532521e+00
-6.441568634485592e+00 +2.702551430076172e+00
-7.002180218503611e+00 +4.200654802822340e+00
-5.806614225138405e+00 +4.450689075169032e+00
-4.976956810420233e+00 +5.017898483654350e+00
-4.248363178444998e+00 +4.391030294687241e+00
-3.109166620857703e+00 +4.909851397926321e+00
-2.033884006907404e+00 +4.304508375183025e+00
-1.313442137853893e+00 +5.536507056579126e+00
+1.817657679476670e-03 +4.048848785911108e+00
+7.146228474291756e-01 +4.777144566686193e+00
+1.258679883849429e+00 +4.657853156559186e+00
+1.744742020908438e+00 +3.537042237725916e+00
+1.794034213388293e+00 +2.021904846887216e+00
+2.639458220153538e+00 +1.235905968723717e+00
+9.693829251465382e-01 +4.576930075450987e-01
+1.847666164750239e+00 -8.602263741859397e-01
+1.930369281814160e+00 -5.016232387694695e-01
+1.925354512413747e+00 -6.978683920247403e-02
+1.471923684888115e+00 +3.098085523539292e-01
+1.872452841045246e+00 +1.462423237624730e-01
+1.499885135047744e+00 +5.292011389774798e-02
+1.927190898390502e+00 +5.590553233234375e-01
+1.797448970592443e+00 +7.778495622618778e-01
+1.894351311856877e+00 +1.465372037339322e+00
+2.527579204768744e+00 +2.727745896201935e+00
+1.159718920816417e+00 +4.409506987660409e+00
+3.764119372019023e-01 +5.621372631388243e+00
-1.590202798956966e-01 +4.841457168683859e+00
-7.262474627392059e-01 +4.521611511847804e+00
-2.276295167827014e+00 +4.634724706203775e+00
-3.324662415214879e+00 +4.198319703866732e+00
-4.102495228694117e+00 +4.482794414231633e+00
-4.921318191262737e+00 +4.320047839089857e+00
-5.583623927327953e+00 +4.972882906610631e+00
-6.655081034964394e+00 +4.176810546683169e+00
-6.546077692762243e+00 +2.813299775094283e+00
-6.820045504673098e+00 +2.056088738170380e+00
-7.413596408290086e+00 +1.128414766403023e+00
-7.416389776779365e+00 +3.453189154280039e-01
-7.006649008533885e+00 -3.380139344164785e-01
-7.460372064515105e+00 -1.400061973760684e+00
-7.178025116050052e+00 -2.376721987151214e+00
-6.555941474078435e+00 -3.113319762054136e+00
-6.676746448943308e+00 -3.969006807740672e+00
-6.522869021238845e+00 -5.093255684714369e+00
-5.852308980660111e+00 -6.075443562899910e+00
-7.097984142591698e+00 -7.300919912391397e+00
-6.788990958729159e+00 -6.083700938313759e+00
-7.864638173031759e+00 -6.386285751445590e+00
-6.126767017254618e+00 -7.011647282483709e+00
-4.823513137552041e+00 -6.875259637384238e+00
-3.715019306392084e+00 -6.698404024850979e+00
-2.686638902054006e+00 -5.882439399074999e+00
-1.613894879360688e+00 -6.687960985883548e+00
+2.119938425109373e-01 -6.548344973757495e+00
+8.528033591146658e-01 -5.934838002077560e+00
+1.565317710060678e+00 -6.264241633481555e+00
+2.327008607929426e+00 -6.341322142835821e+00
+3.050052660744395e+00 -6.789325919259316e+00
+4.089448037736641e+00 -6.525259398587920e+00
+5.108924568658319e+00 -6.699240961621445e+00
+4.499623611976911e+00 -6.812100830392377e+00
};
\addplot [only marks, mark options = {scale = 0.5}, draw=black, fill=black]
table{%
x                      y
-5.000000000000000e+00 -7.500000000000000e+00
-5.811112638384163e+00 -6.511883850922254e+00
-6.173545114304214e+00 -5.888293196480565e+00
-6.600811637225125e+00 -4.784903205863253e+00
-6.404143266917858e+00 -4.257471248430599e+00
-7.025802811448859e+00 -3.531951684901331e+00
-7.137938165730178e+00 -2.285625484660645e+00
-7.441281920167435e+00 -1.035674749146029e+00
-7.974298097150348e+00 -1.508859951975011e-01
-7.217539548135225e+00 +9.506757338264772e-01
-6.679247671018913e+00 +2.117169780203930e+00
-6.852358806602715e+00 +3.212873333561176e+00
-6.466539259091748e+00 +4.100554912890588e+00
-5.744382610485858e+00 +4.622986090930254e+00
-4.437569250764225e+00 +4.017006369697624e+00
-3.324490069245163e+00 +4.807230042137422e+00
-2.844500746332794e+00 +4.591081977144801e+00
-1.900377003628560e+00 +4.336710588449461e+00
-1.133319077923016e+00 +4.727988302318053e+00
-1.132328792339500e-01 +3.786394245637895e+00
+6.978022003159062e-01 +4.858844996670578e+00
+1.731822285089845e+00 +4.165221298284609e+00
+2.639010102790442e+00 +3.905490156118073e+00
+2.415128212627676e+00 +2.820834030929852e+00
+2.081280216572272e+00 +1.588820312004769e+00
+2.375531352861387e+00 +1.168131961061778e+00
+2.189601490957548e+00 +2.013791110684253e-01
+1.303992533149803e+00 -1.527831101903622e-01
+2.367981463873154e+00 -7.018904805965018e-01
+8.493367060642112e-01 -5.287857611544422e-02
+1.961711028968282e+00 -2.445572409114425e-01
+2.429124664201978e+00 +3.821724916943326e-02
+2.748205144692719e+00 +4.682831957873700e-01
+2.830377314666067e+00 +1.807492809123992e+00
+3.010319174673771e+00 +2.606131426366598e+00
+2.599521339847612e+00 +3.258431260774413e+00
+1.635395649935059e+00 +4.083171442517266e+00
+2.656685584996924e-01 +4.806322038257077e+00
-1.149332690032712e+00 +3.996126605479692e+00
-2.185927994622484e+00 +4.949750877866339e+00
-3.007468494648045e+00 +3.658270905344571e+00
-4.183438981321008e+00 +3.782782265830686e+00
-5.161480378870837e+00 +4.101713300537027e+00
-6.134455201678292e+00 +4.509879047450164e+00
-7.004849484857603e+00 +4.343842046036446e+00
-7.252205926975292e+00 +3.514274581932657e+00
-7.088995745108271e+00 +2.475471025512407e+00
-6.281638416127552e+00 +9.075208851389954e-01
-6.644137965630708e+00 -3.333879518841588e-01
-7.000760035961813e+00 -8.524814947892705e-01
-7.229834028067650e+00 -1.911442219251484e+00
-6.573922663203309e+00 -2.737971954685529e+00
-6.338075188453966e+00 -3.988276290109428e+00
-6.556354772049994e+00 -4.649865056913201e+00
-6.415487897634268e+00 -5.602539477760986e+00
-7.228333981555011e+00 -6.263651606386997e+00
-5.999589462461429e+00 -6.878978805168403e+00
-4.671275950452986e+00 -6.934178832348363e+00
-3.674027974644444e+00 -6.979759069775956e+00
-2.124956177736443e+00 -7.001545813684735e+00
-1.844327493465779e+00 -6.713221733176442e+00
-6.129789240086454e-01 -6.470029580285637e+00
+2.887505196356884e-01 -6.055999293722058e+00
+1.269943740875576e+00 -6.831173116229751e+00
+2.358093050377411e+00 -6.876865320694244e+00
+3.337723503408935e+00 -6.998810482999557e+00
+4.898322060744987e+00 -7.426054256480143e+00
+5.162325151936533e+00 -6.792925814410204e+00
+4.799093155451151e+00 -6.361682592777756e+00
};
\addplot [only marks, mark options = {scale = 0.5}, draw=black, fill=black]
table{%
x                      y
-7.500000000000000e+00 +5.000000000000000e+00
-6.744635595641749e+00 +3.951883167148214e+00
-5.694247501657154e+00 +4.191434207902267e+00
-4.399611691756552e+00 +4.963945311053652e+00
-3.935216685031376e+00 +4.655061460007345e+00
-3.178595201498936e+00 +3.496066168718052e+00
-2.121927799820796e+00 +3.806623936797868e+00
-1.362675754060444e+00 +4.715291609984459e+00
-7.815823856090359e-02 +4.120637078503496e+00
+9.267798853975181e-01 +4.679048458491646e+00
+1.842010587691883e+00 +3.874538516859276e+00
+2.209617803209373e+00 +3.700554374615677e+00
+1.855428333651860e+00 +2.898041670530837e+00
+2.082536985438045e+00 +2.020686944808542e+00
+2.638247643731904e+00 +3.653140982797928e-01
+1.778840652884586e+00 -8.590263068321714e-01
+1.300018996240846e+00 +1.168308243398948e-01
+2.547964874486007e+00 -6.328818930018097e-01
+1.049224251964856e+00 -6.865073062365362e-02
+2.073335359634027e+00 -4.636357538409999e-01
+1.069023344478246e+00 +4.116210776889679e-01
+1.655124291216647e+00 +2.889243257259194e-01
+1.453219072717866e+00 +1.992948484577794e-01
+2.610047904532526e+00 -7.838552136887778e-01
+1.729526143333209e+00 +8.063540657005824e-01
+2.437279158594599e+00 +3.454711356755221e-01
+2.008848048103003e+00 +6.746085118049781e-01
+1.529031081193366e+00 +2.193560547535566e+00
+2.729856916879323e+00 +3.324479700835936e+00
+2.135427182459973e+00 +3.711482728813535e+00
+9.574616935770732e-01 +3.861039625143300e+00
+6.031551862262788e-02 +5.095321505029112e+00
-7.393417245654660e-01 +4.142495580113890e+00
-2.075118269821744e+00 +4.771166382688340e+00
-3.033103785891619e+00 +4.068009251432748e+00
-4.170339873662440e+00 +4.490142993019834e+00
-5.131181404178468e+00 +5.536462542267790e+00
-6.061823963046185e+00 +4.740988328972980e+00
-6.788498375552925e+00 +4.313625872575203e+00
-7.112167327165586e+00 +3.385173570040083e+00
-6.915540265478951e+00 +2.130338698923011e+00
-6.870927026422585e+00 +1.582224480796535e+00
-6.775975735411712e+00 +2.173018317760878e-01
-6.272850648011650e+00 -7.246932754934171e-01
-7.157698143555420e+00 -2.050461202850248e+00
-6.478556194666530e+00 -3.193027228483281e+00
-6.795974535416466e+00 -3.673766816530847e+00
-6.124128925234746e+00 -4.902865969432911e+00
-7.027314977004234e+00 -5.221863501058183e+00
-7.050335250649336e+00 -5.968136533565166e+00
-6.248345038205984e+00 -7.041855067651455e+00
-5.755015476260694e+00 -7.336102880154775e+00
-4.754591647839272e+00 -7.624896456228073e+00
-3.461199565463953e+00 -7.367828831779341e+00
-2.485112417713903e+00 -7.409907177432645e+00
-1.437703979828895e+00 -7.216821135859081e+00
-1.056031494247419e-01 -6.864869947386469e+00
+5.770765880197990e-01 -6.322234192745085e+00
+1.786674349215392e+00 -6.690655353789629e+00
+2.338536296079307e+00 -6.688050920488941e+00
+3.232795499503864e+00 -6.894517922058264e+00
+4.132257575556788e+00 -7.583927387223652e+00
+5.135184390710029e+00 -6.459654096307801e+00
+4.556633840217510e+00 -6.857433948995480e+00
};
\addplot [only marks, mark options = {scale = 0.5}, draw=black, fill=black]
table{%
x                      y
-7.500000000000000e+00 +5.000000000000000e+00
-6.460374901234796e+00 +4.186643054925650e+00
-5.724416775994434e+00 +4.997142451984187e+00
-4.887167277295058e+00 +3.957618727336911e+00
-4.300498299034541e+00 +5.043403913988515e+00
-3.093611797987879e+00 +3.813850207836557e+00
-2.643219883551826e+00 +4.310200652205110e+00
-2.000977065206044e+00 +5.125104723830856e+00
-1.109730631446529e+00 +4.078548787179320e+00
+8.423012252778669e-02 +4.299783197755758e+00
+6.054099963391451e-01 +5.055757856473071e+00
+1.700617123619491e+00 +5.314779255262220e+00
+2.965928659627616e+00 +4.273292379830370e+00
+2.632393906330515e+00 +3.174273630976788e+00
+2.357159132439224e+00 +2.325970487289698e+00
+2.666320637824318e+00 +1.098724310486789e+00
+1.373939720591312e+00 +2.951340060716936e-01
+1.779198132517356e+00 -5.258388520294366e-01
+2.328164718448904e+00 -3.953947623697839e-01
+2.330891220578048e+00 -3.930536474063802e-01
+2.096902370570359e+00 -3.272719319082890e-01
+1.824754065368463e+00 +5.076657566762934e-02
+1.287380304164972e+00 +3.520626599682992e-01
+1.642858756778841e+00 -2.589119768729184e-01
+1.582634264848863e+00 +8.098801331224003e-01
+1.649182132651292e+00 -1.696570609982962e-01
+1.894620302228766e+00 +2.390281771873475e-01
+1.009423413174312e+00 +2.948760407772939e-01
+1.557361989507233e+00 -8.439394896186454e-02
+1.487963559577443e+00 +4.427663196393619e-01
+2.191067650806398e+00 -1.970864574176149e-01
+2.208550754173991e+00 -9.384944604444861e-03
+1.988502778468961e+00 +3.792859073018909e-01
+2.121733663849699e+00 +1.283895644165796e+00
+2.249511894815482e+00 +2.478255762184927e+00
+1.950572167030088e+00 +3.057796178064995e+00
+1.658183876057843e+00 +4.526919923022539e+00
+1.558426057631897e+00 +4.665837215089279e+00
+3.031134593568487e-01 +4.136422028262808e+00
-7.443323209002732e-01 +4.605789592855665e+00
-1.787422315381889e+00 +3.858784012389710e+00
-2.622064095166166e+00 +4.315062337310946e+00
-3.581943661892748e+00 +5.171004536692792e+00
-3.947823980577772e+00 +4.810425916335737e+00
-5.212193809737201e+00 +3.979275816123426e+00
-6.055340084003253e+00 +4.460897388595922e+00
-6.241847020816854e+00 +4.256025838787300e+00
-6.732903231050363e+00 +4.575222026822567e+00
-7.371143431693726e+00 +4.161542702671159e+00
-7.714157167422160e+00 +3.091073256788626e+00
-7.599144263438880e+00 +2.130541277002405e+00
-7.534254579350401e+00 +1.372974872406331e+00
-7.344695398951696e+00 +4.310126050341694e-01
-7.437449846983741e+00 +9.938020925561730e-03
-7.469868633946795e+00 -9.405991113857645e-01
-7.588172014201900e+00 -1.814551803663025e+00
-7.553207558659611e+00 -2.497547837061705e+00
-7.593717146756132e+00 -2.730970785397519e+00
-7.348764791853907e+00 -4.010620993676305e+00
-7.450222447845568e+00 -4.842219839075984e+00
-6.667780980591486e+00 -6.136601743037335e+00
-6.769860642202385e+00 -6.436508314154872e+00
-6.879207489232853e+00 -7.036138917085869e+00
-6.951828748618581e+00 -6.439551217160282e+00
-6.102959630382283e+00 -7.353629543775288e+00
-4.960733039354712e+00 -7.948182818343650e+00
-4.598436246832422e+00 -7.558039376854166e+00
-3.246729612510323e+00 -7.427528534169164e+00
-1.842930660652901e+00 -7.719574772935593e+00
-6.300064979012577e-01 -7.565203527949445e+00
+1.382212539299433e-01 -7.202173305502508e+00
+9.224657169378696e-01 -7.065547548489343e+00
+1.800870426434312e+00 -6.574600782543151e+00
+2.699822502051220e+00 -6.622597466345415e+00
+3.447841996492921e+00 -7.114903131470005e+00
+4.382036114658102e+00 -6.958638693459712e+00
+5.445118261700264e+00 -7.596035945326426e+00
+4.666135269445613e+00 -6.701139217757017e+00
};
\addplot [only marks, mark options = {scale = 0.5}, draw=black, fill=black]
table{%
x                      y
-7.500000000000000e+00 +5.000000000000000e+00
-6.926789549948845e+00 +4.496167989804123e+00
-6.282646153011134e+00 +4.199693913537823e+00
-5.294998424291075e+00 +4.776838866651555e+00
-3.753411329592676e+00 +4.379997466971223e+00
-2.726367482249068e+00 +4.897741551713962e+00
-1.905472615252590e+00 +3.856897371718992e+00
-1.210376855191549e+00 +4.404074876759680e+00
-2.110057224957921e-01 +4.814165529868961e+00
+7.981713107969873e-01 +4.022351122430810e+00
+1.550741208657060e+00 +3.717670833965778e+00
+1.689578745564128e+00 +2.865146968633304e+00
+2.018150448301460e+00 +1.806873045637492e+00
+1.647576053899371e+00 +7.320005012057313e-01
+1.363321782115180e+00 +5.344686193965997e-03
+1.441480266563452e+00 -7.979525807341117e-01
+1.921783365931007e+00 -5.126004389912773e-01
+1.728826374121236e+00 -7.248401943081962e-01
+2.006463071817182e+00 -1.975697881224114e-01
+2.492799403332450e+00 +1.187710020550070e-01
+2.555104354881887e+00 +4.542492236553097e-01
+2.714404172892429e+00 +1.643705408908319e+00
+2.512198949861723e+00 +3.129926876550536e+00
+1.088117141680509e+00 +3.987568968410807e+00
+3.504974475655009e-02 +4.210660160972866e+00
-9.962203944067103e-01 +5.047663298931527e+00
-2.214532446008159e+00 +4.715617002344946e+00
-2.906138522888148e+00 +4.851797323973128e+00
-3.655136636293490e+00 +4.502631371704387e+00
-5.158175014156092e+00 +4.464197000170433e+00
-6.143092380735662e+00 +3.899792635872896e+00
-6.793318900668146e+00 +3.407873482887240e+00
-6.513810590645152e+00 +2.483306888168670e+00
-6.323914748267847e+00 +1.650877737240451e+00
-7.030858708401772e+00 +1.057006809038508e+00
-7.563092915367824e+00 +4.906793136953849e-01
-7.733743007837491e+00 -4.159248425885813e-01
-7.536097806498229e+00 -1.625609240147883e+00
-7.815005626254231e+00 -2.536609761076992e+00
-7.658892620054782e+00 -3.726942059400761e+00
-7.010920836997744e+00 -4.633284023958007e+00
-6.952642340622654e+00 -5.449941881373901e+00
-6.889999551434446e+00 -6.305917397138657e+00
-5.926106080058394e+00 -7.099410769707294e+00
-4.588841825176424e+00 -7.017737575752716e+00
-3.741637893206209e+00 -7.089204065083478e+00
-2.238580694498270e+00 -7.444892763410461e+00
-1.388526624843780e+00 -6.545137986440811e+00
-4.759046139085142e-01 -6.893171728177657e+00
+2.878429672279988e-01 -6.648861503864071e+00
+1.322508243616187e+00 -6.430669989347013e+00
+2.512712114274011e+00 -6.935966327315828e+00
+3.626747985827134e+00 -6.853195070688815e+00
+3.724696548410744e+00 -7.273672137038064e+00
+4.769100792937478e+00 -7.588553808758162e+00
+5.200375957498510e+00 -6.577427432371008e+00
+4.811786293199703e+00 -6.683389593021383e+00
};
\path [fill=color0, draw opacity=0] (axis cs:-9.25,-10)
--(axis cs:-9.25,10)
--(axis cs:-9.75,10)
--(axis cs:-9.75,-10)
--cycle;

\path [fill=color0, draw opacity=0] (axis cs:9.75,-10)
--(axis cs:9.75,10)
--(axis cs:9.25,10)
--(axis cs:9.25,-10)
--cycle;

\path [fill=color0, draw opacity=0] (axis cs:10,-9.75)
--(axis cs:10,-9.25)
--(axis cs:-10,-9.25)
--(axis cs:-10,-9.75)
--cycle;

\path [fill=color0, draw opacity=0] (axis cs:10,9.25)
--(axis cs:10,9.75)
--(axis cs:-10,9.75)
--(axis cs:-10,9.25)
--cycle;

\path [fill=color0, draw opacity=0] (axis cs:4.75,-5)
--(axis cs:4.75,4.75)
--(axis cs:4.25,4.75)
--(axis cs:4.25,-5)
--cycle;

\path [fill=color0, draw opacity=0] (axis cs:4.75,6.25)
--(axis cs:4.75,6.75)
--(axis cs:-5,6.75)
--(axis cs:-5,6.25)
--cycle;

\path [fill=color0, draw opacity=0] (axis cs:0.75,2.25)
--(axis cs:0.75,2.75)
--(axis cs:-5,2.75)
--(axis cs:-5,2.25)
--cycle;

\path [fill=color0, draw opacity=0] (axis cs:-4.25,-5)
--(axis cs:-4.25,2.5)
--(axis cs:-5,2.5)
--(axis cs:-5,-5)
--cycle;

\path [fill=color0, draw opacity=0] (axis cs:2.75,-5)
--(axis cs:2.75,-4.5)
--(axis cs:-5,-4.5)
--(axis cs:-5,-5)
--cycle;

\path [fill=color1, draw opacity=0] (axis cs:3,-4)
--(axis cs:3,-0.5)
--(axis cs:0,-0.5)
--(axis cs:0,-4)
--cycle;

\path [fill=color2, draw opacity=0] (axis cs:10,-10)
--(axis cs:10,-5)
--(axis cs:5,-5)
--(axis cs:5,-10)
--cycle;

\addplot [semithick, color0, forget plot]
table {%
-5 -7.5
-5.72527455117708 -7.27436309197445
-6.91670608519179 -5.89516915379396
-7.05441348824396 -5.548738601977
-6.61700386453569 -4.6352895988555
-6.27016713121165 -3.63965464351144
-6.70041121268907 -2.82831628022295
-6.78624811639654 -1.38572363106233
-6.94508694598749 -0.305757210047881
-7.46888512545075 0.382513630028795
-7.53615855245031 1.59783090526423
-7.34547453056586 2.45826704055432
-6.89744357414658 3.60659955138373
-5.8582624232412 4.4422553775325
-5.13895113675875 4.85684251786016
-3.61742607415105 4.48781150992182
-2.03638677714246 4.61009242572371
-0.739398636650141 3.4708499742708
-0.100392047397369 4.20314785655377
0.977575173535266 4.52764048777215
2.00777513738994 4.192225791086
1.36265900016179 3.66478461422267
2.32777559187061 4.11510415365266
2.58024517632946 2.63602891387616
2.73521774864617 1.28358730686442
1.63553211475195 0.238648824169557
1.98288783550398 -0.846876802781452
2.55252850265496 -0.290701830639552
2.30175753904615 -0.0221567617574019
2.13060612472125 0.362561371158131
2.17658084786824 0.646334823167165
2.63849358619669 1.31649701547882
3.06609507750622 2.58040687568636
2.4593374044851 3.52162738832213
1.81513595336072 3.99251512414446
1.21181028811383 4.79769617644045
-0.05300247851012 3.82139376428296
-1.14646259504601 4.40134150804106
-2.2632827154119 4.92916542198465
-3.2630832492029 4.23872835282116
-4.15264923880038 4.9757145829054
-5.19220604865577 3.91775932052697
-6.46345910119974 4.24386754050945
-7.32015037137456 3.26267754037183
-7.22246934886955 2.13584255539188
-7.97841676656441 1.12898244114679
-7.29836140053881 0.614476004960262
-7.56334714769974 -0.125549553472438
-7.49814738604139 -0.96228457859425
-7.81439884916586 -1.78580412620881
-7.61045291273523 -2.77198967893778
-7.62656390264148 -3.46464846888789
-7.24908818181917 -4.26473153896952
-7.65624372318165 -4.81091611017174
-6.66141639970416 -5.80296316545433
-7.26692882518356 -6.33728529347998
-6.10349871673653 -7.24015600239545
-5.12793568255865 -7.39158862346409
-3.60679895152252 -7.41454253641856
-2.27070763112831 -7.19328489310109
-2.12322856497669 -7.53514633300908
-0.945369323815717 -6.81047496062374
-0.438821940804032 -6.79812466673037
0.473496955932098 -6.81568012612148
1.27020241762527 -6.75217715083389
2.26795803007807 -6.76411766819577
2.86411089240189 -6.83866005762911
4.01551763287381 -6.56243326616763
5.05992345753465 -7.27064223213436
};
\addplot [semithick, color1, forget plot]
table {%
-5 -7.5
-5.84012933370614 -6.92543667184327
-6.71304750678741 -6.41882142582505
-6.72188352289179 -5.9131427753113
-6.70444186629203 -5.30888479165999
-6.81123196671692 -4.14787470192606
-6.76085090044273 -2.9541911336301
-6.42155826991624 -1.10162338780859
-6.66181214311697 -0.330538591899205
-6.83044625038195 0.430404081877513
-6.8300541996617 1.56737523053252
-6.44156863448559 2.70255143007617
-7.00218021850361 4.20065480282234
-5.80661422513841 4.45068907516903
-4.97695681042023 5.01789848365435
-4.248363178445 4.39103029468724
-3.1091666208577 4.90985139792632
-2.0338840069074 4.30450837518303
-1.31344213785389 5.53650705657913
0.00181765767947667 4.04884878591111
0.714622847429176 4.77714456668619
1.25867988384943 4.65785315655919
1.74474202090844 3.53704223772592
1.79403421338829 2.02190484688722
2.63945822015354 1.23590596872372
0.969382925146538 0.457693007545099
1.84766616475024 -0.86022637418594
1.93036928181416 -0.501623238769469
1.92535451241375 -0.069786839202474
1.47192368488812 0.309808552353929
1.87245284104525 0.146242323762473
1.49988513504774 0.052920113897748
1.9271908983905 0.559055323323438
1.79744897059244 0.777849562261878
1.89435131185688 1.46537203733932
2.52757920476874 2.72774589620193
1.15971892081642 4.40950698766041
0.376411937201902 5.62137263138824
-0.159020279895697 4.84145716868386
-0.726247462739206 4.5216115118478
-2.27629516782701 4.63472470620377
-3.32466241521488 4.19831970386673
-4.10249522869412 4.48279441423163
-4.92131819126274 4.32004783908986
-5.58362392732795 4.97288290661063
-6.65508103496439 4.17681054668317
-6.54607769276224 2.81329977509428
-6.8200455046731 2.05608873817038
-7.41359640829009 1.12841476640302
-7.41638977677936 0.345318915428004
-7.00664900853389 -0.338013934416479
-7.4603720645151 -1.40006197376068
-7.17802511605005 -2.37672198715121
-6.55594147407843 -3.11331976205414
-6.67674644894331 -3.96900680774067
-6.52286902123885 -5.09325568471437
-5.85230898066011 -6.07544356289991
-7.0979841425917 -7.3009199123914
-6.78899095872916 -6.08370093831376
-7.86463817303176 -6.38628575144559
-6.12676701725462 -7.01164728248371
-4.82351313755204 -6.87525963738424
-3.71501930639208 -6.69840402485098
-2.68663890205401 -5.882439399075
-1.61389487936069 -6.68796098588355
0.211993842510937 -6.5483449737575
0.852803359114666 -5.93483800207756
1.56531771006068 -6.26424163348156
2.32700860792943 -6.34132214283582
3.0500526607444 -6.78932591925932
4.08944803773664 -6.52525939858792
5.10892456865832 -6.69924096162145
};
\addplot [semithick, color2, forget plot]
table {%
-5 -7.5
-5.81111263838416 -6.51188385092225
-6.17354511430421 -5.88829319648057
-6.60081163722513 -4.78490320586325
-6.40414326691786 -4.2574712484306
-7.02580281144886 -3.53195168490133
-7.13793816573018 -2.28562548466064
-7.44128192016743 -1.03567474914603
-7.97429809715035 -0.150885995197501
-7.21753954813522 0.950675733826477
-6.67924767101891 2.11716978020393
-6.85235880660272 3.21287333356118
-6.46653925909175 4.10055491289059
-5.74438261048586 4.62298609093025
-4.43756925076423 4.01700636969762
-3.32449006924516 4.80723004213742
-2.84450074633279 4.5910819771448
-1.90037700362856 4.33671058844946
-1.13331907792302 4.72798830231805
-0.11323287923395 3.7863942456379
0.697802200315906 4.85884499667058
1.73182228508984 4.16522129828461
2.63901010279044 3.90549015611807
2.41512821262768 2.82083403092985
2.08128021657227 1.58882031200477
2.37553135286139 1.16813196106178
2.18960149095755 0.201379111068425
1.3039925331498 -0.152783110190362
2.36798146387315 -0.701890480596502
0.849336706064211 -0.0528785761154442
1.96171102896828 -0.244557240911442
2.42912466420198 0.0382172491694333
2.74820514469272 0.46828319578737
2.83037731466607 1.80749280912399
3.01031917467377 2.6061314263666
2.59952133984761 3.25843126077441
1.63539564993506 4.08317144251727
0.265668558499692 4.80632203825708
-1.14933269003271 3.99612660547969
-2.18592799462248 4.94975087786634
-3.00746849464805 3.65827090534457
-4.18343898132101 3.78278226583069
-5.16148037887084 4.10171330053703
-6.13445520167829 4.50987904745016
-7.0048494848576 4.34384204603645
-7.25220592697529 3.51427458193266
-7.08899574510827 2.47547102551241
-6.28163841612755 0.907520885138995
-6.64413796563071 -0.333387951884159
-7.00076003596181 -0.85248149478927
-7.22983402806765 -1.91144221925148
-6.57392266320331 -2.73797195468553
-6.33807518845397 -3.98827629010943
-6.55635477204999 -4.6498650569132
-6.41548789763427 -5.60253947776099
-7.22833398155501 -6.263651606387
-5.99958946246143 -6.8789788051684
-4.67127595045299 -6.93417883234836
-3.67402797464444 -6.97975906977596
-2.12495617773644 -7.00154581368473
-1.84432749346578 -6.71322173317644
-0.612978924008645 -6.47002958028564
0.288750519635688 -6.05599929372206
1.26994374087558 -6.83117311622975
2.35809305037741 -6.87686532069424
3.33772350340893 -6.99881048299956
4.89832206074499 -7.42605425648014
5.16232515193653 -6.7929258144102
};
\addplot [semithick, color3, forget plot]
table {%
-7.5 5
-6.74463559564175 3.95188316714821
-5.69424750165715 4.19143420790227
-4.39961169175655 4.96394531105365
-3.93521668503138 4.65506146000734
-3.17859520149894 3.49606616871805
-2.1219277998208 3.80662393679787
-1.36267575406044 4.71529160998446
-0.0781582385609036 4.1206370785035
0.926779885397518 4.67904845849165
1.84201058769188 3.87453851685928
2.20961780320937 3.70055437461568
1.85542833365186 2.89804167053084
2.08253698543805 2.02068694480854
2.6382476437319 0.365314098279793
1.77884065288459 -0.859026306832171
1.30001899624085 0.116830824339895
2.54796487448601 -0.63288189300181
1.04922425196486 -0.0686507306236536
2.07333535963403 -0.463635753841
1.06902334447825 0.411621077688968
1.65512429121665 0.288924325725919
1.45321907271787 0.199294848457779
2.61004790453253 -0.783855213688778
1.72952614333321 0.806354065700582
2.4372791585946 0.345471135675522
2.008848048103 0.674608511804978
1.52903108119337 2.19356054753557
2.72985691687932 3.32447970083594
2.13542718245997 3.71148272881354
0.957461693577073 3.8610396251433
0.0603155186226279 5.09532150502911
-0.739341724565466 4.14249558011389
-2.07511826982174 4.77116638268834
-3.03310378589162 4.06800925143275
-4.17033987366244 4.49014299301983
-5.13118140417847 5.53646254226779
-6.06182396304619 4.74098832897298
-6.78849837555293 4.3136258725752
-7.11216732716559 3.38517357004008
-6.91554026547895 2.13033869892301
-6.87092702642259 1.58222448079653
-6.77597573541171 0.217301831776088
-6.27285064801165 -0.724693275493417
-7.15769814355542 -2.05046120285025
-6.47855619466653 -3.19302722848328
-6.79597453541647 -3.67376681653085
-6.12412892523475 -4.90286596943291
-7.02731497700423 -5.22186350105818
-7.05033525064934 -5.96813653356517
-6.24834503820598 -7.04185506765146
-5.75501547626069 -7.33610288015478
-4.75459164783927 -7.62489645622807
-3.46119956546395 -7.36782883177934
-2.4851124177139 -7.40990717743265
-1.4377039798289 -7.21682113585908
-0.105603149424742 -6.86486994738647
0.577076588019799 -6.32223419274508
1.78667434921539 -6.69065535378963
2.33853629607931 -6.68805092048894
3.23279549950386 -6.89451792205826
4.13225757555679 -7.58392738722365
5.13518439071003 -6.4596540963078
};
\addplot [semithick, color4, forget plot]
table {%
-7.5 5
-6.4603749012348 4.18664305492565
-5.72441677599443 4.99714245198419
-4.88716727729506 3.95761872733691
-4.30049829903454 5.04340391398851
-3.09361179798788 3.81385020783656
-2.64321988355183 4.31020065220511
-2.00097706520604 5.12510472383086
-1.10973063144653 4.07854878717932
0.0842301225277867 4.29978319775576
0.605409996339145 5.05575785647307
1.70061712361949 5.31477925526222
2.96592865962762 4.27329237983037
2.63239390633052 3.17427363097679
2.35715913243922 2.3259704872897
2.66632063782432 1.09872431048679
1.37393972059131 0.295134006071694
1.77919813251736 -0.525838852029437
2.3281647184489 -0.395394762369784
2.33089122057805 -0.39305364740638
2.09690237057036 -0.327271931908289
1.82475406536846 0.0507665756676293
1.28738030416497 0.352062659968299
1.64285875677884 -0.258911976872918
1.58263426484886 0.8098801331224
1.64918213265129 -0.169657060998296
1.89462030222877 0.239028177187348
1.00942341317431 0.294876040777294
1.55736198950723 -0.0843939489618645
1.48796355957744 0.442766319639362
2.1910676508064 -0.197086457417615
2.20855075417399 -0.00938494460444486
1.98850277846896 0.379285907301891
2.1217336638497 1.2838956441658
2.24951189481548 2.47825576218493
1.95057216703009 3.057796178065
1.65818387605784 4.52691992302254
1.5584260576319 4.66583721508928
0.303113459356849 4.13642202826281
-0.744332320900273 4.60578959285566
-1.78742231538189 3.85878401238971
-2.62206409516617 4.31506233731095
-3.58194366189275 5.17100453669279
-3.94782398057777 4.81042591633574
-5.2121938097372 3.97927581612343
-6.05534008400325 4.46089738859592
-6.24184702081685 4.2560258387873
-6.73290323105036 4.57522202682257
-7.37114343169373 4.16154270267116
-7.71415716742216 3.09107325678863
-7.59914426343888 2.13054127700241
-7.5342545793504 1.37297487240633
-7.3446953989517 0.431012605034169
-7.43744984698374 0.00993802092556173
-7.46986863394679 -0.940599111385764
-7.5881720142019 -1.81455180366302
-7.55320755865961 -2.49754783706171
-7.59371714675613 -2.73097078539752
-7.34876479185391 -4.01062099367631
-7.45022244784557 -4.84221983907598
-6.66778098059149 -6.13660174303733
-6.76986064220239 -6.43650831415487
-6.87920748923285 -7.03613891708587
-6.95182874861858 -6.43955121716028
-6.10295963038228 -7.35362954377529
-4.96073303935471 -7.94818281834365
-4.59843624683242 -7.55803937685417
-3.24672961251032 -7.42752853416916
-1.8429306606529 -7.71957477293559
-0.630006497901258 -7.56520352794944
0.138221253929943 -7.20217330550251
0.92246571693787 -7.06554754848934
1.80087042643431 -6.57460078254315
2.69982250205122 -6.62259746634541
3.44784199649292 -7.11490313147
4.3820361146581 -6.95863869345971
5.44511826170026 -7.59603594532643
};
\addplot [semithick, color5, forget plot]
table {%
-7.5 5
-6.92678954994884 4.49616798980412
-6.28264615301113 4.19969391353782
-5.29499842429107 4.77683886665155
-3.75341132959268 4.37999746697122
-2.72636748224907 4.89774155171396
-1.90547261525259 3.85689737171899
-1.21037685519155 4.40407487675968
-0.211005722495792 4.81416552986896
0.798171310796987 4.02235112243081
1.55074120865706 3.71767083396578
1.68957874556413 2.8651469686333
2.01815044830146 1.80687304563749
1.64757605389937 0.732000501205731
1.36332178211518 0.005344686193966
1.44148026656345 -0.797952580734112
1.92178336593101 -0.512600438991277
1.72882637412124 -0.724840194308196
2.00646307181718 -0.197569788122411
2.49279940333245 0.118771002055007
2.55510435488189 0.45424922365531
2.71440417289243 1.64370540890832
2.51219894986172 3.12992687655054
1.08811714168051 3.98756896841081
0.0350497447565501 4.21066016097287
-0.99622039440671 5.04766329893153
-2.21453244600816 4.71561700234495
-2.90613852288815 4.85179732397313
-3.65513663629349 4.50263137170439
-5.15817501415609 4.46419700017043
-6.14309238073566 3.8997926358729
-6.79331890066815 3.40787348288724
-6.51381059064515 2.48330688816867
-6.32391474826785 1.65087773724045
-7.03085870840177 1.05700680903851
-7.56309291536782 0.490679313695385
-7.73374300783749 -0.415924842588581
-7.53609780649823 -1.62560924014788
-7.81500562625423 -2.53660976107699
-7.65889262005478 -3.72694205940076
-7.01092083699774 -4.63328402395801
-6.95264234062265 -5.4499418813739
-6.88999955143445 -6.30591739713866
-5.92610608005839 -7.09941076970729
-4.58884182517642 -7.01773757575272
-3.74163789320621 -7.08920406508348
-2.23858069449827 -7.44489276341046
-1.38852662484378 -6.54513798644081
-0.475904613908514 -6.89317172817766
0.287842967227999 -6.64886150386407
1.32250824361619 -6.43066998934701
2.51271211427401 -6.93596632731583
3.62674798582713 -6.85319507068882
3.72469654841074 -7.27367213703806
4.76910079293748 -7.58855380875816
5.20037595749851 -6.57742743237101
};
\end{axis}

\end{tikzpicture}
 \end{minipage}  
  \begin{minipage}[b]{\columnwidth}
\includegraphics[width = \columnwidth]{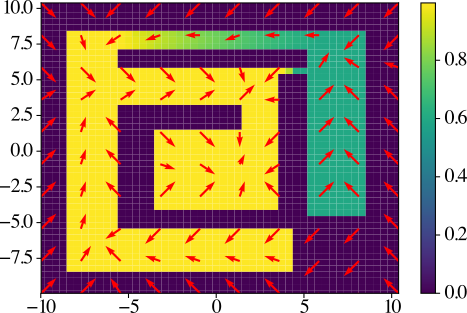} \\[.7em] \mbox{ }
  \end{minipage}	
  \\[-2em]
 	\caption{Left: Environment of the robot with obstacles ({\color{color0} $\bullet$}), a package ({\color{color1} $\bullet$}), and a client collection point  ({\color{color2} $\bullet$}). Closed-loop executions of robot fulfils the specification $\psi$ in \eqref{eq:spec}. Right: Robust probabilities computed for the abstract model.}
	\label{fig:regions}
\end{figure*}
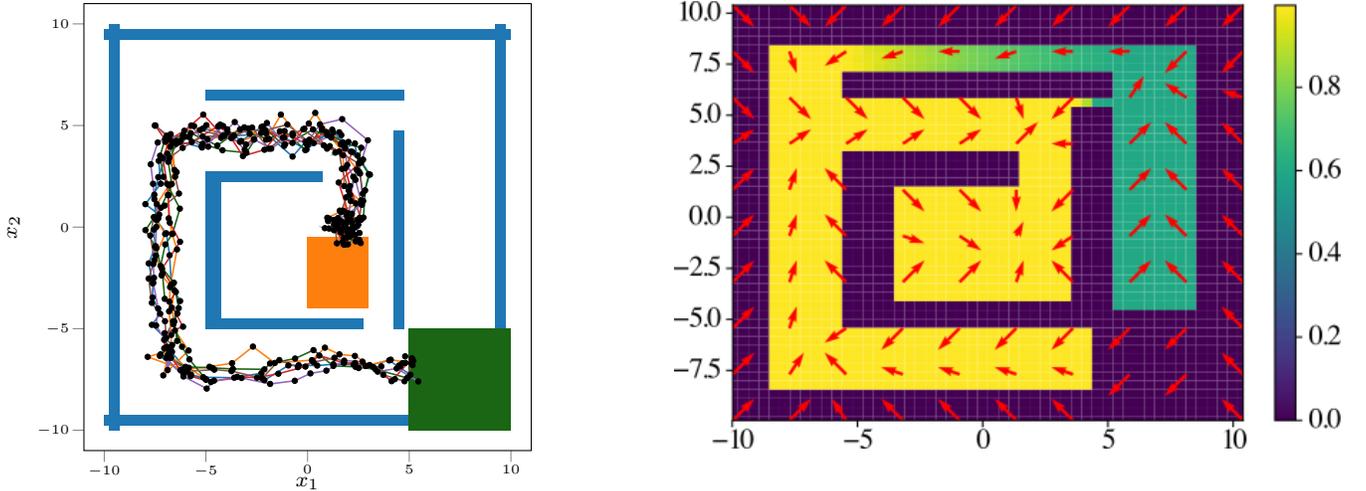

\smallskip
\noindent\textbf{Robot example.}
As a second example, we consider the model
\begin{equation*}
\begin{cases}
x_{t+1} =x_t +u_t + w_t,\quad w_t\sim \mathcal N(0,0.1\mathbb I_2)\\
y_t = x_t, \qquad\qquad x_t\in[-10,10]^2,\,\, u_t\in[-1,1]^2.
\end{cases}
\end{equation*}
We choose the specification
\begin{equation}
\label{eq:spec}
	\psi:= \left( (\neg \textsf{obs} \wedge \neg \textsf{col} ) \until \textsf{pac} \right) \wedge (\neg \textsf{obs} \until \textsf{col}),
\end{equation}
for which the atomic propositions $\textsf{obs}, \textsf{pac}, \textsf{col}$ refer
respectively to \textsf{obstacles}, a \textsf{package}, and a client \textsf{collection point}, and are depicted in Fig.~\ref{fig:regions} in
blue, orange (middle), and green (bottom right) regions.
We want to evaluate the probability that the robot can go and pick up the package and bring it to the collection point for the client, without running into any obstacle. 
\\
We abstract the model without order reduction ($P = \mathbb I_2$) and with space discretisation $\boldsymbol{\delta} = [0.41576,0.4326]^T$.
For bisimulation relation we choose precision $\epsilon=0.6$, $\delta= 0$.
The input space is partitioned into $49$ squares.
The control refinement $u = \hat u + (\hat x -x)$ fully compensates for the incurred errors in the previous step. 
Closed-loop executions of the robot with the synthesised robust controller is simulated thrice for initial states $x_0 = [-5,-7.5]^T$ and $x_0 = [-7.5,5]^T$.
In all cases, the robot fulfils the task expressed via $\psi$ in \eqref{eq:spec}.
Robust probability of satisfying the specification is computed based on the abstract model and plotted on the right in Fig.~\ref{fig:regions} as a function of initial state of the robot.
The robot starting from right-side passage has smaller probabilities of satisfying $\psi$ because it needs to move in the upper passage that is narrower, and thus increases the probability of hitting the obstacles.


\section{Conclusions and future work}
In this paper, we have introduced a new robust way of synthesising control strategies and verifying probabilistic temporal logic properties. Beyond this theoretical contribution, future work will focus on the computational aspects of this approach to prepare for application on more realistic problems.




\section*{Acknowledgement}
The authors would like to acknowledge Alessandro Abate for his contributions  to the preceding article presented at the ADHS conference \cite{DBLP:conf/adhs/HaesaertSA18}.

\bibliographystyle{abbrv}
\bibliography{library}

\appendices


 \section{Proofs}
 \subsection{Proof of Lemma \ref{lem:rob_op}}
 
 \noindent{\bfseries Proof of  inequalities \eqref{eq:mono} and \eqref{eq:mono_rob}.}
 First we show that for any two functions $V$ and $W$ satisfying $V(\hat x,q)\geq W(\hat x, q)$, the inequality is preserved for operators  $\mathbf T^{\pol{}}_\delta$ and $\mathbf T^*_\delta$, that is,  \eqref{eq:mono} and \eqref{eq:mono_rob} hold.
These operators a composed of function $\max(\cdot,a)$, $a\in\mathbb R$, conditional expectation, subtraction by the constant $\delta$, the truncation function $\Lim(\cdot)$, and $\sup_{\pol{}}(\cdot)$; each of these preserve the inequality. Thus this shows that \eqref{eq:mono} and \eqref{eq:mono_rob} hold. 

\noindent{\bfseries Proof of convergence in \eqref{eq:sat_rob_sol} and \eqref{eq:sat_rob_sol_opt}.}
Consider $\mathbf T^{\pol{}}_\delta(V_0)(\hat x,q)\ge 0 = V_0(\hat x,q)$, by induction on  \eqref{eq:mono}   it follows that 
the series \(\{(\mathbf T^{\pol{}}_\delta)^{l}(V_0)\}_{l\ge 0}\) with $V_0 = 0$ is point-wise monotonically increasing. Since it is also  bounded by one, this ensures point-wise convergence of the series to a function $V^{\pol{}}_\infty$. The same reasoning holds for the series \(\{(\mathbf T^\ast_\delta)^{l}(V_0)\}_{l\ge 0}\).

\noindent{\bfseries Proof of existence of a solution for \eqref{eq:fixed_point}.} We show that $V^{\pol{}}_\infty$ defined in \eqref{eq:sat_rob_sol} is a solution for the fixed-point equation:
\begin{align*}
\mathbf T^{\pol{}}_{\delta} (V_{\infty}^{\pol{}})(\hat x,q) & = 
\Lim\left(\mathbf T^{\pol{}} \left(\lim_{l\rightarrow\infty} (\mathbf T^{\pol{}}_\delta)^{l}(V_0)\right)(\hat x ,q)-\delta\right)\\
&  = \lim_{l\rightarrow\infty}\Lim\left(\mathbf T^{\pol{}} \left( (\mathbf T^{\pol{}}_\delta)^{l}(V_0)\right)(\hat x ,q)-\delta\right)\\
& = \lim_{l\rightarrow\infty} (\mathbf T^{\pol{}}_\delta)^{l+1}(V_0)
= V_{\infty}^{\pol{}}(\hat x,q),
\end{align*}
where we have used the definition of $\mathbf T^{\pol{}}_{\delta}$ and the bounded convergence theorem for interchanging the limit and the expectation operators.

In order to prove {\bfseries uniqueness of the solution} of the fixed-point equation \eqref{eq:fixed_point}, we need the notion of absorbing sets as in \eqref{eq:absorbing} and two propositions presented next.
%

Next proposition provides an upper bound on the series \(\{(\mathbf T^{\pol{}}_\delta)^{l}(V)\}_{l\ge 0}\) over an absorbing set $I$.
 \begin{proposition}
       \label{prop:V_upper_bound}
       Any function $V: \hat \X\times Q\rightarrow [0,1]$ satisfies the following inequality on the absorbing set $I$, 
\begin{align}
	0\leq (\mathbf T_\delta^{\pol{}})^l (V)(\hat x,q)\leq \mathbf L (\|V\|_{\infty}-l\delta),\,\,\, \forall (\hat x,q)\in I,\label{eq:inv_set_cvg}
\end{align}
which implies that $ (\mathbf T_\delta^{\pol{}})^l (V)(\hat x,q)=0$ for all
$(\hat x,q)\in I$ and any $l\geq 1/\delta$.
 \end{proposition}
 
 \begin{IEEEproof}
 The proof is based on induction. The inequality \eqref{eq:inv_set_cvg} holds for $l=1$. Take any $(\hat x,q)\in I$ to get
\begin{align}
	&(\mathbf T_\delta^{\pol{}}) (V)(\hat x,q)\notag\\
	&= \mathbf L\left(\int_{\X\times Q} 
	\!\!\!\!\max\{
	\ind{F}{q'}, V(\hat x', q')\} \bar\Tr(d\hat x'\times q'|\hat  x,q,\mu )-\delta\right)\notag\\
	&=\mathbf L\bigg(\int_{\X\times Q} 
	\left[\ind{F}{q'}+\ind{Q\setminus F}{q'}V(\hat x', q')\right] \notag \\ &\hspace{3cm}\times \bar\Tr(d\hat x'\times q'|\hat  x,q,\mu )- \delta\bigg) \notag\\
	&=\mathbf L\left(\int_{I}  V(\hat x', q') \bar\Tr(d\hat x'\times q'|\hat  x,q,\mu )- \delta\right)\notag\\&\leq  \mathbf L (\|V\|_{\infty}-\delta).
	\label{eq:induction1}
\end{align}
Now suppose \eqref{eq:inv_set_cvg} holds for an $l$. We have the following for any $(\hat x,q)\in I$:
\begin{align*}
		&(\mathbf T_\delta^{\pol{}})^{l+1} (V)(\hat x,q) \leq
		\mathbf T_\delta^{\pol{}} \left[\mathbf L (\|V\|_{\infty}-l\delta)\right](\hat x,q)\notag \\
		&\quad\le \mathbf L\big(\mathbf L (\|V\|_{\infty}-l\delta) - \delta\big)  = \mathbf L\big(\|V\|_{\infty}-(l+1)\delta\big).\notag
\end{align*}
The first inequality holds due to the monotonicity of the operator $\mathbf T_\delta^{\pol{}}$. For the second inequality, we have applied \eqref{eq:induction1} to the function $\mathbf L (\|V\|_{\infty}-l\delta)$.
Then \eqref{eq:inv_set_cvg} also holds for $l+1$. This completes the proof.
 \end{IEEEproof}
 
 We now introduce the {\bfseries transient states} and use this set to prove the contractivity of the fixed point.
 \begin{proposition}
 \label{prop:transient}
 Let $I$ be the largest absorbing set with property \eqref{eq:absorbing}. The set of states $R:=(\hat \X\times (Q\setminus F))\setminus I$, includes all the \emph{transient} states of the product gMDP $ \M\otimes\mathcal{A}_\psi$, that is   for all $(\hat x,q)\in R$:
  \begin{equation*}
  \mathbb P_{\pols\times (  \M\otimes \mathcal{A}_\psi)}\left[( \xh{t},q_{t})\in R, \  \forall t\geq 0 |\xh{0}=\hat x, q_0=q\right]=0.
  \label{eq:setR_prob}
  \end{equation*}
 \end{proposition}
 
 In words, the state trajectories starting from $R$ will eventually leave $R$ and either go to the absorbing set $I$ or to the set of satisfying states $\hat \X\times F$.

\noindent{\bfseries Proof of uniqueness.} 
We show that for any function $V: \X\times Q\rightarrow [0,1]$, it holds that
\begin{equation*}
\lim_{l\rightarrow\infty }(\mathbf T^{\pol{}}_\delta)^{l}(V)(\hat x,q) = V^{\pol{}}_\infty(\hat x,q),
\end{equation*}
for all $(\hat x,q)\in \hat \X\times Q$ with $V^{\pol{}}_\infty$ defined in \eqref{eq:sat_rob_sol}.
%
Partition $\hat \X\times Q$ into $\hat \X\times F$, $I$, and $R$ which are respectively the set of target states for the property, the largest absorbing set, and the set of transient states defined in Prop.~\ref{prop:transient}.  Take any $\ell\geq 1/\delta$ and define $ V_\ell:=(\mathbf T_{\delta}^{\pol{}})^{\ell}$. According to Prop.~\ref{prop:V_upper_bound}, $V_\ell=0$ on the absorbing set $I$.
For $V_0=0$ and for all $(\hat x,q)\in \hat \X\times Q$, we have
\begin{align*}
	&\mathbf T_\delta^{\pol{}} (V_\ell)(\hat x,q) \\
	&= \mathbf L\big(\int_{\X\times Q} 
	\left[\ind{F}{q'}+\ind{Q\setminus F}{q'} (V_0+V_\ell)(\hat x', q')\right]\notag \\
	&\hspace{3cm}\times \bar\Tr(d\hat x'\times q'| \hat x,q,\mu )- \delta\big)\\
	&=\mathbf T_\delta^{\pol{}} (V_0)(\hat x,q) + \int_{\X\times Q}  \!\!\!\ind{Q\setminus F}{q'}V_\ell(\hat x', q') \bar\Tr(d\hat x'\times q'| \hat x,q,\mu ) \notag\\
		&=\mathbf T_\delta^{\pol{}} (V_0)(\hat x,q) + \int_{R}  V_\ell(\hat x', q') \bar\Tr(d\hat x'\times q'|\hat  x,q,\mu ) \notag\\
		&\leq\mathbf T_\delta^{\pol{}} (V_0)(\hat x,q) + \|V_\ell\|_{\infty}\int_{R}   \bar\Tr(d\hat x'\times q'| \hat x,q,\mu ) \notag.
\end{align*}  
Since $\mathbf T_\delta^{\pol{}} (V_\ell)(\hat x,q)=0$ for all $(\hat x,q)\in I$, we can inductively show that
\begin{align}
	&(\mathbf T_\delta^{\pol{}})^p (V_\ell)(\hat x,q) \leq  (\mathbf T_\delta^{\pol{}})^p  (V_0)(\hat x,q) \notag\\ &\hspace{1cm}+ \|V_\ell\|_{\infty} \mathbb P\left[(\xh{t},q_{t})\in R, \  \forall t\leq p |\xh{},q\right].\label{eq:seq_right}
\end{align}
According to Prop.~\ref{prop:transient}, the bounded invariance probability in the right-hand side of \eqref{eq:seq_right} converges to zero, i.e.,
$$\lim_{p\rightarrow\infty}\mathbb P\left[(\xh{t},q_{t})\in R, \  \forall t\leq p |\xh{0}=\xh{},q_0=q\right] =  0.$$  
Thus the right hand side of \eqref{eq:seq_right} converges to $V_\infty$. Additionally, $(\mathbf T_\delta^{\pol{}})^p (V_\ell)(\hat x,q)$ is trivially lower bounded by the series $ (\mathbf T_\delta^{\pol{}})^p  (V_0)(\hat x,q)$ converging to $V_\infty$. Therefore $(\mathbf T_\delta^{\pol{}})^p (V_\ell)(\hat x,q)$ will also converge to $V_\infty$.
This proves that $V_\infty$ is the unique fixed point.

\subsection{Proof of Theorem~\ref{thm:opt:mu}}
\begin{IEEEproof}
	Consider the stationary Markov policy $\mu^*$, then $V^\ast_\infty$ is the solution of the fixed point equation associated to $\mu^\ast$, that is 
	\begin{equation*}
		V^*_\infty = \mathbf T^{\mu^\ast}_\delta V^*_\infty\,\,. 
	\end{equation*}
	By Lem.~\ref{lem:rob_op}, this fixed point has a unique solution. Since $\lim_{l\rightarrow\infty}(\mathbf T^{\mu^\ast}_\delta)^l (V_0)$ is also a solution of this fixed point, it has to hold that 
\begin{equation}\notag
	V_\infty^{\ast} = 
\lim_{l\rightarrow\infty}(\mathbf T^{\mu^\ast}_\delta)^l (V_0)\mbox{ with }V_0=0,\end{equation}
which proves the theorem.
\end{IEEEproof}
\subsection{Proof of Theorem~\ref{thm:delreach_infty}}
We  need to show that $(0,\delta)$-robust satisfaction holds according to Def.~\ref{def:epdelrob}.
In order to prove this, we present a lemma which states that
the robust value functions computed on $\widehat\M$ give a lower bound on the value functions computed on the coupled system $\widehat \M||_{\rel} \M$.
\begin{lemma}
\label{lem:preserv_ineq}
Suppose $\widehat\M\preceq^{\delta}_0\M$ with simulation relation $\rel$ and a mapping 
 $\mu:\hat\X\times Q\rightarrow \hat \A$
is given.
Let $V(\hat x, q)\leq V_{||}(\hat x, x, q)$ for all $(\hat x, x)\in \rel$. Then
\begin{equation}
\label{eq:preserv_ineq}
\mathbf T^{\pol{}}_{\delta} (V)(\hat x,q) \leq \mathbf T^{\pol{}} (V_{||})(\hat x,x,q),\quad \forall(\hat x, x)\in \rel,
\end{equation}
where $\mathbf T^{\pol{}}_{\delta}$ is the robust operator \eqref{eq:kronmap} with respect to stochastic transitions of $\widehat\M$ and $\mathbf T^{\pol{}}$ is the Bellman operator \eqref{eq:T_op} with respect to stochastic transitions of $\widehat\M ||_{\rel}\M$.
\end{lemma}

\begin{IEEEproof}
We start from $\mathbf T^{\pol{}} (V)$, which is
	\begin{align*}
\mathbf T^{\pol{}} (V)(\hat x,q) & = \int_{\hat\X}\sum_{q'\in Q} \max\{\mathbf 1_F(q'), V(\hat x',q')\}\\
& \hspace{3cm}\times
\bar \Tr(q'\times d\hat x'|\hat x,q,\mu).
\end{align*}
 For $(\hat x, x)\in \rel$ and the policy $\pol{}:\hat\X\times Q\rightarrow \hat\A$  applied to the lifted transition kernel of the composed system $\widehat\M ||_{\rel} \M$,
condition ({\bfseries L1.}) in Def.~\ref{def:del_lifting} gives the equivalent integral
\begin{equation*}
\int_{\hat\X\times \X}  \!\!\! \max\{\mathbf 1_F(\hat q^+), V(\hat x',\hat q^+)\}
			\Wt(d\hat{{x}}'\times d x'{\mid} \mu,\hat x,x)
 \end{equation*}
with  $\hat q^+=\tau(q,\Lab\circ \hat h(\hat x))$.
The above integral is equal to
\begin{align}
	&\int_{\rel}  \! \max\{\mathbf 1_F(\hat q^+), V(\hat x',\hat q^+)\}
			\Wt(d\hat{{x}}'\times d x'{\mid} \pol{},\hat x,x) \notag \\&\hspace{-.2cm}+\int_{(\hat\X\times \X)\setminus \rel}  \hspace{-.9cm}\max\{\mathbf 1_F(\hat q^+), V(\hat x',\hat q^+)\}
			\Wt(d\hat{{x}}'\times d x'{\mid} \pol{},\hat x,x)\notag\\
			&\leq \int_{\rel}  \!\! \max\{\mathbf 1_F(\hat q^+), V(\hat x',\hat q^+)\}
			\Wt(d\hat{{x}}'\!\times\! d x'{\mid} \hat u,\hat x,x)+\delta.\!\! \notag
\end{align}
The inequality holds due to $\Wt((\hat\X\times \X)\setminus \rel{\mid} \hat u,\hat x,x)\leq \delta$.
Moreover, 
it holds that $\hat q^+=q^+$ for $(\hat x',x')\in \rel$, and according to the assumption of the lemma, $V(\hat x',\hat q^+)\leq V_{||}(\hat x',x',\hat q^+)$. Therefore, the integral over $\rel$ is equal to
\begin{align}
	&\int_{\rel}  \! \max\{\mathbf 1_F(q^+), V_{||}(\hat x',x', q^+)\}
			\Wt(d\hat{{x}}'\times d x'{\mid} \hat u,\hat x,x)\notag\\
		&	\leq
		\int_{\hat \X\times \X}  \! \max\{\mathbf 1_F(q^+), V_{||}(\hat x',x', q^+)\}
			\Wt(d\hat{{x}}'\times d x'{\mid} \hat \mu,\hat x,x)\notag\\
			& = \mathbf T^{\mu} (V_{||})(\hat x,x, q).\notag
\end{align}
The left-hand side of \eqref{eq:preserv_ineq} is
\begin{align*}
    \mathbf T^{\pol{}}_\delta &(V)(\hat x,q)  = \Lim\left(\mathbf T^{\pol{}} (V)(\hat x,q)-\delta\right)\\
    & \le \Lim\left(\mathbf T^{\mu} (V_{||})(\hat x,x, q)+\delta-\delta\right)
    = \mathbf T^{\mu} (V_{||})(\hat x,x, q),
\end{align*}
since 
$\mathbf T^{\mu} (V_{||})$ only takes values in $[0,1]$.
%

We can now continue the proof of Thm.~\ref{thm:delreach_infty}.
We apply the operator $\mathbf T^{\pol{}}$ inductively to both side of \eqref{eq:preserv_ineq} and use Thm.~\ref{thm:Bellman1}, Lem.~\ref{lem:rob_op}, and equation~\eqref{eq:sat_rob} to get
\begin{equation*}
	\mathbb R^{\delta}_{\pols{}\times ( \widehat \M\otimes \mathcal{A}_\psi )} \leq \mathbb P_{\pols{}\times ((\widehat \M||_{\rel} \M )\otimes \mathcal{A}_\psi )}(\exists t\ge 0:q_t\in F).
\end{equation*}
Furthermore, we know from Prop.~\ref{prop:refine_controller} that for each $\pols{}$ we can construct a corresponding $\hat\Ca(\pols{},\psi)$ such that 
\begin{align*}
	 \mathbb P_{\pols{}\times ((\widehat \M||_{\rel} \M )\otimes \mathcal{A}_\psi )}&(\exists t\ge 0:q_t\in F)\\
	 & = \mathbb P_{\hat\Ca(\pols{},\psi)\times (\widehat \M||_{\rel} \M )}(\word\vDash\psi).
\end{align*}
Furthermore, based on Prop.~\ref{prop:lifting_prob}, we can construct a control strategy $\Ca$ such that 
\[\mathbb P_{\hat\Ca(\pols{},\psi)\times (\widehat \M||_{\rel} \M )}(\word\vDash\psi)=\mathbb P_{\Ca(\pols{},\psi)\times  \M }(\word\vDash\psi),
\]
which completes the proof.
 
\end{IEEEproof}

\subsection{Proof of Theorem \ref{thm:1sthit}}
\begin{IEEEproof}
%
%
%
%
%
%
Consider the \emph{un-truncated robust operator} defined
\begin{equation*}
\label{eq:T_unconstrained}
	\hat{\mathbf{T}}^\mu_{\delta} (V)(\hat x,q):= \mathbf{T}^\mu (V)(\hat x,q) -\mathbf 1_{Q\setminus F} (q)\delta.
\end{equation*}
Remark that the $\delta$ deviation is only subtracted over the complement of $F$. In compare with the robust operator $\mathbf{T}^\mu_{\delta}$ in \eqref{eq:kronmap}, this minor change in formulation of the $\delta$ subtraction has no impact over the domain $\X\times (Q\setminus F)$. Therefore, it follows that \[\mathbf{T}^\mu_{\delta}(V)(\hat x,q)\geq \hat{\mathbf{T}}^\mu_{\delta} (V)(\hat x,q).\]
For ease of notation, we denote
$V^l : = \big(\mathbf{T}^\mu\big)^l (V_0)$ and
$\hat{V}^l:=\big(\hat{\mathbf{T}}^\mu_{\delta}\big)^l (V_0)$ with  $V^0=\hat{V}^0=0$.
We have that 
\begin{align*}
\hat{V}^1(\hat x,q) &= 	\mathbf{T}^\mu V^0(\hat x,q) -\mathbf 1_{Q\setminus F}(q)
\delta\\
&=\mathbf{T}^\mu V^0(\hat x,q) -\delta \mathbb P\left(H^F(\hat x,q)\ge 1\right).
\end{align*}
Note that $H^F(\hat x,q)$ is zero if $q\in F$, otherwise it is positive with probability one. 
Now suppose the equality \eqref{eq:hitting_time} holds for $l$:
$$
\hat{V}^l  = V^l-	 \delta \sum_{n=1}^{l} \mathbb P\left(H^F(\hat x,q)\ge n\right).
$$
Then for $\hat{V}^{l+1}$ we get
\begin{align*}
&\hat{V}^{l+1}(\hat x,q)  = \hat{\mathbf{T}}^\mu_{\delta} (\hat{V}^l)(\hat x,q) = 
\mathbf{T}^\mu (\hat{V}^{l})(\hat x,q) -\mathbf 1_{Q\setminus F} (q)\delta\\
&=\int_{F\times \X} 1 \bar\Tr(d\hat x'\times q'|\mu,q,\hat x) -\mathbf 1_{Q\setminus F} (q)\delta
\\&\qquad	+\int_{Q\setminus F\times \X} \hat{V}^{l}(\hat x',q') \bar\Tr(d\hat x'\times q'|\mu,q,\hat x)
\\&=\mathbf{T}^\mu \left(V^l\right)(\hat x,q) -\mathbf 1_{Q\setminus F} (q)\delta
\\&-	\int_{(Q\setminus F)\times \X} \delta \sum_{{n=1}}^{l} \mathbb P\left(H^F(\hat x',q')\ge n\right)  \bar\Tr(dx',q'|\mu,x,q)
\\&=V^{l+1}(\hat x,q)-	 \delta \sum_{{n=2}}^{l+1} \mathbb P\left(H^F(\hat x,q)\ge n\right) 
\\& -\delta\mathbb P\left(H^F(\hat x,q)\ge 1\right)
\\&=V^{l+1}(\hat x,q)- \delta \sum_{{n=1}}^{l+1} \mathbb P\left(H^F(\hat x,q)\ge n\right).
\end{align*}
This proves the theorem by induction.
 \end{IEEEproof}

\subsection{ Proof of Theorem \ref{thm:delepsreach_infty}}
We present a lemma which states that
the $(\epsilon,\delta)$-robust value functions computed on $\widehat\M$ give a lower bound on the value functions computed on the coupled system $\widehat \M||_{\rel} \M$.
This will help us in proving the use of the $(\epsilon,\delta)$-robust operator. In a similar fashion to Lemma \ref{lem:preserv_ineq}, the following lemma is presented.
\begin{lemma}
\label{lem:rob_inequality}
Suppose $\widehat\M\preceq_{\eps}^{\delta} \M$ with simulation relation $\rel$ and a mapping 
 $\mu:\hat\X\times Q\rightarrow \hat \A$
is given.
Let $V(\hat x, q)\leq V_{||}(\hat x, x, q)$ for $(\hat x, x)\in \rel$, then
\begin{equation}
\mathbf T^{\mu}_{\eps,\delta} (V)(\hat x,q) \leq \mathbf T^{\mu} (V_{||})(\hat x,x,q)\quad \forall  (\hat x, x)\in \rel
\label{thm:eps_del_ineq}
\end{equation}
where $\mathbf T^{\mu}_{\eps,\delta}$ is the $(\epsilon,\delta)$-robust operator  with respect to stochastic transitions of $\widehat\M$ and $\mathbf T^{\pol{}}$ is the Bellman operator \eqref{eq:T_op} with respect to stochastic transitions of $\widehat\M ||_{\rel}\M$.
\end{lemma}
\begin{IEEEproof}[Proof of Lemma \ref{lem:rob_inequality}]
For $(\hat x, x)\in \rel$:
	\begin{align}
	&\int_{\hat\X} \min_{q'\in \bar\trans(q,\hat x')}\!\!\max\left\{\mathbf 1_{F}(q'), V(\hat x',q')\right\}   \Tr(d\hat x'|\hat x,\mu)-\delta\notag\\
&= \int_{\hat\X\times \X}  \min_{q'\in \bar\trans(q,\hat x')}\!\!\max\left\{\mathbf 1_{F}(q'), V(\hat x',q')\right\}\\ &\hspace{1cm}\times
			\Wt(d\hat{{x}}'\times d x'{\mid} \hat u,\hat x,x) -\delta.\notag
\shortintertext{
Then the above integral is equal to
}
	&\int_{\rel}  \! \min_{q'\in \bar\trans(q,\hat x')} \max\{\mathbf 1_F(q'), V(\hat x',q')\}
			\Wt(d\hat{{x}}'\times d x'{\mid} \hat u,\hat x,x) \notag \\&+\notag\int_{(\hat\X\times \X)\setminus \rel} \min_{q'\in \bar\trans(q,\hat x')}\max\{\mathbf 1_F(q'), V(\hat x',q')\}\\ &\hspace{1cm}\times
			\Wt(d\hat{{x}}'\times d x'{\mid} \hat u,\hat x,x) -\delta \notag\\
			&\leq \int_{\rel}  \! \min_{q'\in \bar\trans(q,\hat x')} \max\{\mathbf 1_F(q'), V(\hat x',q')\}\notag\\ &\hspace{3cm}\times
			\Wt(d\hat{{x}}'\times d x'{\mid} \hat u,\hat x,x) \!\!\label{eq:RXXv2}
\end{align}
For $(\hat x',x')\in \rel$, it holds that $q'=\trans(q,\Lab(h(\hat x)))\in \bar\trans(q, \hat x) $ and  $V(\hat x',q')\leq V_{||}(\hat x',x',q')$. Therefore, the integral over $\rel$ in \eqref{eq:RXXv2} is equal to
\begin{align}
	&\int_{\rel}  \! \max\{\mathbf 1_F(q'), V_{||}(\hat x',x', q')\}
			\Wt(d\hat{{x}}'\times d x'{\mid} \hat u,\hat x,x)\notag\\
		&	\leq
		\int_{\hat \X\times \X}  \! \max\{\mathbf 1_F(q'), V_{||}(\hat x',x', q')\}
			\Wt(d\hat{{x}}'\times d x'{\mid} \hat \mu,\hat x,x)\notag\\
			& = \mathbf T^{\mu} (V_{||})(\hat x,x, q)\notag
\end{align}
The truncation operator $\mathbf L$ to the $[0,1]$ interval preserves the point-wise inequality
and $\mathbf T^{\mu} (V_{||})(\hat x,x, q)$ only takes values in $[0,1]$, therefore the inequality \eqref{thm:eps_del_ineq} holds for all $ (\hat x, x)\in \rel$.
\end{IEEEproof}
Replacing Lemma \ref{lem:preserv_ineq} with Lemma \ref{lem:rob_inequality} in the proof of Theorem \ref{thm:delreach_infty} yields the proof for Theorem \ref{thm:delepsreach_infty}.

\subsection{Proof of Theorem~\ref{thm:maxprob}}
The proof of Theorem~\ref{thm:maxprob} differs substantially from the proofs of Theorems \ref{thm:delreach_infty} and \ref{thm:delepsreach_infty}. Still, we also here precede the proof with a lemma that states how (upper)-bounds on value functions are propagated over the dynamic programming recursions. 

\begin{lemma}
If $\M\preceq_{\eps}^{\delta} \widehat\M$
 and if $\hat V(\hat x, q)\geq V(x, q)$ for $(x,\hat x)\in \rel$, then
\begin{equation*}\label{eq:eps_del_ineq_opt}
\mathbf T^\ast_{-\eps,-\delta} (\hat V)(\hat x,q) \geq \mathbf T^\ast (V)(x,q)
\end{equation*}
for $(x,\hat x)\in \rel$ and for the respective stochastic transitions of $\widehat\M$ and $ \M$.
\end{lemma}

\begin{IEEEproof}
For any policy $\nu:\X\rightarrow\A$, and for $(x,\hat x) \in \rel$, consider the operator
\begin{align*}
&\mathbf T^{\nu} (V)(x,q)\\ &= \int_{\X} \ \max \left(\mathbf 1_{F}(q'), V(x', q')\right) {\Tr}(dx'|x,\nu(x,q))\nonumber
 \shortintertext{ with $q'=\tau(q,\Lab\circ  h( x))$. }
&= \int_{\X\times \hat\X} 
\max\{\mathbf 1_F(q'), V(x', q')\}\\&\qquad \times
			\Wt(d\hat{{x}}'\times d x'{\mid} \nu(x,q),\hat x,x)\notag
\shortintertext{ with the lifting $\Wt$ of $\M\preceq_{\eps}^{\delta} \widehat\M$. Since $V(x', q')\leq 1$:}
&\leq \int_{\rel} 
\max\{\mathbf 1_F(q'), V(x', q')\}\\&\qquad \times
			\Wt(d\hat{{x}}'\times d x'{\mid} \nu(x,q),\hat x,x)+\delta\notag\\
\shortintertext{ for all $(x,\hat x) \in \rel$: $V( x, q)\leq \hat V(\hat x, q)$}
&\leq \int_{\rel} 
\max\{\mathbf 1_F(q'), \hat V(\hat x', q')\}\\&\qquad \times
			\Wt(d\hat{{x}}'\times d x'{\mid} \nu(x,q),\hat x,x)+\delta\notag\\
			\shortintertext{Since $q'\in \bar\trans(q,x)$}
&\leq \int_{\X\times \hat\X}\max_{q'\in \bar\trans(q,x)}
\max\{\mathbf 1_F(q'), \hat V(\hat x', q')\}\\&\qquad \times
			\Wt(d\hat{{x}}'\times d x'{\mid} \nu(x,q),\hat x,x)+\delta\notag\\
			&= \int_{\hat\X}\max_{q'\in \bar\trans(q,x)}
\max\{\mathbf 1_F(q'), \hat V(\hat x', q')\}\\
&\qquad \times
			\hat\Tr(d\hat{{x}}'\mid \InF( \nu,x,\hat x),\hat x)+\delta\notag
\end{align*}
with $ \InF$ the interface function for  $\M\preceq_{\eps}^{\delta} \widehat\M$.
Since the truncation preserves the inequality it follows that for any given policy $\nu:\X\rightarrow\A$,
\begin{equation*}
\mathbf T^{\nu} (V)(x,q)\leq \mathbf T^\ast_{-\eps,-\delta} (\hat V)(\hat x,q)\qquad \forall ( x,\hat x)\in \rel,
\end{equation*}
and it also follows that the inequality \eqref{eq:eps_del_ineq_opt} holds.
\end{IEEEproof}

\begin{IEEEproof}[\textbf{Proof of Theorem~\ref{thm:maxprob}}]
The satisfaction probabilities are computed iteratively with
\begin{equation*}
V_{l+1}^\ast = \mathbf T^\ast (V_l)
\mbox{ and }
	\hat V_{l+1}^\ast  = \mathbf T_{-\eps,-\delta}^\ast (V_l).
\end{equation*}
and initialised as $V_0* =0 $ and $\hat V_0*  =0 $.
Since $\hat V_0*\geq V_0*$ for all $(x,\hat x)\in\rel$, it holds that
\begin{align*}
	\hat V_{l+1}^\ast(\hat x,q)\geq V_{l+1}^\ast( x,q), \qquad \forall (x,\hat x)\in \rel, l>0.\end{align*}
Therefore using i.a. Prop.~\ref{prop:sup_satisfaction}, the proof follows \emph{mutatis mutandis} the proof of Thm.~\ref{thm:delreach_infty}.
%
\end{IEEEproof}

 \ifCLASSOPTIONcaptionsoff
  \newpage
\fi

\end{document}